\documentclass[12pt,letterpaper]{article}

\usepackage{graphicx,array}
\usepackage{url}
\usepackage{color}
\usepackage{latexsym}
\usepackage{amsthm}
\usepackage{amsmath}
\usepackage{amsfonts}
\usepackage{amssymb}
\usepackage{mathtools}
\usepackage{bbold}
\usepackage{amsmath,amssymb}
\usepackage[numbers,sort&compress]{natbib}
\usepackage{slashed}
\usepackage{mathrsfs}
\usepackage{enumerate}
\usepackage{tikz}
\usepackage{mdframed}
\usepackage{setspace}  
\usepackage{esvect}
\usepackage{esint}

\usepackage{subcaption}
\usepackage[normalem]{ulem}
\usepackage{setspace}
\usepackage{multirow,makecell}

\usepackage[utf8x]{inputenc}%
\usepackage{tcolorbox}%

\usepackage{hyperref} 

\hypersetup{
    colorlinks=true,       
    linkcolor=red,          
    citecolor=blue,        
    filecolor=magenta,   
    urlcolor=blue           
}
\usepackage[all]{hypcap} 

\usepackage{natbib}
\newcommand{\nc}{\newcommand}

\nc{\beq}{\begin{equation}}  
\nc{\eeq}{\end{equation}}  
\nc{\beqa}{\begin{eqnarray}}  
\nc{\eeqa}{\end{eqnarray}}  
\nc{\bit}{\begin{itemize}}  
\nc{\eit}{\end{itemize}}

\usepackage{floatrow}
\newfloatcommand{capbtabbox}{table}[][\FBwidth]

\usepackage{blindtext}

\newcommand{\be}{\begin{equation}}
\newcommand{\ee}{\end{equation}}          
\newcommand{\bea}{\begin{eqnarray}}
\newcommand{\eea}{\end{eqnarray}}
\newcommand{\bc}{\begin{center}}
\newcommand{\ec}{\end{center}}

\def\cale{{\mathcal{E}}}

\def\PT{\textsc{Pythia}8}

\newcommand{\ab}[1]{\langle #1 \rangle}
\newcommand{\sq}[1]{[ #1 ]}
\newcommand{\la}[1]{\langle #1 |}

\newcommand{\ra}[1]{| #1 \rangle }

\def\eps{\epsilon}

\newcommand{\ds}{\displaystyle}

\begin{document}

\begin{flushright}
CERN-TH-2025-260
\end{flushright}

\hfill

\begin{center}
	{\LARGE \bf {Energy Correlators of Spinning Sources}} 
\end{center}

\vspace{1pt}

\begin{center}
{\large Marc Riembau$^{\, a,\, b}$ and Minho Son$^{\, c}$} \\
\vspace{15pt}
 $^{a}$\textit{\small Theoretical Physics Department, CERN, 1211 Geneva 23, Switzerland} \\     
 $^{b}$\textit{\small Theoretical Particle Physics Laboratory, Institute of Physics, EPFL, Lausanne, Switzerland} \\ 
 $^{c}$\textit{\small Department of Physics, Korea Advanced Institute of Science and Technology, \\
 291 Daehak-ro, Yuseong-gu, Daejeon 34141, Republic of Korea}
\end{center}

\vspace{5pt}

\begin{abstract}
The $N$-point energy correlator measures the energy flux through $N$ detectors. 
We present a general framework that characterizes its full angular dependence in a series of \textit{spinning energy correlators}. 
These spinning correlators resurrect the angular momentum structure of both the source and the detector configuration, lost otherwise in inclusive measurements.  
We demonstrate that unitarity and energy positivity confine these correlators to a sharply bounded region, with the boundary realized by extremal correlators generated by pure spin states. 
We present a first calculation of spinning energy correlators in QCD  as well as spinning energy-charge correlators. 
Their enhanced insensitivity to infrared dynamics opens up a new set of observables that directly probe the hard part of the scattering. 
Finally, we provide generalized sum rules, extended to spinning correlators and to conserved charges beyond energy.
\end{abstract}

\thispagestyle{empty}  
\newpage  
  
\setcounter{page}{1}

\begingroup
\hypersetup{linkcolor=black,linktocpage}
\tableofcontents
\endgroup
\newpage


\section{Introduction}
Energy correlations provide a simple yet powerful way to probe the dynamics of quantum field theories. In a collider setting, they measure how energy is distributed in idealized detectors at infinity and therefore capture collective properties of the multiparticle state without the requirement of fully reconstructing the detailed S-matrix. This makes them ideal observables in QCD \cite{PhysRevD.17.2298,PhysRevLett.41.1585}, where the full multiparticle dynamics is intractable. The theoretical robustness and experimental cleanness of correlator observables have attracted recent interest, see \cite{Moult:2025nhu} for a recent review. 
However, existing analyses of energy correlators focus almost exclusively on inclusive measurements where all information about the global orientation of detectors is integrated over. When the state is generated by an operator with a spin, as it is the case in collider experiments, the inclusive correlator erases the information of the detector orientation and of the density matrix of the operator acting as a source. Consequently, the inclusive correlator is blind to the angular momentum structure of the process and to crucial physical information. 
The goal of this work is to systematically restore and classify the full angular dependence of the $N$-point energy correlators of spinning sources. In particular, our main findings are as follows:
\begin{itemize}\itemsep-0.15em 
	\item The dependence on the Euler angles is universal and fully constrained by symmetries, captured entirely in Wigner $D$-matrices. 
	\item The remaining dynamical information, that depends on the $2N-3$ internal angles $z_{ij}$, is fully encoded in a set of \textit{spinning energy correlators}
	\[ 	H^J_{h^\prime-h,m^\prime-m}(z_{ij})\,\,\, \]
	which describe the dynamics of the theory in specific angular momentum channels, with $h,h^\prime$ being the spin projections of the source, and $m,m^\prime$ being the spin projections of the detector geometry.
	\item The spinning correlators $H^J_{h^\prime-h,m^\prime-m}(z_{ij})$ obey a series of positivity constraints. These are the spin-$J$, $N$-point generalization of the 
	average null energy condition (ANEC) bounds on the parameter controlling the one-point energy correlator \cite{Hofman:2008ar}.
	\item The ratios of the spinning correlators to the inclusive one have enhanced insensitivity to infrared (IR)-dynamics. They are truly probes of the hard dynamics.
\end{itemize}
The spinning energy correlators provide a framework to fully capture the kinematical and polarization information carried by the multiparticle states. Therefore, they characterize both the source and the dynamics generating the final state.

Before starting with the body of the paper, we briefly overview the previous points and the recent activities in 
literature.  The energy correlator is equivalent to computing correlators of the energy flow operator $\cale_n$, which consists of an idealized model for a calorimeter. In theories with an S-matrix it is an operator acting on asymptotic states $|\alpha\rangle$ as
\be
\cale_n |\alpha \rangle \,=\, \sum_{i\in \alpha} E_i \delta^{(2)}(\Omega_i-\Omega_n)\,|\alpha\rangle~,
\label{eq:Enonalpha}
\ee
where $E_i$ is the energy of a particle moving through the solid angle $\Omega_i$. 
The energy flow operator $\cale_n$ admits a natural representation in terms of a light-ray operator
\be
\cale_n = \lim_{r\to \infty} \int_0^\infty dt\,r^2 n^i T_{0i}(t,r\hat{n})~,
\label{eq:calendef}
\ee
where $T^{\mu\nu}$ corresponds to the stress-energy tensor. In theories with asymptotic states, this operator reproduces the action on multiparticle states in Eq.~\ref{eq:Enonalpha}, measuring the energy flux through a direction $\vec{n}$. 
Remarkably, this representation of $\cale_n$ is well defined in a conformal field theory where it plays a central role in the light-ray and conformal collider formalisms~\cite{Sveshnikov:1995vi,Cherzor:1997ak,Korchemsky:1997sy,BELITSKY2001297,Bauer:2008dt,Hofman:2008ar,Belitsky:2013bja,Belitsky:2013xxa,Kravchuk:2018htv,Kologlu:2019mfz}. 
Therefore, energy correlators provide a common language for conformal field theory (CFT) and collider QCD studies~\cite{Belitsky:2013xxa,Komiske:2022enw,Chen:2022jhb,Chen:2024nyc,Lee:2024esz,Chen:2025rjc,Chang:2025zib,Chang:2025kgq,Chen:2025ffl}, allowing insights and techniques to interplay between them. 
On the other hand, being able to perform measurements on tracks~\cite{Chen:2020vvp,Jaarsma:2025tck}, experimental measurements of these observables have provided the best determination of the strong coupling constant using jet substructure~\cite{Chen:2023zlx,CMS:2024mlf} and an accurate reinterpretation of the ALEPH data~\cite{Electron-PositronAlliance:2025wzh,Electron-PositronAlliance:2025fhk}. 
This combination of theoretical robustness with experimental cleanness has motivated their use in a variety of phenomenological studies, which include top-quark mass measurements~\cite{Holguin:2022epo,Holguin:2023bjf}, deep inelastic scattering and future lepton-hadron facilities~\cite{Liu:2022wop,Liu:2023aqb,Liu:2024kqt,Mantysaari:2025mht,Gao:2025evv,Cao:2025icu,Huang:2025ljp,Zhu:2025qkx,Gao:2025cwy}, heavy-ion and quark-gluon plasma studies~\cite{Andres:2022ovj,Barata:2023bhh}, and electroweak jet measurements \cite{Ricci:2022htc,Alipour-fard:2025dvp}. 
This illustrates how energy correlators provide an observable ideally suited to probe a wide range of physical phenomena at colliders.

In this work we extend this program by classifying the fully differential kinematical structure of the $N$-point energy correlator. 
Continuing with the physically relevant example of the QCD vacuum being excited by the electromagnetic current, the complete information is encoded in the fully differential density matrix of the $N$-point energy correlator,
\be
\langle \cale_{n_1}\dots \cale_{n_N}\rangle_{hh^\prime} \,=\, \frac{1}{\mathcal{N}}\int d^4x\,e^{iq\cdot x} \,\langle \epsilon_{h^\prime}^*\cdot J(x)\, \cale_{n_1}\dots \cale_{n_N} \, \epsilon_h\cdot J\rangle~.
\label{eq:Npointcorr}
\ee
In general, the correlator depends on the total momentum injected $q^\mu$ as well as on the positions of the $N$ detectors. Each detector is specified by a null vector $n_i^\mu = (1,\vec{n}_i)$ with $\vec{n}^2=1$. The $N$ detectors are then fully determined by the coordinates of $N$ points on the sphere, denoted by $\vec{n}_1,\,\dots, \vec{n}_N$, amounting to $2N$ coordinates.

\begin{figure}
	\centering
	\includegraphics[width=0.40\linewidth]{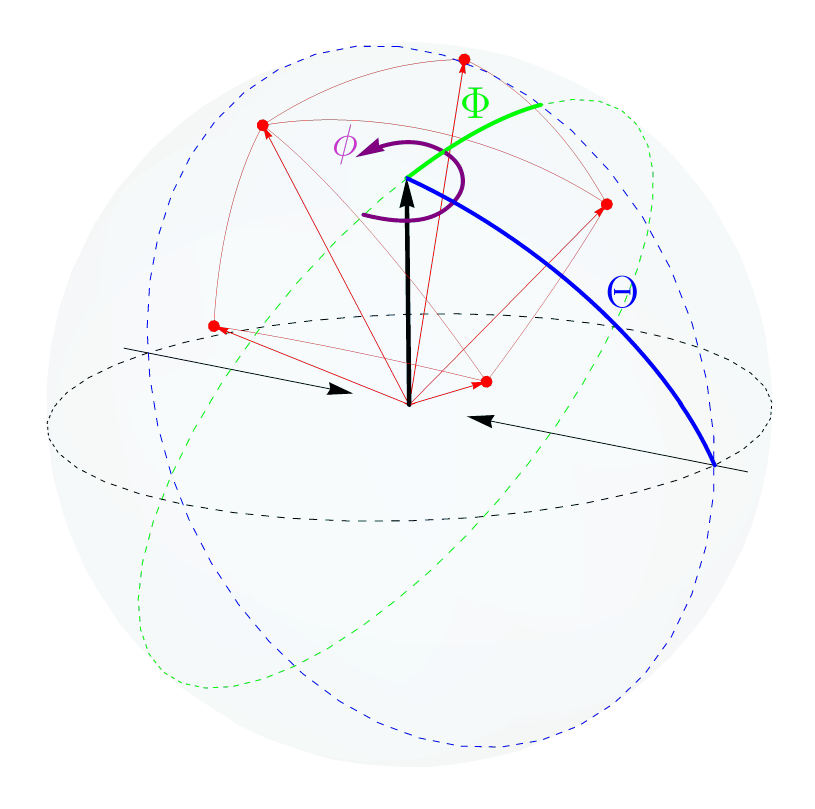}
	\caption{\small Example of the kinematics of a 5 point correlator. The $2\times 5=10$ coordinates that specify the location of the detectors can be split into $2\times 5-3=7$ internal angles fixing the configuration, and 3 Euler angles locating the \textit{rigid body} of detectors. The 7 internal angles are obtained by a triangulation of the detectors. Measuring the Euler angles requires an external coordinate system to embed the rigid body. The angle $\Theta$ and $\phi$ require an axis and an orientation within the detectors and can always be measured. The azimuthal $\Phi$ requires an external reference frame, as it is the case when the operator exciting the vacuum is boosted.}
	\label{fig:figure5pt}
\end{figure}

It is convenient to split the $2N$ coordinates in two different sets. The first set of coordinates determine the relative position of the detectors among themselves, thereby fixing the \textit{rigid body} of detectors. The minimal set of coordinates needed to completely fix the relative distances among $N$ points is $2N-3$ coordinates, as can be seen by considering a minimal triangulation of the points. 
Those independent $2N-3$ coordinates are specified by a subset of the $N^2$ angular distances $z_{ij}$, defined as
\be
z_{ij}\,\equiv\, \frac{1-\cos\theta_{ij}}{2}~,
\label{eq:zijdef}
\ee
where $\cos\theta_{ij}=\vec{n}_i\cdot \vec{n}_j$. It should be noted that, for the case with $N>3$, none of any subset 
uniquely specifies the detector configuration. The remaining 3 degrees of freedom are the $SO(3)$ Euler angles, that will be denoted by $\Phi, \Theta$ and $\phi$, and they orient the rigid body of detectors in three-dimensional space. An example for the 5-point correlator is illustrated in Fig.~\ref{fig:figure5pt}.

Keeping the information on the Euler angles reveals a rich structure of the spinning energy correlators. We will find that the density matrix of correlators has the form, schematically,
\be
\langle \cale_{n_1}\dots \cale_{n_N}\rangle_{hh^\prime}
\,=\,
H(z_{ij})\delta_{hh^\prime}\,+\,
\sum_{J,m,m^\prime} H^J_{h^\prime-h,m^\prime-m}(z_{ij}) D^J_{h^\prime-h,m^\prime-m}(\Phi,\Theta,\phi)~,
\label{eq:Npointdensitymatrix}
\ee
where $D^J_{h^\prime-h,m^\prime-m}(\Phi,\Theta,\phi)$ are the Wigner-$D$ matrices. The function $H(z_{ij})$ is the usual inclusive energy correlator considered through the literature, whereas the spinning correlators $H^J_{h^\prime-h,m^\prime-m}(z_{ij})$ are nontrivial functions of the internal angles $z_{ij}$ that specify the configuration of detectors. These spinning correlators can be observed only after being differential on the Euler angles. The structure arises from decomposing the rigid body of detectors on a polarization basis and projecting it on the operator's polarization basis. 
The behavior of the correlator on the azimuthal angle $\phi$ is characterized by the spinning structure of the detectors.  
The dependence on $\Phi$ necessitates to embed the detectors in an external coordinate frame, and can be uncovered whenever the current is injected with a boost with respect to the lab frame. For instance, this can be done by measuring the angle with respect to the scattering angle in electroweak boson production, as in \cite{Ricci:2022htc}, or by measuring the transverse momentum of the hadronic system, as in \cite{Kang:2023big,Gao:2025evv}. In all these examples, the spinning energy correlators $H^J_{h^\prime-h,m^\prime-m}(z_{ij})$ carry crucial information of the source's polarization that can be recovered with suitable observables. A discussion on characterizing the orientation of detectors in the context of the two-point correlator can be first found in \cite{Fox:1980tz}. 
The dependence on the angular distribution of the thrust axis was explored by experimental collaborations \cite{OPAL:1998tla,DELPHI:2000uri}, see also \cite{Mateu:2013gya}.
It should be noted that the collinear limit of the spinning correlators with $m^\prime-m\neq 0$ is controlled by the transverse spin part of the operator product expansion \cite{Chang:2020qpj}.

The density matrix in Eq.~\ref{eq:Npointdensitymatrix} is a density matrix of the energy flow which should be positive definite. This implies that the spinning correlators $H^J_{h^\prime-h,m^\prime-m}(z_{ij})$ are in fact constrained in terms of the inclusive one $H(z_{ij})$. This set of theoretical constraints stemming from unitarity and energy positiveness are the spin-$J$, $N$-point generalization of the ANEC bounds \cite{Hofman:2008ar,Faulkner:2016mzt,Hartman:2016lgu,Hofman:2016awc,Li:2015itl}. While the one-point energy correlator depends only on a single parameter, and therefore ANEC constraints bound the values of such parameter, the constraints on the higher point spinning energy correlators put a constraint on the functional dependence of the internal angles $z_{ij}$. We shall see how theories living in the boundaries of the parameter space are those generating free states with a definite angular momentum. It is reminiscent to the case of the one-point correlator \cite{Zhiboedov:2013opa}. 
It implies that theories with a nontrivial renormalization group (RG) flow between two free regimes might transition from one boundary to the other. 
A striking realization of such a phenomenon is QCD, as it transitions between extremal energy correlators as the theory flows from the ultraviolet (UV) to the IR, as explicitly shown in \cite{Riembau:2025wjc}. 
We will demonstrate that this transition between boundaries is a generic property which also presents in higher points.

\smallskip

The paper is structured as follows. In Section~\ref{sec:structureNpoint}, we present the generic structure of the $N$-point energy correlator for spinning sources in various aspects, including the theoretical constraints from the positivity. While our discussion is largely applicable to the center of mass frame of states, we also provide how to get the boosted correlator by appropriate transformations. Additionally, we briefly discuss the collinear limit of spinning energy operators. In Section~\ref{sec:QCD}, we not only discuss the analytical properties of the spinning two-point energy correlator and the energy-charge correlator in QCD in great detail, but also demonstrate them with our numerical simulation.  In Section~\ref{sec:sumrules}, we discuss the sum rules connecting the spinning parts of the $N$-point correlator of general detectors with the spinning $N-1$-point. 
We conclude in Section~\ref{sec:conclusions}.  In Appendix~\ref{app:sec:spinning:sumrules:QCD}, we demonstrate the sum rules by explicitly deriving one-point energy or charge correlator from various two-point correlators in QCD.

\section{Structure of the $N$-point energy correlator}
\label{sec:structureNpoint}

In this section we present the classification of the $N$-point energy correlator for spinning sources in terms of Wigner $D$-matrices. We start the discussion with the phenomenologically relevant case of a vector current in Section~\ref{sec:spinningvector}. The theoretical constraints on the spinning correlators stemming from unitarity and energy positiveness are presented in Section~\ref{sec:positivity}. In Section~\ref{sec:onepointspinning}, we recover known results for the one-point correlator of a vector current and a stress energy tensor, and we generalize it to the case of an arbitrary spin. In Section~\ref{sec:twopointspinning}, we discuss the constraints for the two-point correlator. We comment on the $N$-point generalization in Section~\ref{sec:higherpoints}, on recovering the correlator for an arbitrary boosted current from the one at rest in Section~\ref{sec:boosted}, and on the collinear limit in Section~\ref{sec:collinearspinning}.

\subsection{Spinning correlators of a vector current}
\label{sec:spinningvector}

The main object considered in this subsection is the density matrix associated with the $N$-point correlator sourced by a vector current, as written in Eq.~\ref{eq:Npointcorr}. 
The energy correlator evaluated in the state generated by a vector current is determined by the correlator
\be
H^{\mu\nu}_{\cale^{(N)}} \,=\, \int d^4x\,e^{iq\cdot x} \,\langle  J^\mu(x)\, \cale_{n_1}\dots \cale_{n_N} \, J^\nu(0) \rangle~,
\label{eq:Npointhadronictensor}
\ee
which we will refer to as the hadronic tensor of the $N$-point energy correlator, indicated by the subscript ${}_{\cale^{(N)}}$. This hadronic tensor encodes all the information of both the source and the dynamics. The projection onto an helicity basis, together with the proper normalization by the total rate, leads to the density matrix of correlators in Eq.~\ref{eq:Npointcorr}. 
It is particularly convenient to study the correlator in the center of mass frame, where the injected momentum in the operator is given by $q^\mu=(Q,\vec{0})$ and the null vectors associated with the detectors directions are given by $n^\mu_i = (1,\vec{n}_i)$. 
In the rest frame, the temporal components of the hadronic tensor vanish due to the Ward identity, and it is determined by the spatial components. 
The spatial part of the hadronic tensor can be decomposed into the identity plus a traceless symmetric tensor as
\be
H^{ab}_{\cale^{(N)}}\,=\, H^{\mathbb{1}}_{\cale^{(N)}}(z_{ij})\,\frac{\delta^{ab}}{3}\,+\, \sum_k H^{(k)}_{\cale^{(N)}}(z_{ij})\,S_{(k)}^{ab}~,
\label{eq:tensordecompositionNpoint}
\ee
where $a,b=1,\dots,3$ are spatial indices. 
The tensors $S_{(k)}^{ab}$, in the case of a vector current and energy operators, are traceless symmetric and $k$ runs over all independent tensor structures. As will become clear below, the number of independent traceless symmetric tensors depends on the number of detectors $N$. 

The crucial observation is that the functions $H^{\mathbb{1}}_{\cale^{(N)}}(z_{ij})$ and $H^{(k)}_{\cale^{(N)}}(z_{ij})$ appearing in Eq.~\ref{eq:tensordecompositionNpoint} only depend on the internal angles $z_{ij}$ among the detectors, as defined in Eq.~\ref{eq:zijdef}. Instead, all the dependence on the Euler angles is entirely contained in $S_{(k)}^{ab}$. 
In particular, this implies that integrating the hadronic tensor over the Euler angles annihilates the traceless symmetric part and one is left with 
\be
\int\frac{d\phi}{2\pi}\frac{d\Phi}{2\pi}\frac{d\cos\Theta}{2}H^{ab}_{\cale^{(N)}}=H^{\mathbb{1}}_{\cale^{(N)}}(z_{ij}) \frac{\delta^{ab}}{3}~,
\ee
which is the inclusive part of the correlator. This object is the usual inclusive energy correlator commonly 
discussed in the literature. 
Consequently, the inclusivity in the Euler angles implies that the information encoded in the \textit{spinning energy correlations} $H^{(k)}_{\cale^{(N)}}(z_{ij})$ is integrated out. To recover, classify, and understand the structure of such spinning energy correlations is the main goal of this work. 
We shall see that the spinning correlators $H^{(k)}_{\cale^{(N)}}(z_{ij})$ are direct probes of the angular momentum structure of both the source and the final state. They exhibit a nontrivial dependence on the variables $z_{ij}$ that can be reliably computed at fixed order in a perturbation theory. Moreover, the endpoint divergences are connected to the ones already contained in $H^{\mathbb{1}}_{\cale^{(N)}}(z_{ij})$, and they are subject of positivity constraints stemming from unitarity and energy positiveness.

The density matrix in Eq.~\ref{eq:Npointcorr} is obtained by projecting the hadronic tensor in Eq.~\ref{eq:tensordecompositionNpoint} onto an helicity basis,
\be
\langle \cale_{n_1}\dots \cale_{n_N}\rangle_{h^\prime h}
\,=\, \frac{1}{\mathcal{N}}\, \epsilon_{h'}^{*a} H^{ab}_{\cale^{(N)}} \epsilon_h^b~,
\ee
where $\mathcal N$ is the normalization by the total rate, given by the trace of the hadronic tensor without energy insertions $\mathcal{N} = -\frac13 \eta_{\mu\nu}H^{\mu\nu}$ with the mostly minus metric convention. 
The polarization vectors $\epsilon_h$ are defined with respect to the quantization axis used to project the spin of the current $J^\mu$. In an $e^+e^-$ collision, this axis can naturally be identified with the beam direction. 
For any two detector configurations with the same set of internal angles $z_{ij}$, it is always possible to rotate one into the other via an Euler rotation. Equivalently, given a rigid body of detectors specified by a suitable set of $2N-3$ internal angles $z_{ij}$, its orientation in space can be chosen arbitrarly through an Euler rotation $R$. The corresponding transformation of the hadronic tensor is
\be
H^{ab}_{\cale^{(N)},(\Phi^\prime,\Theta^\prime,\phi^\prime)} \,=\, R^{ac} H^{cd}_{\cale^{(N)},(\Phi,\Theta,\phi)} (R^T)^{db}~,
\label{eq:canonicalbasisNpoint}
\ee
where $T$ denotes the transpose of the matrix~\footnote{The angles of the Euler rotation are defined as in Fig.\ref{fig:figure5pt}: a $\phi$ rotation around the $\hat{z}$ axis, a $\Theta$ rotation around the $\hat{y}$ axis, and last a $\Phi$ rotation around the $\hat{z}$ axis.}.  
Equivalently, one might regard the network of detectors as fixed and instead rotate the spin axis associated to the polarization vectors $\epsilon_h$.  
This assumes that it is possible to establish a coordinate system in the rigid body of detectors based only on their geometry, and we can associate a unique set of Euler angles to each configuration. This will be further discussed in Section~\ref{sec:higherpoints}. 
Assuming this, we can identify all configurations modulo an Euler rotation. Thus, with full generality we can choose to compute the hadronic tensor in the frame defined by $\Phi=\Theta=\phi=0$ so that the hadronic tensor $H^{ab}_{\cale^{(N)}}$ depends solely on the relative angles among detectors. 
In this canonical configuration, we can further project the hadronic tensor onto a polarization basis aligned with the intrinsic axis of the detector frame,
\be
(H_{\cale^{(N)}})_{\lambda^\prime\lambda} \,=\, \epsilon_{\lambda^\prime}^{*a} H^{ab}_{\cale^{(N)},(0,0,0)} \epsilon_\lambda^{b}~,
\label{eq:HNpolarizedbasis}
\ee
where now $\lambda=-,+,0$ labels the spin component along this axis. 
We may naturally choose this intrinsic axis to coincide with the beam axis, so that both the detector configuration (modulo Euler rotations) and the source current are projected along the same direction. 
With this choice, the density matrix in Eq.~\ref{eq:Npointcorr} can be written as the hadronic tensor in the canonical frame projected on a spin basis, dressed by a rotation of the polarization basis given by the Wigner $D$-matrices,
\be
\langle \cale_{n_1}\dots \cale_{n_N}\rangle_{h^\prime h}\,=\, \frac{1}{\mathcal{N}}\,
D^{J=1}_{h^\prime\lambda^\prime}
(H_{\cale^{(N)}})_{\lambda^\prime\lambda}
(D^{J=1})^\dagger_{\lambda h}\,.
\label{eq:NpointcorrINpolarizedbasis}
\ee
All dependencies on the Euler angles are isolated into the Wigner $D$-matrices for $J=1$, $D^{J=1}_{h\lambda}$, given by
\footnote{Using the convention of $\epsilon_\pm =\frac{1}{\sqrt{2}} (\pm 1, i, 0)$, $\epsilon_0= (0, 0, -1)$ for the polarization, matching Eq.~\ref{eq:NpointcorrINpolarizedbasis:Wigner} to the definition for Wigner $D$-matrices in literature, which are
	$d^{J=1}_{1,\pm 1}(\Theta) = \frac{1}{2} (1 \pm \cos\Theta)$ and $d^{J=1}_{0,\pm 1} = -d^{J=1}_{\pm 1,0} = \mp \frac{1}{\sqrt{2}}\sin\Theta$. We have $(D^{J=1})^\dagger_{\lambda h} (\Phi,\Theta,\phi)  =\, (D^{J=1}_{h\lambda})^* \,=\, \epsilon^{*a}_\lambda (R^T)^{ab} \epsilon^b_h$}
\be
\begin{split}
	D^{J=1}_{h\lambda} (\Phi,\Theta,\phi) &=\, e^{-i(\lambda\phi + h \Phi)} d^{J=1}_{h\lambda}(\Theta) \,=\, \epsilon^{*a}_{h} R^{ab} \epsilon^b_{\lambda}~,
\end{split}
\label{eq:NpointcorrINpolarizedbasis:Wigner}
\ee
The phase structure reflects the fact that the spin component $\lambda$,  associated with the rigid detector configuration, couples to the azimuthal angle \(\phi\), while the spin component \(h\), associated with the source current, couples to the azimuthal angle \(\Phi\). 
This becomes manifest upon using $(D^{J=1}_{h\lambda})^* = (-1)^{h-\lambda}D^{J=1}_{-h,-\lambda}$, and rewriting  Eq.~\ref{eq:NpointcorrINpolarizedbasis} as
\be
\langle \cale_{n_1}\dots \cale_{n_N}\rangle_{h^\prime h}\,=\,
\sum_{J=0}^{J=2} \sum_{\lambda,\lambda^\prime}
c^J(h,h^\prime,\lambda,\lambda^\prime)\,\,(H_{\cale^{(N)}})_{\lambda^\prime\lambda}(z_{ij})\,\,
D^{J}_{h^\prime-h,\lambda^\prime-\lambda}(\Phi,\Theta,\phi)\,,
\label{eq:NpointcorrFinalForm}
\ee
where $c^J(h,h^\prime,\lambda,\lambda^\prime)$ are the appropriate Clebsch-Gordan coefficients. 
In this form, the dependence on the Euler angles is fixed by symmetry. All non-trivial, dynamical information is encoded in the spinning correlators $(H_{\cale^{(N)}})_{\lambda^\prime\lambda}(z_{ij})$, which do depend on the internal angles $z_{ij}$.  
In this form it is clear that the two azimuthal phases $\Phi$ and $\phi$ depend on the spinning structure of the source and of the detector configuration, respectively. 
The azimuthal angle \(\phi\) encodes rotations of the rigid detector configuration around its intrinsic axis, whereas the angle \(\Phi\) probes the spin structure of the source and contributes only through interference terms in the density matrix. From a phenomenological perspective, sensitivity to \(\Phi\) requires an external reference frame. This dependence can be accessed, for example, when the current is produced with a boost relative to the laboratory frame, such as by measuring the transverse momentum of the hadronic system \cite{Kang:2023big,Bhattacharya:2025bqa}, or by correlating with the scattering plane in electroweak boson production \cite{Ricci:2022htc}. In all such cases, spinning energy correlators encode essential information that becomes accessible through observables sensitive to the full angular configuration of the detectors.

The number of independent spinning energy correlators depends on the number of detectors $N$, as we will discuss in detail in the subsequent sections. For $N=1$, corresponding to the case of the one-point energy correlator, there is a single spinning structure with $J=2$. In this case, there is no internal structure and therefore no internal angle $z_{ij}$. The correlator depends only on the Euler angles, and the associated spinning energy correlator reduces to a single coefficient determined by the dynamics and the invariant momentum transfer $q^2$. 
For $N=2$, there are two independent $J=2$ structures, that correspond to the spin projections $|m|=0,2$ of the spin of the state projected along the spin axis of the detector configuration. The coefficients controlling these structures, the spinning energy correlators, now have a nontrivial dependence on the single internal angle $z$.

Although our discussion has focused on spinning correlators sourced by a vector current, the extension to more general sources is straightforward. In that case, the sum over the total angular momentum in Eq.~\ref{eq:NpointcorrFinalForm} runs up to $J=2J_S$, with $J_S$ being the spin of the source. The rest of the discussion immediately follows.

\subsection{Positivity of spinning correlators}
\label{sec:positivity}

The spinning energy correlators $H^{(k)}_{\cale^{(N)}}(z_{ij})$ appearing in Eq.~\ref{eq:tensordecompositionNpoint} obey a set of constraints relative to the unpolarized correlator $H^{\mathbb{1}}_{\cale^{(N)}}(z_{ij})$, which follow from unitarity and positivity of the energy. 
This is readily seen by observing that $H^{ab}_{\cale^{(N)}}$ can be written in terms of the spectral density of the two-point correlator, convoluted with a positive measure. Explicitly, the hadronic tensor can be written as
\be
H^{ab}_{\cale^{(N)}}\,=\,
\sum_\alpha \sum_{i_1\dots i_N\in\alpha} (2\pi)^4\delta^{(4)}(p-p_\alpha) w_{i_1,1}\cdots w_{i_N,N} \langle 0 | J^{a\dagger } |\alpha\rangle \langle \alpha| J^b|0\rangle \ ,
\ee
where the first sum runs over intermediate multiparticle states, denoted by $\ra{\alpha}$, and the second over the particles within each state. 
The $w_{i_k,k}$ refers to the measurement of the $i_k$-th particle in the state by the $k$-th detector, which is given by $w_{i_k,k} = E_{i_k} \delta^{(2)}(\Omega_{i_k}-\Omega_{n_k})$ where $E_{i_k}$ denotes the energy of the detected hadron and $\Omega_{i_k}$ is its solid angle, and it forces to coincide with the one determined by the detector $\Omega_{n_k}$, see Eq.~\ref{eq:Enonalpha}. Given that $w_{i,j} \geq 0$ and that $\langle 0 | J^{a\dagger} |\alpha\rangle \langle \alpha| J^b|0\rangle$ is positive definite, the hadronic tensor $H^{ab}_{\cale^{(N)}}$  must be a positive definite matrix. Note that this positiveness only requires the positivity of the measure, and thus it applies not only to energy correlators but also to correlators of any positive-definite quantity, belonging to a class of positive observables.

The eigenvalues of the hadronic tensor $H^{ab}_{\cale^{(N)}}$ are invariant under Euler rotations. Consequently, the positivity constraints are independent of them.  
Eigenvalues do depend on the internal angles $z_{ij}$ though, and therefore positivity constrains the functional form of the spinning correlators in terms of these internal angles. 
To make these constraints explicit, it is convenient to decompose the correlator in a basis of traceless symmetric tensors constructed from polarization vectors along the $\hat{z}$ axis. A convenient basis is $S_0\equiv \frac{-1}{\sqrt{6}}(\epsilon^*_-\epsilon_-+\epsilon^*_+\epsilon_+-2\epsilon^*_0\epsilon_0) = \frac{1}{\sqrt{6}}(\epsilon_+\epsilon_-+\epsilon_-\epsilon_++2\epsilon_0\epsilon_0)$, $S_{\pm 1}= \frac{1}{\sqrt{2}}(\epsilon^*_\pm \epsilon_0 - \epsilon^*_0\epsilon_\mp)$ and $S_{\pm 2} = \epsilon^*_\pm \epsilon_\mp$.
In this basis, the normalized hadronic tensor can be written as
\be
 \displaystyle\frac{H_{\cale^{(N)}} }{H^{\mathbb{1}}_{\cale^{(N)}}/3} = \mathbb{1}  + \sum_{m=-2}^{m=+2} c_m(z_{ij})\, S_m~,
\label{eq:normspinning}
\ee
where the coefficients $c_m$ depend only on the internal angles $z_{ij}$ and can be identified with the spinning correlators $(H_{\cale^{(N)}})_{\lambda'\lambda}$ in Eq.~\ref{eq:HNpolarizedbasis}.  
Since the Euler-averaged hadronic tensor is proportional to $H^{\mathbb{1}}_{\cale^{(N)}}$, which is therefore strictly positive, it is natural to use this quantity to normalize all functions multiplying the traceless symmetric tensors. 
With this normalization, the coefficients $c_m(z_{ij})$ in Eq.~\ref{eq:normspinning} are ratios of the spinning correlators with respect to the inclusive one, and have order one constraints for any value of $z_{ij}$ \footnote{It should be noted that the spinning correlators are distributions, and therefore the $c_m(z_{ij})$ coefficients in Eq.~\ref{eq:normspinning} need to be carefully interpreted. As we shall see, in the case of the two-point correlator, for the regular part of the correlators in $0<z<1$ the ratio is well defined, while in the endpoints we define them as the ratio of the coefficients of the delta functions.}. 
Hermiticity of the hadronic tensor, under which $S^\dagger_{\pm 2} = S_{\mp 2}$ and $S^\dagger_{\pm 1} = - S_{\mp 1}$, implies that the coefficients must satisfy $c_2 = c_{-2}^*$ and $c_1 = - c_{-1}^*$, and thus they can be parametrized as $c_{\pm 2} = c_2 e^{\pm i\phi_2}$ and $c_{\pm 1} = i c_1 e^{\pm i\phi_1}$. 
Furthermore, by writing $\phi_1=\xi+\alpha$ and $\phi_2=2\xi+\alpha$, one finds that the eigenvalues are independent of $\xi$ and only depend on $\alpha$. This reflects the fact that $\xi$ can be rotated away by an Euler rotation along $\hat{z}$ and is therefore degenerate with the azimuthal angle $\phi$.

The positive definiteness of Eq.~\ref{eq:normspinning} constrains the five independent normalized spinning correlators $c_i(z_{ij})$ to lie within a bounded region of parameter space. 
For $\alpha=0$, the region is a cone with the vertex at negative $c_0$ and $c_2$. For $\alpha=\pi$, $c_2\to -c_2$, while $c_0$ is invariant, the cone is reflected while intermediate values of $\alpha$ merge the two regimes. The boundary of the allowed region is generated by hadronic tensors of reduced rank, corresponding to states that are pure in their spin projection along the detector axis. More generally, a co-dimension-$d$ boundary is generated by tensors of rank $3-d$. Positivity further implies that any point in the interior of the allowed region corresponds to a convex combination of such boundary configurations.

It should be noted that the constraints follow from the representation of the correlator in terms of matrix elements. Whenever such representation is available,  the bounds are automatically satisfied and do not impose additional constraints. For instance, no further bounds on effective field theory coefficients can be obtained from these whenever the EFT allows the calculation of scattering amplitudes in terms of such coefficients. Their utility becomes apparent in situations where this representation is unknown or inaccessible, such as in conformal field theories, where energy correlators are computed via null integrals of local correlators expressed in terms of CFT data~\cite{Hofman:2008ar,Hofman:2016awc,Li:2015itl,Chowdhury:2017vel,Cordova:2017zej,Meltzer:2018tnm,Hartman:2023qdn,Dempsey:2025yiv,Mecaj:2025ecl}. 
In such cases, the positivity bounds provide genuinely nontrivial information. The bounds presented here extend the known positivity constraints to the spinning parts of the $N$-point correlator. We leave the study of their implications on CFT data to future work.

\subsection{One-point correlator of spinning sources}
\label{sec:onepointspinning}

It is instructive to recover known results for the one-point correlator of a vector current in the language of the previous section. 
The one-point correlator only depends on the two angles specifying the unit vector $\vec{n}$, which corresponds to the detector direction, and there is no internal structure. 
Consequently, its functional form is entirely determined by an Euler rotation. 
Since a single vector $\vec{n}$ admits a single symmetric traceless tensor, the hadronic tensor is given by
\be
H^{ab}_\cale \,=\, \frac{H^{\mathbb{1}}_\cale}{3} \,\left( \delta^{ab} \,+\, a_\cale \left( 3n^an^b-\delta^{ab} \right) \right)~,
\label{eq:Honepointtensor}
\ee
where $a_\cale$ is a parameter that depends on the dynamics of the theory~\footnote{This normalization differs by a factor of 3 from the standard convention in the literature; it is chosen so that the resulting bounds are of order one.}. 
The coefficient $H^{\mathbb{1}}_\cale/3$ corresponds to the energy times the total rate, and together with the normalization $1/{\mathcal{N}}$ it is fixed to be the total energy of the state. Indeed, the integral over the detector direction $\vec{n}$ annihilates the traceless tensor, leaving an inclusive correlator which is just the total energy when dotted with a polarization state. 
Positivity of the hadronic tensor implies a constraint on $a_\cale$. 
Choosing the canonical orientation $n^a=(0,0,1)$, the positiveness of the matrix in Eq.~\ref{eq:Honepointtensor} implies
\be
-\frac12\leq a_\cale \leq 1~,
\label{eq:1pt:positivity}
\ee
which is the well known Hofman-Maldacena bound \cite{Hofman:2008ar}. 
Within the perspective presented in the previous section, $a_\cale$ admits a natural interpretation as the average spin projection of the state along the detector axis $\vec{n}$. 
This is manifest by decomposing the hadronic tensor in Eq.~\ref{eq:Honepointtensor} in the polarization basis,
\be
\frac{H_\cale}{H^{\mathbb{1}}_\cale/3}\,=\,
\frac{3}{2}(\epsilon_+^*\epsilon_++\epsilon_-^*\epsilon_-)
\left(1-\frac{a_\cale+1/2}{3/2} \right)\,+\, 
3\epsilon_0\epsilon_0
\left(\frac{a_\cale+1/2}{3/2} \right)\,,
\label{eq:onepointPoldecomp}
\ee
where, given the bounds on $a_\cale$, one has $0\leq \frac{a_\cale+1/2}{3/2} \leq 1$. 
This expression in Eq.~\ref{eq:onepointPoldecomp} shows that the hadronic tensor  is a convex combination of a purely longitudinal state with $\lambda=\lambda^\prime=0$, and an unpolarized state with $\lambda=-\lambda^\prime=\pm$, where the spin is projected along $\vec{n}$. 
The boundaries of the allowed region correspond to states with a definite angular momentum, either $\lambda=\lambda^\prime=0$ or $\lambda=-\lambda^\prime=\pm$. 
The upper bound $a_{\cale}=1$ is saturated by $\lambda=\lambda^\prime=0$ configurations, as obtained by a current of minimally coupled scalars $J^\mu = \phi^* \overleftrightarrow{D}^\mu \phi$ or by fermions coupling only through a dipole interaction $\mathcal{L} \supset \bar{\psi}\sigma^{\mu\nu}\psi F_{\mu\nu}$, since in the rest frame they only interpolate with two-particle states that are back-to-back and have zero spin along $\vec{n}$. Instead, the lower bound $a_{\cale}=-\frac12$ is saturated by states with $\lambda=\pm 1$, as obtained by a current of minimally coupled fermions $J^\mu = \bar{\psi}D^\mu \psi$, or by a term coupling a photon to a scalar and another vector as $\mathcal{L}\supset \phi F^{\mu\nu}\widetilde{F}_{\mu\nu}$. The $a_{\cale}=-\frac12$ boundary is also saturated by a Wess-Zumino-Witten (WZW) term, which in momentum space leads to a current coupling the photon to three pions proportional to $J^\mu\sim \epsilon^{\mu\nu\rho\sigma}p_{1\nu}p_{2\rho}p_{3\sigma}$. All these currents saturate the boundary whenever the fields interpolate free particles. That free theories live in the boundary of the parameter space was shown in \cite{Zhiboedov:2013opa}. 
Interactions push the values of $a_{\cale}$ in the interior of the region. The $a_\cale$ parameter has therefore a nontrivial flow in strongly coupled theories that transmute their degrees of freedom as they flow from the UV to the IR. QCD is a striking realization of this phenomenon \cite{Riembau:2025wjc}.

The density matrix of physical photon polarizations is recovered by multiplying the hadronic tensor by the Wigner-$d$ matrices as in Eq.~\ref{eq:NpointcorrINpolarizedbasis},
\be \label{eq:densitymatrix:onept:photonpolarization}
\ab{\cale_n}_{h^\prime h}\,=\, \frac{Q}{4\pi}
\begin{pmatrix}
	1+a_\cale \Big (\ds\frac32\sin^2\Theta-1 \Big ) & -a_\cale e^{2i\Phi } \ds\frac32\sin^2\Theta 
	\\[10pt]
	-a_\cale e^{-2i\Phi } \ds\frac32\sin^2\Theta & 	1+a_\cale \Big (\ds\frac32\sin^2\Theta-1 \Big )
\end{pmatrix}~,
\ee
where $h,h' = -, +$ are the photon physical polarizations. 
The diagonal terms are the well known form of the one-point correlator for an unpolarized beam. The off-diagonal term can only be measured by being differential in the angle $\Phi$, which can be done whenever the current injected is with a nonzero boost in the lab frame. For instance, this can be done by measuring the transverse momentum of the hadronic system, as in \cite{Kang:2023big,Gao:2025evv}, or by measuring the angle with respect to the scattering angle in electroweak boson production, as in \cite{Ricci:2022htc}.

The generalization to sources of spin-$J$ is straightforward and phenomenologically relevant, since the one-point correlator of a generic source can be decomposed into a sum of definite-spin contributions. 
For instance, the fact that $e^+e^-\to\text{hadrons}$ is captured by only $J=0$ and $J=2$ spinning correlators holds only at leading order in $\alpha_{\textit{em}}$.  More generally, the positivity constraints on spin-$J$ sources allow to set constraints on the space of CFTs, e.g. see \cite{Meltzer:2018tnm}. In the following we present a compact and systematic way to present the results. 

The density matrix for a symmetric traceless operator of spin-$J$, labeled as $\mathcal{O}_{\mu_1\dots\mu_J}$, is given by
\be
\langle \cale_{n}\rangle_{hh^\prime} \,=\, \epsilon_{h^\prime}^{*\mu_1 \cdots \mu_J}\,
\int d^4x\,e^{iq\cdot x} \,\langle \mathcal{O}_{\mu_1\dots\mu_J}(x)\, \cale_{n} \, \mathcal{O}_{\nu_1\dots\nu_J}(0)\rangle\,\epsilon_{h}^{\nu_1 \cdots \nu_J}~,
\label{eq:NpointcorrSpinJ}
\ee
where $\epsilon_h^{\nu_1 \cdots \nu_J}$ is the polarization tensor for spin-$J$ of the helicity $h$ (similarly for $\epsilon_{h'}^{*\mu_1 \cdots \mu_J}$). In the center of mass rest frame, the associated hadronic tensor with $2J$ indices can be decomposed into tensors in irreducible representations of appropriate spin quantum numbers which are constructed out of Kronecker $\delta$'s and single detector vector $\vec{n}$, as in
\be
H^{a_1\dots a_J b_1\dots b_J}_\cale \,=\, \sum_{j=0}^J H_{2j} S^{a_1\dots a_J b_1\dots b_J}_{2j,m=0}~,
\label{eq:NpointcorrSpinJ:irreps}
\ee
where $H^{a_1\dots a_J b_1\dots b_J}_\cale$ is the hadronic tensor of a spin-$J$ source and $S^{a_1\dots a_J b_1\dots b_J}_{2j,m=0}$ is a tensor in the irreducible representation of spin-$2j$ and zero azimuthal quantum number, $m=0$. The scalar coefficients $H_{2j}$ are the higher-spin analog of the $a_\cale$ parameter. 
There are total $J+1$ tensors since all tensors in the right-hand side of Eq.~\ref{eq:NpointcorrSpinJ:irreps} have $m=0$ for the one-point correlator. For a vector current, tensors from odd representations cannot appear in the energy correlator although they are in general present in the charge correlators, like in the case of the charge correlators for a chiral current  \cite{Riembau:2024tom}. 
Denoting the state for the total hadronic tensor in Eq.~\ref{eq:NpointcorrSpinJ:irreps} with $|H_\cale \rangle$, $S^{a_1\dots a_J b_1\dots b_J}_{2j,m=0}$ with $|2j,0 \rangle$, and the traceless symmetric polarization tensor for spin-$J$ with polarization $m$ $|J,m \rangle$, the contraction of the hadronic tensor with polarization tensors is realized as the matrix element $\langle Jm_1 Jm_2 | H_\cale \rangle$ which is given by a sum of Clebsch-Gordan coefficients,
\be
\epsilon_{\lambda^\prime}^{*a_1 \cdots a_J} H^{a_1\dots a_J b_1\dots b_J}_\cale \epsilon_{\lambda}^{b_1 \cdots b_J} 
=  \langle Jm_1 Jm_2 | H_\cale \rangle \,=\, \sum_{j=0}^J H_{2j} \langle Jm_1 Jm_2 | 2j,0 \rangle~,
\label{eq:NpointcorrSpinJ:irreps:ME}
\ee
where $m_1 = - m_2 =  \lambda = \lambda'$ for non-vanishing elements. Since the integral over the Euler angles only leaves the singlet, the coefficient $H_{0}$ is fixed by energy conservation to be the total energy. This leaves a total of $J$ model-dependent coefficients $H_{2j}$. In the same way $a_\cale$ was controlling the relative size between the spin-0 and spin-1 projections along $\vec{n}$, the $H_{2j}$ coefficients control the relative contribution of the different spin projections of the state along $\vec{n}$. 
The positiveness of the energy $\langle \cale \rangle \geq 0$ implies that the matrix elements in Eq.~\ref{eq:NpointcorrSpinJ:irreps:ME} should be positive for any $\lambda$. For a spinning source of spin $J$, there are $J+1$ independent positivity constraints, leading to a $J$-dimensional bounded space.

As an illustration, we can use this approach to re-derive, once again, positivity bounds already discussed for the vectorial source. For the vector current with $J=1$, the selection rule of the angular momentum allows three non-vanishing matrix elements with $m_1 = - m_2 = 0, \pm1$. There are however two distinct sum rules which are given by
\be
\begin{split}
	\frac{\langle 1010|H_\cale \rangle}{\langle 1010|00 \rangle} \frac{1}{H_0} &= 1 + \frac{H_2}{H_0} \frac{\langle 1010|20 \rangle}{\langle 1010|00 \rangle} 
	= 1 + 2 \left ( - \frac{1}{\sqrt{2}} \frac{H_2}{H_0} \right )~\geq 0\,\,\,,
	\\[5pt]
	\frac{\langle 111\text{-}1|H_\cale \rangle}{\langle 111\text{-}1|00 \rangle} \frac{1}{H_0} &= 1 + \frac{H_2}{H_0} \frac{\langle 111\text{-}1|20 \rangle}{\langle 111\text{-}1|00 \rangle} 
	= 1 - \left (- \frac{1}{\sqrt{2}} \frac{H_2}{H_0} \right )~\geq 0\,\,\,,
	\label{eq:onepointCG}
\end{split}
\ee
where the matrix element with $m_1 = - m_2 = 1$ is the same as the one with $m_1 = - m_2 = -1$. 
The normalization by $H_0$ is equivalent to the normalization by the total energy in Eq.~\ref{eq:Honepointtensor}. The two sum rules mean that observers with a polarized source should measure positive energy flux. The three non-vanishing matrix elements construct the diagonal matrix $(H_{\cale})_{\lambda'\lambda}$ in Eq.~\ref{eq:NpointcorrINpolarizedbasis}.
Renaming the ratio $H_2/H_0$ and demanding the positivity of two matrix elements successfully reproduces the bound in Eq.~\ref{eq:1pt:positivity},
\be
-\frac{1}{2} \leq -\frac{1}{\sqrt{2}}\frac{H_2}{H_0} = a_\cale \leq 1~.
\ee
This result is equivalent to the observation in Eq.~\ref{eq:onepointPoldecomp} where the bounds come from the convex sum of two states with different spin with the convexity due to the positivity of Eq.~\ref{eq:onepointCG}.

For a spinning source of $J=2$, the energy-momentum tensor, the five non-vanishing matrix elements in Eq.~\ref{eq:NpointcorrSpinJ:irreps:ME} allowed by the selection rule are matrix elements with $m_1 = - m_2 = 0, \pm 1, \pm 2$, and the three distinctive ones are given by
\be
\label{eq:matrixelement:positivity:J2}
\begin{split}
	\frac{\langle 2020|H_\cale \rangle}{\langle 2020|00 \rangle} \frac{1}{H_0} 
	&= 
	1 - \sqrt{\frac{10}{7}} \frac{H_2}{H_0} + 3\sqrt{\frac{2}{7}} \frac{H_4}{H_0}~\geq 0\,\,\,,
	\\[5pt]
	\frac{\langle 212\text{-}1|H_\cale \rangle}{\langle 212\text{-}1|00 \rangle} \frac{1}{H_0} 
	&= 
	1 -  \sqrt{\frac{5}{14}} \frac{H_2}{H_0}   - 2\sqrt{\frac{2}{7}} \frac{H_4}{H_0}  ~\geq 0\,\,\,,
	\\[5pt] 
	\frac{\langle 222\text{-}2|H_\cale \rangle}{\langle 222\text{-}2|00 \rangle} \frac{1}{H_0} 
	&= 
	1 + \sqrt{\frac{10}{7}}\frac{H_2}{H_0} + \sqrt{\frac{1}{14}}\frac{H_4}{H_0}~\geq 0\,\,\,,
\end{split}
\ee
where we made the Clebsch-Gordan coefficients explicit. 
The interpretation of these bounds is exactly the same as for the vector source. The parameter space depends on $H_2/H_0$ and $H_4/H_0$. In this plane, the bounds in Eq.~\ref{eq:matrixelement:positivity:J2} determine a triangle of allowed space. Any point inside is therefore a convex sum of states with spin projection 0, 1 and 2 along $\vec{n}$. In order to saturate the bounds one can have a pair of back-to-back scalars, opposite helicity fermions or opposite-helicity gluons, but it is clear that the specific matter content is irrelevant, and the crucial part is the spin projection of the state.

Renaming the ratios as $-3\sqrt{\frac{10}{7}}\frac{H_2}{H_0} = t_2$ and $\frac{35}{2} \sqrt{\frac{1}{14}}\frac{H_4}{H_0} = t_4$ for the comparison with the results of \cite{Hofman:2008ar} and demanding the positivity of the matrix elements in Eq.~\ref{eq:matrixelement:positivity:J2}, one obtains
\be \label{eq:onepoint:spin2PositivityT2T4}
\begin{split}
	&1 + \frac{1}{3}t_2 + \frac{12}{35}t_4 = \left ( 1 - \frac{1}{3}t_2  + \frac{2}{35} t_4 \right ) + \frac{2}{3} t_2  + \frac{2}{7} t_4  \geq 0~,
	\\[5pt]
	&1   + \frac{1}{6}t_2  - \frac{8}{35} t_4  = \left ( 1 - \frac{1}{3}t_2  + \frac{2}{35} t_4 \right ) + \frac{1}{2} t_2 - \frac{2}{7} t_4  \geq 0~,
	\\[5pt] 
	&1 - \frac{1}{3}t_2 + \frac{2}{35}t_4 \geq 0~.
\end{split}
\ee 
The explicit form of the tensor that matches to the Clebsch-Gordan coefficients is given by
\be
\label{eq:1ptcorrelator:spintwo:polarizationbasis}
\epsilon_{\lambda^\prime}^{*ab} H^{ab, cd}_\cale \epsilon_{\lambda}^{cd} 
=
Q\left [ 1+ 
t_2 \left ( \epsilon^{*ab} \epsilon^{ac} n^b n^c -\frac{1}{3} \right ) +
t_4 \left ( |\epsilon^{ab} n^a n^b|^2 - \frac{4}{7} \epsilon^{*ab} \epsilon^{ac} n^b n^c  + \frac{2}{35} \right ) \right ]~,
\ee 
where polarization indices are left implicit and each term is traceless symmetric.  
This tensor can be properly decomposed into irreducible representations~\footnote{
Explicitly, three irreducible representations are given by
$S_{2j=0,m=0}^{ab,cd} = \frac{2}{15} \left [ \frac{1}{2}\left (  \delta^{ac} \delta^{bd} +  \delta^{ad} \delta^{bc}  \right ) - \frac{1}{3}\delta^{ab} \delta^{cd} \right ]$,\\ 
$S_{2j=2,m=0}^{ab,cd} = \frac{4}{7} \Big [ \frac{1}{4}\left ( \delta^{ac} n^{b}n^d + \delta^{ad} n^{b} n^c + \delta^{bc} n^{a} n^d + \delta^{bd} n^{a} n^c \right )
- \frac{1}{3} \left ( \delta^{ab} n^{c}n^d + \delta^{cd} n^{a}n^b \right )
- \frac{1}{6} \left ( \delta^{ac} \delta^{bd} + \delta^{ad} \delta^{bc} \right )
+ \frac{2}{9}\delta^{ab} \delta^{cd} \Big ]$,
and 
$S_{2j=4,m=0}^{ab,cd}
=n^a n^b n^c n^d - \frac{1}{7} \left ( \delta^{ab} n^c n^d + \delta^{ac} n^b n^d + \delta^{ad} n^b n^c + \delta^{bc} n^a n^d
+ \delta^{bd} n^a n^c + \delta^{cd} n^a n^b \right )
 + \frac{1}{35}  \left (  \delta^{ac} \delta^{bd} +  \delta^{ad} \delta^{bc}  + \delta^{ab} \delta^{cd} \right )$~\cite{SPENCER1970475}. 
$S_{2j=0,m=0}^{ab,cd}$ and $S_{2j=2,m=0}^{ab,cd}$ match to those in~\cite{Belin:2019mnx}, whereas $S_{2j=4,m=0}^{ab,cd}$ appears different from~\cite{Belin:2019mnx}. 
}. 
This is different from the tensor decomposition found in the literature \cite{Hofman:2008ar,Cordova:2017zej,Meltzer:2018tnm,Belin:2019mnx} which can be obtained by redefining the coefficients as $t_2 - \frac{4}{7} t_4 \rightarrow t_2$ while keeping $t_4$ as in Eq.~\ref{eq:1ptcorrelator:spintwo:polarizationbasis}. It can be understood by noticing that the expression in Eq.~\ref{eq:1ptcorrelator:spintwo:polarizationbasis} can be reorganized as
\be
\epsilon_{\lambda^\prime}^{*ab} H^{ab, cd}_\cale \epsilon_{\lambda}^{cd} 
=
Q\left [ 1+ 
\left ( t_2  - \frac{4}{7} t_4 \right ) \left ( \epsilon^{*ab} \epsilon^{ac} n^b n^c -\frac{1}{3} \right ) +
t_4 \left ( |\epsilon^{ab} n^a n^b|^2 - \frac{2}{15} \right ) \right ]~.
\ee

\begin{figure}
	\centering
	\includegraphics[width=0.5\linewidth]{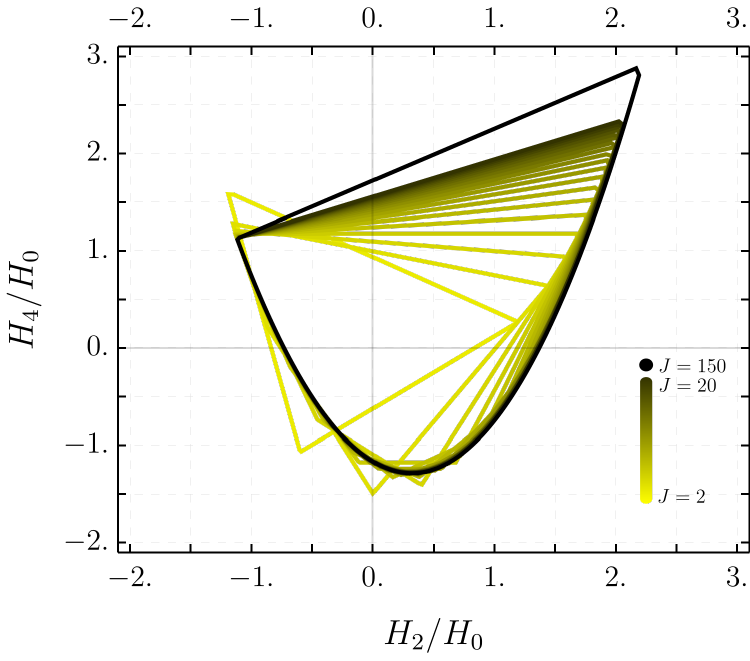}
	\caption{\small Bounds on the normalized coefficients  $H_2/H_0$ and $H_4/H_0$ of the spinning energy correlator for various values of the spin-$J$ of the source operator.}
	\label{fig:SpinJbounds}
\end{figure}

Using the representation of the dotted hadronic tensors in terms of the Clebsch-Gordan coefficients in Eq.~\ref{eq:NpointcorrSpinJ:irreps:ME}, constraints for arbitrary $J$ can be immediately derived. 
We show in Fig.~\ref{fig:SpinJbounds} the allowed parameter space for the normalized coefficients $H_2/H_0$ and $H_4/H_0$ of the spinning energy correlator for various values of the spin-$J$ of the source operator. For a given spin $J$, we consider the $J+1$ constraints,
\be
\langle Jm, J-m | H_\cale \rangle\,=\, H_0+\sum_{j=1}^J H_{2j}\,\frac{\langle J m, J -m| 2j,0\rangle}{\langle J m, J -m| 0,0\rangle}\,\geq\,0~,
\label{eq:linearconstraintsJ}
\ee
for $m=0,\dots,J$. The allowed values of $H_2/H_0$ and $H_4/H_0$ are obtained by a linear programming routine incorporating all the $J+1$ bounds in Eq.~\ref{eq:linearconstraintsJ} and finding the extremal values of $H_2/H_0$ and $H_4/H_0$ allowing $H_{2j}$ with $j\geq 2$ to vary within the constraints. Explicitly, we write $H_2/H_0$ and $H_4/H_0$ in radial coordinates and for a discrete set of angles between 0 and 2$\pi$ we find the maximal value for the radius under the $J+1$ constraints in Eq.~\ref{eq:linearconstraintsJ}. 
For $J=2$, the allowed region is a triangle which coincides with the one generated from the bounds in Eq.~\ref{eq:matrixelement:positivity:J2}, the boundaries given by the requirements that the operator with polarization $m=0,1,2$ has positive energy. At a higher spin, the $J+1$ linear constraints in Eq.~\ref{eq:linearconstraintsJ} generate a polygon with a number of edges increasing with $J$. As can be seen in Fig.~\ref{fig:SpinJbounds}, at large values of $J$, the upper bounds remain flat while the lower bounds tend to  smooth curves. The curve is actually a conic section, strongly suggesting the existence of an analytical approach to the $J\to \infty$ constraints. It should be noted that a simple linear transformation maps the bounds to those obtained from the moment problem.

\subsection{The spinning two-point energy correlator}
\label{sec:twopointspinning}

While the kinematical dependence of the one-point correlator is entirely captured by the Euler angles, the two-point correlator has a non-trivial dependence on the internal angle $z$ which is defined as
\be
z=\frac{1-\vec{n}_1\cdot\vec{n}_2}{2}~.
\ee 
The hadronic tensor of Eq.~\ref{eq:tensordecompositionNpoint} for the two-point energy correlator of a vector current can be written in terms of the two detector positions, $\vec{n}_1$ and $\vec{n}_2$. 
The most general tensor structure is given by
\be
\label{eq:tensordec2point}
\begin{split}
	H^{ab}_{\cale\cale}\,=\, H^{\mathbb{1}}_{\cale\cale}(z)\,\frac{\delta^{ab}}{3} 
	&+ H^c_{\cale\cale}(z)\left( \frac{1}{1-z}\frac{n_1^a+n_2^a}{2}\frac{n_1^b+n_2^b}{2}-\frac{\delta^{ab}}{3}\right)
	\\[5pt]
	&+ H^b_{\cale\cale}(z)\left( \frac{1}{z}\frac{n_1^a-n_2^a}{2}\frac{n_1^b-n_2^b}{2}-\frac{\delta^{ab}}{3}\right)~.
\end{split}
\ee
The singlet coefficient $H^{\mathbb{1}}_{\cale\cale}(z)$ is the quantity most commonly computed in the literature, as it is the only contribution that survives upon integrating over the Euler angles in a polarized scattering. 
The remaining two terms in Eq.~\ref{eq:tensordec2point} are traceless symmetric tensors and their coefficients $H^c_{\cale\cale}(z)$ and $H^b_{\cale\cale}(z)$ define the spinning energy correlators and can only be accessed through measurements differential on the Euler angles. 
The superscripts ${}^c$ and ${}^b$ in the spinning correlators stand for \textit{collinear} and \textit{back-to-back}, and denote the structures that survive in the $\vec{n}_1=\vec{n}_2$ and $\vec{n}_1=-\vec{n}_2$ limits, respectively. 
As will be shown below, with an appropriate choice of frame the tensor structures become independent of the relative angle $z$.  
The only two independent symmetric traceless tensor structures which are also symmetric under $n_1\leftrightarrow n_2$ are those appearing in Eq.~\ref{eq:tensordec2point}. For instance, the combination $n_1^an_1^b+n_2^an_2^b$ can be written in terms of those in Eq.~\ref{eq:tensordec2point}, and including it amounts to a redefinition of $H^c_{\cale\cale}(z)$ and $H^b_{\cale\cale}(z)$. 
While the structure of the energy-energy correlator is determined by Eq.~\ref{eq:tensordec2point}, the two-point correlator of other quantities may present a  more general structure. For instance, they can present parity-odd terms  similar to the case of one-point correlator of conserved charges of a chiral current, as shown in \cite{Riembau:2024tom} and as we will discuss in Section~\ref{sec:spinning:beyondenergy} in the context of energy-charge correlators. 

\begin{figure}
	\centering
	\includegraphics[width=0.4\linewidth]{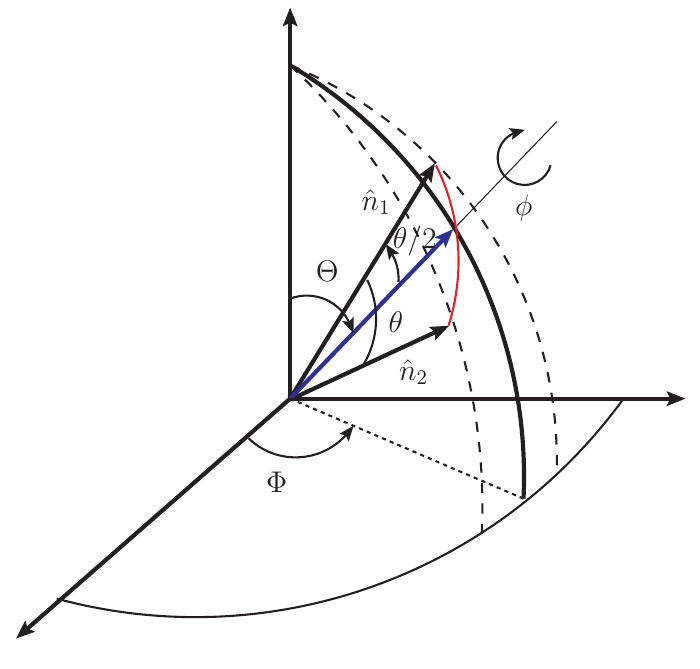}
	\caption{\small Generic configuration of the two-point energy correlator. The single rigid body angle is given by $\theta$. The remaining angles $\Theta$, $\Phi$ and $\phi$ are Euler angles.}
	\label{fig:twopointcorrdiagram}
\end{figure}

\begin{figure}
	\centering
	\includegraphics[width=0.5\linewidth]{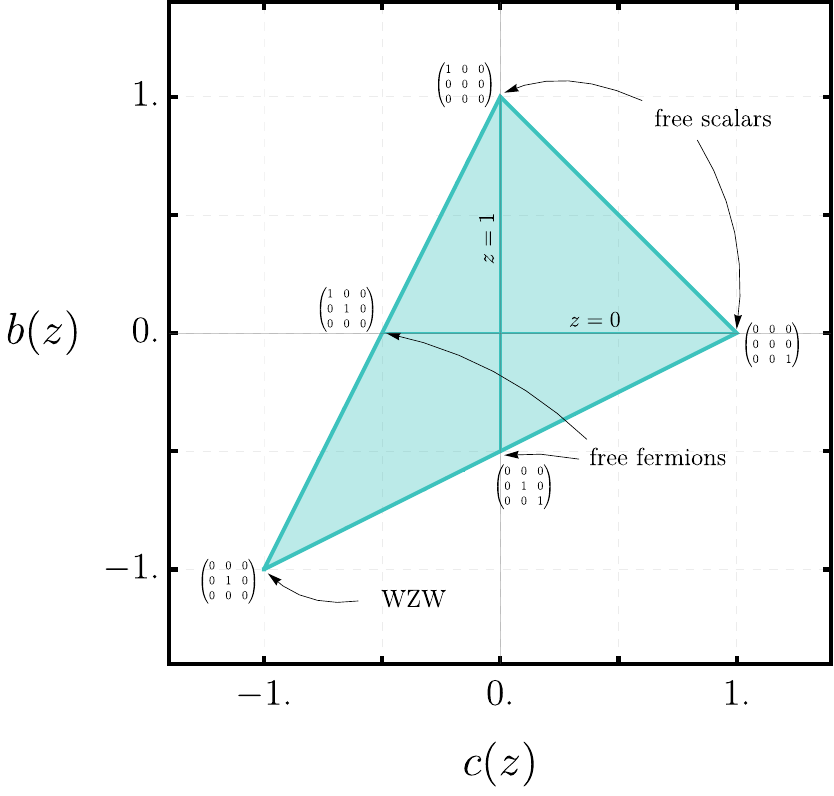}
	\caption{\small Allowed space for the $c(z)$ and $b(z)$ functions of the two-point correlator. 
		The hadronic tensor at special points are explicitly shown, together with some example of theories that generates them. In the collinear $(z\to 0)$ and back-to-back $(z\to 1)$ limits, the allowed space collapses to the lines at $b(0)=0$ and $c(1)=0$, respectively.}
	\label{fig:bdries2pt}
\end{figure}

The identification of the spinning correlators $H^c_{\cale\cale}(z)$ and $H^b_{\cale\cale}(z)$ with particular coefficients of the expansion in Wigner $D$-matrices depends on the choice of axis used to project the angular momentum of the rigid body of detectors. Different choices are related by Euler rotations and correspond to a linear combination of Wigner $D$-matrices at fixed $\ell$. 
For the case of the two-point correlator, it is particularly convenient to work in a frame where the $\hat{z}$-axis coincides with the center of mass direction of the detector configuration, given by $\frac{1}{2}(n_1^\mu+n_2^\mu)$. In this frame, the two detector directions are written as
\be
\vec{n}_1\,=\,(\sqrt{z},0,\sqrt{1-z})\,,\quad\quad
\vec{n}_2\,=\,(-\sqrt{z},0,\sqrt{1-z})~,
\label{eq:twopointcanonicalbasis}
\ee
where we chose to orient relative separation between both detectors along the $x$-axis, any other choice is related by a redefinition of the azimuthal angle $\phi$. 
With this choice of the frame, the sum and difference of the detector directions are given by $\frac{\vec{n}_1+\vec{n}_2}{2} = (0,0,\sqrt{1-z})$ and $\frac{\vec{n}_1-\vec{n}_2}{2} = (\sqrt{z},0,0)$, respectively, so that the tensors in the decomposition in Eq.~\ref{eq:tensordec2point} become independent of the internal angle $z$.  
A generic detector configuration obtained by an Euler rotation is illustrated in Fig.~\ref{fig:twopointcorrdiagram}. 
Projecting the hadronic tensor in the polarization basis gives
\be
\frac{H_{\cale\cale}}{H^{\mathbb{1}}_{\cale\cale}/3} \,=\,
(1-c+\frac12 b) H_{\pm\mp}\,+\, \frac32 b H_{\pm\pm}\,+\, (1+2c-b)H_{00}~,
\ee
where $H_{\pm\mp} = -(\epsilon_+\epsilon_-+\epsilon_-\epsilon_+)$, $H_{\pm\pm} = \epsilon_+\epsilon_++\epsilon_-\epsilon_-$, and $H_{00}=\epsilon_0\epsilon_0$. 
The functions $c(z)$ and $b(z)$ carry all non-trivial dynamical information of the spinning correlators and they are the equivalent of the $a_\cale$ parameter for the one-point correlator of a vector current. They are defined as
\be
c(z)=\frac{H^{c}_{\cale\cale}(z)}{H^{\mathbb{1}}_{\cale\cale}(z)}\,\hspace{1cm}\text{and}\hspace{1cm} 
b(z)=\frac{H^{b}_{\cale\cale}(z)}{H^{\mathbb{1}}_{\cale\cale}(z)}~.
\label{eq:candbdefinition}
\ee
In the canonical frame in Eq.~\ref{eq:twopointcanonicalbasis}, the hadronic matrix is diagonal, and the positivity constraints immediately imply the two-point correlators $c(z)$ and $b(z)$ to lie within the triangle shown in Fig.~\ref{fig:bdries2pt}, whose boundaries are given by the inequalities
\be
1+2c(z) -b(z)\geq 0~,\quad\quad
1-c(z) +2b(z)\geq 0~,\quad\quad
1-c(z) -b(z)\geq 0~.
\label{eq:constraintsoncandb}
\ee
Three vertices at $(c,b)=(1,0)$, $(0,1)$, $(-1,-1)$ correspond to the situations where the hadronic tensor is proportional to $\epsilon_z\epsilon_z$, $\epsilon_x\epsilon_x$ and $\epsilon_y\epsilon_y$, respectively. The triangle is generated by the convex sum of those three matrices.
The opposite boundary to a vertex is given by the positivity of the corresponding eigenvalue, associated to an eigenvector proportional to $\vec{n}_1+\vec{n}_2$, $\vec{n}_1-\vec{n}_2$ and $\vec{n}_1\wedge\vec{n}_2$, respectively. 
Points in the interior of the triangle arise from convex sums of these contributions, while points on a boundary of codimension $k$ correspond to hadronic tensors with $3-k$ nonvanishing components.

At the central point $c=b=0$ the hadronic matrix reduces to the identity and the dependence on the Euler angles is trivial as only the singlet component survives and the operator behaves like a scalar operator. 
For $b=0$ and $c\neq 0$, the hadronic matrix becomes equivalent to the one point decomposition in Eq.~\ref{eq:onepointPoldecomp}, with a convex sum of $H_{\pm\mp}$ and $H_{00}$, saturated at $c=-1/2$ and $c=1$. In fact, in the collinear limit $z\to 0$ it must be that $b(z)\to 0$ as the structure of the one-point correlator is recovered. 
In the strict $z\to 0$ limit and in gapped theories, the two-point correlator becomes equivalent to the one-point correlator of the energy-squared operator $\langle \cale^2\rangle$, which obeys the same tensor decomposition and positivity properties as $\langle\cale\rangle$, parameterized this time by a coefficient $a_{\cale^2}$. 
A nonzero $b(z)$ controls the size of $H_{\pm\pm}$, which by the positivity must be, in absolute value, smaller than the $H_{\pm\mp}$ contribution.

In the back-to-back limit $z\to 1$, one must instead have that $c(z)\to 0$ as the symmetric traceless part of the tensor should be proportional to $\text{diag}(2,-1,-1)$. 
In this regime, the hadronic tensor is a convex sum of the tensors along $\epsilon_x\epsilon_x$, corresponding to the direction of the back-to-back detectors (see Eq.\ref{eq:twopointcanonicalbasis}), and the tensor $\epsilon_y\epsilon_y+\epsilon_z\epsilon_z$. For $c(z)=0$, each structure dominates at the boundaries at $b(z) = 1$ and $b(z)=-1/2$, respectively. 

Finally, the vertex at $c(z)=b(z)=-1$ corresponds to a tensor proportional to $(n_1\wedge n_2)^a(n_1\wedge n_2)^b$ direction. This configuration is realized by a Wess-Zumino-Witten current, for which in the momentum space the photon couples to three pions as $\epsilon^{\mu\nu\rho\sigma}p_1^\nu p_2^\rho p_3^\sigma$. Writing $p_3^\mu=q^\mu-p_1^\mu-p_2^\mu$, this structure reduces in the rest frame to a $(n_1\wedge n_2)^a$ term. 
The WZW current saturates the vertex for $0<z<1$. In the exact limit $z\to0$ and $z\to 1$, the trace of the hadronic tensor $H^{\mathbb{1}}_{\cale\cale}$ vanishes, together with $H^c_{\cale\cale}$ and $H^b_{\cale\cale}$. For $z=0$, there is an additional contribution to the correlator corresponding of placing both detectors on the same particle, leading to $b(0)=0$ and $c(0)=-1/2$, in accordance with the previous discussion for the one-point correlator.


The dependence on the Euler angles can be made explicit by projecting the hadronic tensor onto an unpolarized transverse state after a rotation. 
Using the canonical basis introduced in Eq.~\ref{eq:twopointcanonicalbasis} and illustrated in Fig~\ref{fig:twopointcorrdiagram}, this leads to the unpolarized two-point energy correlator which is given by
\be
\ab{\cale_{n_1}\cale_{n_2}}\,=\,
\frac{1}{4\pi}\text{EEC}(z)\,\times\,
\left[
1\,+\,
a_{\cale\cale}^{(2,0)}(z)\left(\frac32 \sin^2\Theta-1\right)
\,-\,
a_{\cale\cale}^{(2,2)}(z)\frac34 \sin^2\Theta \cos 2\phi
\right]\,,
\label{eq:en1en2spinningexplicit}
\ee
where $\text{EEC}(z)$ denotes the usual inclusive two-point energy correlator, $\frac{1}{\mathcal{N}}H^{\mathbb{1}}_{\cale\cale}(z)$, obtained by integrating over the polar and azimuthal angles $\Theta$ and $\phi$. The spinning correlators are encoded in $a_{\cale\cale}^{(2,0)}(z)$ and $a_{\cale\cale}^{(2,2)}(z)$ as
\be
a_{\cale\cale}^{(2,0)}(z)\,=\, \frac{H^c_{\cale\cale} (z)- \ds\frac12 H^b_{\cale\cale} (z)}{H^\mathbb{1}_{\cale\cale}(z)}~,\quad \text{and}\quad
a_{\cale\cale}^{(2,2)}(z)\,=\, \frac{ H^b_{\cale\cale} (z)}{H^\mathbb{1}_{\cale\cale}(z)}~,
\label{eq:definitions:c20c22}
\ee
in the same basis as Eq.~\ref{eq:twopointcanonicalbasis}, and thus are written as $a_{\cale\cale}^{(2,0)}(z)=c(z)-\frac12 b(z)$ and $a_{\cale\cale}^{(2,2)}(z)=b(z)$ in terms of functions $c(z)$ and $b(z)$ in Eq.~\ref{eq:candbdefinition}. 
The indices of $a^{(J,m)}$ denote the spin quantum numbers of the configuration along the spin axis given by the detector axis $\vec{n}_1+\vec{n}_2$. Therefore, Eq.~\ref{eq:en1en2spinningexplicit} is similar to a decomposition in spherical harmonics, but while the angle $\Theta$ corresponds to the polar angle between the beam axis and the detector axis, the azimuthal $\phi$ is an azimuthal angle of the detector configuration with respect to the detector axis. In this unpolarized case, the dependence on the azimuthal angle with respect to the beam axis, $\Phi$, vanishes. 
In this normalization, the positivity constraints in Eq.~\ref{eq:constraintsoncandb} are recasted into
\be
-\frac12 \leq a_{\cale\cale}^{(2,0)}(z) \,\leq\, 1\,\,\,,\quad\quad\text{and}\quad\quad 
\left\vert\,a_{\cale\cale}^{(2,2)}(z) \right\vert\,\leq\, \frac23\,\left( 1-a_{\cale\cale}^{(2,0)}(z)\right)~,
\label{eq:constraintsc20c22}
\ee
The constraints in $a_{\cale\cale}^{(2,0)}(z)$, inclusively in the azimuthal angle $\phi$, are the same as the constraints on the $a_\cale$ parameter of the one-point energy correlator. For a nonvanishing $a_{\cale\cale}^{(2,2)}(z)$, the allowed space for $a_{\cale\cale}^{(2,0)}(z)$ shrinks until $a_{\cale\cale}^{(2,2)}=\pm 1$ and $a_{\cale\cale}^{(2,0)}=-1/2$. Notice that if we were to naively expand the correlator in spherical harmonics and demand the positiveness of the energy we would obtain weaker constraints than the ones derived requiring the positiveness of the hadronic tensor. This is because the physical photon does not interpolate the longitudinal components of the density matrix. 

Finally, the existence of the constraints in Eq.~\eqref{eq:constraintsc20c22} implies that spinning correlators must share the same infrared singularities as the inclusive correlator. In QCD, while the inclusive energy correlator $\text{EEC}(z)$ is strongly enhanced near the endpoints at $z\ll 1$ and $1-z\ll 1$ due to soft and collinear radiation, the spinning correlators can exhibit at most the same degree of enhancement in order to remain consistent with positivity.

\subsection{Comments on higher points}
\label{sec:higherpoints}

The extension to the general $N$-point correlator splits into a part that is fixed by symmetry and holds for any $N$, and a part that genuinely depends on the number of detectors, which is the geometry of the moduli space of configurations. We discuss them in turn.

\smallskip

The organizing structure of the spinning correlators is fixed by the spin of the source and is independent of $N$. For a vector current, the hadronic tensor $H^{ab}_{\cale^{(N)}}$ is a symmetric tensor in its two spatial indices, and its decomposition is always of the form in Eq.~\ref{eq:tensordecompositionNpoint}, a singlet and a five-dimensional traceless symmetric part. For a source of spin $J_S$, the angular momentum structures runs up to  $2J_S$, so the number of spinning correlators is bounded by $J_S$ and independent of $N$. The dependence on the Euler angles is always carried by the Wigner-$D$ matrices as in Eq.~\ref{eq:NpointcorrFinalForm}. Moreover, the positivity bounds of Section~\ref{sec:positivity} hold true for any $N$. The role of $N$ is therefore controlling the number of independent nonzero spinning structures. As discussed in the previous section, the one-point correlators has a single spinning structure, and the two-point correlator case can be written in terms of the detector axis of Eq.~\ref{eq:twopointcanonicalbasis}, which leads to two spinning structures. 

The $N=3$ case is the first one that potentially contains all five different spinning structures for $J_S=1$. Indeed, choosing tensors built from the symmetric products $n^{(a}_i n^{b)}_j$ for $i,j=1,2,3$ leads to these six structures, which contain the trace plus the traceless symmetric. 
While constructing the covariant tensors explicitly is convenient for the computation of the spinning correlators, it is not a necessity. From the calculation of the hadronic tensor it is possible to do it with a complete basis of polarization tensors along the detector axis. 
As an explicit example for the $N=3$ case, consider the detector spin axis
\be
\vec{s}\,=\, \vec n_1 \wedge \vec n_2 + \vec n_2 \wedge \vec n_3 + \vec n_3 \wedge \vec{n}_1\,,
\ee
which is orthogonal to the plane defined by three detectors as it is orthogonal to any difference of vectors $\vec n_i-\vec n_j$. 
In the frame where $\vec s = (0,0,1)$, all detectors have the same polar angle. We can use this axis to define a set of polarization vectors, a longitudinal that coincides with $\vec s$ and two transverse which live on the plane. The azimuthal angle $\phi$ can be defined by ordering the vectors from a purely geometrical criteria, e.g. $z_{12}\geq z_{23}\geq z_{13}$. Then, dotting the hadronic tensor with these polarization vectors generates all spinning structures. 

\smallskip

The problem of parametrizing the dependence of the energy correlators on the internal angles $z_{ij}$ naturally separates into two parts. First, one must characterize the moduli space of $N$ points on the sphere 
 modulo $SO(3)$ rotations and modulo $S_N$ permutations. Second, for any given point in this moduli space, one must choose a concrete embedding in $\mathbb{R}^3$ 
in order to define a set of Euler angles describing the rigid-body orientation of the detector configuration.

A convenient way to describe the moduli space is through the Gram matrix, whose entries $G_{ij}$ are the dot product between the $i$-th and $j$-th vectors, $G_{ij}=\vec{n}_i\cdot \vec{n}_j=1-2z_{ij}$, with $G_{ii}=1$. By construction, the Gram matrix is positive definite and, being constructed from vectors in $\mathbb{R}^3$, it has rank 3. It is symmetric, and the $N(N-1)/2$ entries $G_{ij}$ with $i<j$ are determined by a given configuration of detectors. Since the Gram matrix is rotational invariant, two configurations of detectors are equivalent up to an $SO(3)$ rotation if and only if their associated Gram matrices coincide, up to permutations of rows and columns due to the $S_N$ symmetry. 

Positivity of the Gram matrix implies a set of constraints on $G_{ij}$, and therefore on $z_{ij}$. Positivity of the 2-by-2 minors implies that $z_{ij}(1-z_{ij})\geq 0$, which is indeed the constraint to interpret $z_{ij}$ as defined through the dot product of two vectors as in Eq.\ref{eq:zijdef}. Positivity of the 3-by-3 minors implies nonlinear inequalities among the $z_{ij}$, compactly written as $-4e_3+4e_2-e_1^2\geq 0$, where we used the symmetric polynomials $e_1=z_{ij}+z_{jk}+z_{ki}$, $e_2=z_{ij}z_{jk}+z_{jk}z_{ki}+z_{ki}z_{ij}$ and $e_3=z_{ij}z_{jk}z_{ki}$. These conditions are precisely those required for the $z_{ij}$ to arise from three unit vectors on the sphere and already illustrate that the space of allowed configurations is highly constrained.

The rank-three condition on the Gram matrix implies a large set of redundancies. Out of the $N(N-1)/2$ entries $G_{ij}$, the rank of the matrix implies that only $2N-3$ of them are independent. The only cases where the redundancy is absent is for $N=2$ and $N=3$, but for $N\geq 4$ a choice of coordinates becomes nontrivial. It should be noted that not all subset of $2N-3$ coordinates can specify a give configuration of points in $S^2$, as it is clear by considering a subset that does not involve the dot product with a given vector $\vec{n}_j$, therefore leaving it completely undetermined. A practical parametrization is 
provided by the set of elements of the first two rows of the matrix, corresponding to the $2N-3$ angular distances
\be
z_{12},\,z_{13}\,,\dots ,\, z_{1N},\, z_{23}\,,\dots,\, z_{2N}\,,
\ee
which determine all other $z_{ij}$ through rank constraints. 

However, the reconstruction is not unique. Solving the 4-by-4 minors formed by the vectors $\vec{n}_1$, $\vec{n}_2$, $\vec{n}_i$ and $\vec{n}_j$ yields, in general, two solutions for the remaining distance $z_{ij}$. 
The two solutions for $z_{ij}$ indicate the cases where both $\vec{n}_i$ and $\vec{n}_j$ are in the same side of the plane defined by $\vec{n}_1$ and $\vec{n}_2$, and the case where they are in opposite sides. 
Importantly, this ambiguity is discrete rather than continuous. Indeed, the graph defined by the chosen set of distances is a Laman graph (in fact, a $(1,1,N-2)$-complete tripartite graph), meaning that it is locally rigid, given by the fact that the ambiguity left by the vanishing of the minors is a sign ambiguity, not a continuous one. Resolving such ambiguity seems to require additional discrete information to the coordinate choice, such as the sign of $\det(\vec{n}_1,\vec{n}_2,\vec{n}_i)$. 

From the perspective of spinning energy correlators, this structure suggests that the kinematics of higher-point functions is governed by a constrained but rich moduli space, with well-defined continuous coordinates supplemented by discrete choices. While the Gram matrix formulation provides a systematic starting point, identifying an optimal coordinate system for general $N$ and understanding the analytic structure of correlators as functions of these variables, remains open. We leave a detailed investigation of these questions and their implications for spinning correlators to future work.

\subsection{Boosted correlators}
\label{sec:boosted}

The energy correlators break Lorentz symmetry, as they measure the flux not only in a particular direction, but also in a particular frame. The choice of frame is implemented by parametrizing the detector directions as $n^\mu=(1,\vec{n})$, so that the energy detectors defined in Eq.~\ref{eq:calendef} act on asymptotic states as in Eq.~\ref{eq:Enonalpha}. Through this work we have studied and classified the different structures of the energy correlators in the rest frame of the source, where the momentum injected is $q^\mu = (Q,\vec{0})$. Here we discuss how correlators of boosted sources, for which $q^{\prime \mu} = (Q^\prime,\vec{q}^\prime)$, are obtained by performing systematic transformations.

The symmetries of the energy correlators are manifestly restored through the language of light-ray operators~\cite{Sveshnikov:1995vi,Hofman:2008ar,Belitsky:2013bja,Belitsky:2013xxa,Kravchuk:2018htv,Kologlu:2019mfz,Kologlu:2019bco}. 
In lightcone coordinates~\footnote{From here on we neglect the mass of the final state particles.}, the energy operator in Eq.~\ref{eq:calendef} can be written as~\cite{Belitsky:2013bja}
\be
\cale_n = \int_{-\infty}^\infty d(xn)\,\lim_{(x\bar{n})\to \infty}  (x\bar{n})^2  T^{\mu\nu}((x\bar{n}){n}^\mu + (xn)\bar{n}^\mu) \bar n_\mu \bar n_\nu~,
\label{eq:calendeflightcone}
\ee
where two light-like basis vectors are $n^\mu=(1,\vec{n})$ and $\bar{n}^\mu=(1,-\vec{n})$ and the dependence on $n\bar{n}$ is implicit.
In this form it is straightforward to see that 
$\cale_n$ is homogeneous under the dilatations and rescalings of the detector vectors. Under the dilatations $x\to \sigma^{-1} x$, the energy-momentum tensor scales as $T^{\mu\nu}\to \sigma^4 T^{\mu\nu}$, and thus $\cale_n\to \sigma \cale_n$ which is expected as it measures the energy. 
Under the rescaling of the detector directions $n^\mu\to \rho n^\mu$ and $\bar n^\mu\to \rho^{-1}\bar n^\mu$, the energy operator scales as $\cale_n\to \rho^{-3}\cale_n$. 
Under a Lorentz transformation, the detector direction transforms as $n^{\prime\mu} = \Lambda^\mu_{\,\,\nu} n^\nu$. In general, this yields $n^{\prime \mu} = \lambda (1,\vec{n}^\prime)$ where $\lambda=\Lambda^0_{\,\,\nu}n^\nu$ depends on the boost and the detector orientation.  
Restoring the canonical normalization $n^{\prime 0}=1$ requires rescaling by $\lambda^{-1}$, under which the energy operator scales as $\lambda^{3}$. 
As a result, the hadronic tensor of a boosted source is given by

\be
H^{\prime \,\,\mu\nu}_{\cale^{(N)}} \,=\,
\left(\prod_{i=1}^N \lambda_i^3 \right) \, \Lambda^\mu_{\,\,\alpha}\Lambda^\nu_{\,\,\beta}\,H^{\alpha\beta}_{\cale^{(N)}}~,
\label{eq:boostedcorrelators}
\ee
where $\lambda_i = \Lambda^0_{\,\,\nu}n^\nu_i$ is the rescaling of each detector, which in general is different due to the different directions of $\vec{n}_i$ with respect to the boost direction. This shows how to obtain the full correlator for a source with arbitrary momentum from the one in the rest frame.

As a simple example, consider the one-point correlator of a scalar source. In the rest frame the correlator is simply
\be
\ab{\cale_n}\,=\, \frac{m}{4\pi}~,
\label{eq:onepointscalar}
\ee
where we momentarily changed the notation so that $q^\mu = (m,\vec{0})$. Being a scalar operator, there is no dependence on the Euler angles leading to an homogeneous energy distribution. From this one can obtain the one-point correlator for a boosted source, so that the source has energy $E$ with boost $z_\star\equiv m^2/E^2$. 
We can choose the detector $\vec n$ to have an angle $\cos\Theta$ with the boost axis. This induces a nontrivial dependence on the direction between the boost and the detector direction as measured in the original frame rest frame, given by $\cos\Theta$, through the rescaling factor
\be
\lambda = \Lambda^0_{\,\,\nu}n^\nu = \frac{1+\sqrt{1-z_\star}\cos\Theta}{\sqrt{z_\star}}\,=\,
\frac{\sqrt{z_\star}}{1-\sqrt{1-z_\star}\cos\Theta^\prime}~.
\label{eq:lambdafactor}
\ee
In the second equality $\cos\Theta^\prime$ refers to the angle between the correctly normalized detector and the boost direction. 
The cube of this factor allows to express the one-point energy correlator for a scalar source with $q^\mu=(m,\vec{0})$, in Eq.~\ref{eq:onepointscalar}, to the one-point correlator of a scalar source with momentum $q^\mu=(E,E\sqrt{1-z_\star})$,
\be
\ab{\cale_n}^\prime \,=\, \frac{E}{4\pi}\, \frac{z_\star^2}{(1-\sqrt{1-z_\star}\cos\Theta^\prime)^3}~.
\ee
At large boosts, i.e. $z_\star\ll 1$, there is in fact a suppression for the correlator at generic angles. However, for $\cos\Theta^\prime \geq \frac{1-z^{2/3}}{\sqrt{1-z}}\simeq 1-z^{2/3}$, so for $\Theta\lesssim \sqrt{2}z^{1/3}$, one gets an enhancement due to the accumulation of energy in the collinear region. It should be noted that the integral over the Euler angle $\Theta^\prime$ still leads to the total energy, now given by $E$, since one has that indeed $\int_{-1}^1\frac{d\cos\Theta^\prime}{2}\, \frac{z_\star^2}{(1-\sqrt{1-z_\star}\cos\Theta^\prime)^3} = 1$.

This behavior admits a simple physical interpretation. One factor of $\lambda$ is due to the redshift of the measured energy, while the remaining $\lambda^2$ reflects the enhancement of the density of states in the collinear region.  Indeed, since the energy correlator is assumed to be measured within a solid angle $d\cos\Theta d\Phi$, one gets
\be
d \cos\Theta \to d \cos\Theta^\prime \frac{z_\star}{(1-\sqrt{1-z_\star}\cos\Theta^\prime)^2}\,
=\, \lambda^2\,d \cos\Theta^\prime \,~.
\ee

For spinning sources, there is the technical complication of having two directions, one corresponding to the spin axis of the source, and the other corresponding to the direction of the boost. It is technically and conceptually convenient to consider therefore the helicity states of the spinning source, namely the case where both spin and boost axis coincide. In this case it is immediate to apply the discussion of the scalar case to the case of an unpolarized and boosted vector current with respect to the case in Eq.~\ref{eq:densitymatrix:onept:photonpolarization}, obtaining
\be \label{eq:onepointvectorboosted}
\begin{split}
	\ab{\cale_n} &= \frac{m}{4\pi}\left(1+a_\cale \left(\frac{3}{2}\sin^2\Theta - 1 \right)\right)
	\\[5pt]
	&\underset{\text{boost}}{\rightarrow} \frac{E}{4\pi} \frac{z_\star^2}{(1-\sqrt{1-z_\star}\cos{\Theta^\prime})^3}
	\left(1+a_\cale \left(\frac{3}{2}\frac{z_\star \sin^2{\Theta^\prime}}{(1-\cos{\Theta^\prime}\sqrt{1-z_\star})^2} - 1\right)\right)\,.
\end{split}
\ee
The boosted Wigner D-matrix, together with the overall measure, still integrates to zero, as required by the fact that the integral gives the total energy. While the measure is monotonic in $\cos\Theta^\prime$, the $a_\cale$ term has a maximum at $\cos\Theta^\prime = \sqrt{1-z_\star}$. The longitudinal polarization leads to a different functional form, with $\frac12 \sin^2\Theta\to \cos^2\Theta$ in Eq.~\ref{eq:onepointvectorboosted}. The different functional form can be used to discriminate the polarizations of an electroweak boson, as explored in~\cite{Ricci:2022htc}. 

The arguments shown here apply to arbitrary spinning sources and to general multi-point correlators, providing a unified description of fully differential energy measurements in both rest and boosted frames.

\subsection{Collinear limit of spinning energy correlators}
\label{sec:collinearspinning}

For completeness, we briefly discuss the collinear limit of both inclusive and spinning energy correlators. In this regime, the correlators are controlled by an operator product expansion (OPE) governed by light-ray operators carrying definite transverse spin \cite{Chang:2020qpj,Chen:2022jhb}. 
Our goal in this section is to show explicitly how different spinning correlators probe different transverse-spin sectors of this OPE.

The form of the energy in terms of a lightray operator in Eq.~\ref{eq:calendeflightcone} can be generalized. In a free theory, the local operator $\mathcal{O}^{\mu_1\dots\mu_J}$ of scaling dimension $\Delta$ can be mapped to a lightray operator by integrating it along a null direction. For the sector with vanishing transverse spin, this construction takes the form  \cite{Kravchuk:2018htv} 
\be
\mathscr{O}_n^{(J-1,1-\Delta,0)} = \lim_{(x\bar{n})\to \infty} \int_{-\infty}^\infty d(xn)\,(x\bar{n})^{\Delta-J}  \mathcal{O}^{\mu_1\dots\mu_J}((x\bar{n}){n}^\mu+(xn)\bar{n}^\mu) \bar n_{\mu_1} \cdots \bar n_{\mu_J}~,
\label{eq:Ozerotransverse}
\ee
where the superscript denotes the eigenvalues under dilatations, boosts and rotations around the null direction $n^\mu$. 
Light-ray operators with nonzero transverse spin are obtained by contracting some indices with polarization tensors transforming in irreducible representations of the transverse rotation group,
\be
\mathscr{O}_n^{(J-1,1-\Delta,h)} 
= \lim_{(x\bar{n})\to \infty} \int_{-\infty}^\infty d(xn)\,(x\bar{n})^{\Delta-J}  \mathcal{O}^{\mu_1\dots\mu_J\nu_1\dots\nu_k}((x\bar{n}){n}^\mu + (xn)\bar{n}^\mu) \bar n_{\mu_1} \cdots \bar n_{\mu_J}\epsilon^{h}_{\nu_1 \cdots \nu_k}~,
\label{eq:Ononzerotransverse}
\ee
where $\epsilon^{h}_{\nu_1 \cdots \nu_j}$ is the polarization tensor in the irreducible representation of a transverse spin $h$. 
While the light-ray operator in Eq.~\ref{eq:Ononzerotransverse} still has the dimension of $J-1$ and scales as $1-\Delta$ under the boosts, it has a nonvanishing transverse spin $h$.

The light-ray operators appear in the collinear OPE of two energy flow operators. The spin structure of this OPE is particularly transparent when expressed in spinor variables. Rather than organizing the expansion solely in terms of scalar products $n_1\cdot n_2$, we write schematically
\be
\cale_{n_1}\cale_{n_2} \,\sim\, \ab{12}^{k_-}\sq{12}^{k_+} \mathscr{O}_{n}^{(K)}~,
\label{eq:lightrayOPE}
\ee
where $K$ collectively denotes the quantum numbers carried by the light-ray operators and $\ab{12}$ and $\sq{12}$ are the usual spinor products. 
Energy operators scale as $+1$ under the dilatations, which fixes $J=3$ for the light-ray operators appearing in the OPE. Under the boosts, energy operators scale as $-3$, while the light-ray operators as $1-\Delta$ and both $\ab{12}$ and $\sq{12}$ as $+1$. Matching under the boosts in Eq.~\ref{eq:lightrayOPE} requires 
$k_-+k_+ \,=\, \tau-4$ where $\tau=\Delta-J$ is the twist.
The helicity operator annihilates the energy operators, while it contributes $-1$ when acting on $\ab{12}$ and $+1$ when acting on $\sq{12}$. 
Matching the helicity in Eq.~\ref{eq:lightrayOPE} implies 
$
k_--k_+ = h$\,. 
Solving these constraints and expressing the spinor contractions in terms of the relative angle and azimuthal phase,  $\ab{12}\sq{21} = 2n_1\cdot n_2$ and $\ab{12}/\sq{12} = e^{-2i\phi}$,  the OPE in Eq.~\ref{eq:lightrayOPE} takes the form
\be
\ab{12}^{k_-}\sq{12}^{k_+} = \frac{1}{(n_1\cdot n_2)^{2-\frac{\tau}{2}}} e^{-i\, h \phi}~.
\label{eq:OPEexplicit}
\ee
This expression makes explicit that the strength of the collinear singularity is controlled by the twist, while the azimuthal dependence directly encodes the transverse spin. As a result, different spinning energy correlators project onto different transverse-spin sectors of the same light-ray operator tower.

As an illustrative example, we briefly show how light-ray operators with transverse spins arise in a free-field description and how they control the collinear limit of spinning energy correlators. The purpose of this example is not to provide a complete derivation, but to give a concrete interpretation of transverse spin in terms of helicity interference \cite{Chen:2022jhb}.
For gluons, the energy flow operator obtained from the light transform of the stress-energy tensor acts as $
\cale_n \sim \int \frac{d^3p}{2E}\,\delta^2(\Omega_n-\Omega_p)\,E\sum_{h=\pm} a^\dagger_{p,h}a_{p,h}$ 
and it measures the energy carried by gluons in the direction $n^\mu$, inclusively over the helicity. The leading-twist light-ray operators appearing in the collinear OPE have $J=3$ and $\tau=2$. A representative operator takes the schematic form
\be
\mathcal{O}_{\mu\nu\rho}\sim G^A_{\mu\alpha} D_\nu G^A_{\beta\rho}\, \eta^{\alpha\beta}\,,
\label{eq:Omunurhoschematic}
\ee
whose light transform produces a number operator weighted by $E^2$, again summed over helicities. This corresponds to the transverse-spin–zero sector of the OPE. 

Operators with nonzero transverse spin arise by replacing the metric contraction with appropriate polarization tensors. For instance, contracting with $\epsilon^\alpha_- \epsilon^\beta_-$ instead of $\eta^{\alpha\beta}$ in Eq.~\ref{eq:Omunurhoschematic} yields a light-ray operator with transverse spin $h=-2$,
\be\label{eq:twist2m2gluon}
\mathscr{O}_n^{(h=-2)} \sim
\int \frac{d^3p}{2E}\,\delta^2(\Omega_n-\Omega_p) E^2 a^\dagger_{p,-}a_{p,+}\,,
\ee
which lowers the helicity of the state by two units. When inserted in a correlator, this operator probes interference terms between amplitudes with opposite-helicity gluons. At finite angular separation, such interference becomes observable through the azimuthal dependence of the correlator, as discussed in Ref.~\cite{Chen:2021gdk,Chen:2022jhb} and widely exploited in collider observables~\cite{Panico:2017frx}. 

At this scaling dimension, gauge theories do not admit operators with higher transverse spins. In gravity there is a $h=4$ candidate. 
The stress-energy tensor of a gravitational wave is given by $T_{\mu\nu}\sim \frac{1}{G_N} \partial_\mu h_{\alpha\beta}\partial_\nu h_{\alpha\beta}$~\cite{Gonzo:2020xza} which leads to a number operator acting on gravitons. A candidate for a $J=3$ light-ray operator is given by $\mathcal{O}_{\mu\nu\rho}\sim \frac{1}{G_N} \partial_\mu h_{\alpha\beta}\partial_\nu\partial_\rho h_{\gamma\delta}\,\eta^{\alpha\gamma}\eta^{\beta\delta}$. 
By contracting with $\epsilon_\mp^\alpha\epsilon_\mp^\beta$ instead of $\eta^{\alpha\beta}$, one gets a maximally spinning $|h|=4$ operator. This appears, for instance, in the interference of the energy-energy OPE of a graviton splitting into two gravitons and has a similar interpretation as in the case of gluons, namely, a contribution from amplitudes with opposite helicity gravitons.

The connection with spinning energy correlators is direct. The azimuthal angle $\phi$ controlling the collinear OPE coincides with the Euler angle appearing in the general form of the correlator. Consequently, the collinear limit of a spinning energy correlator with angular momentum $m$ is governed by a tower of light-ray operators with transverse spin 
$h=m$. The positivity bounds derived for spinning correlators therefore constrain the relative size of the corresponding spinning and non-spinning contributions in the OPE.

\section{Spinning Correlators in QCD}
\label{sec:QCD}

In this section we discuss spinning energy correlators in perturbative QCD in detail. We show that each spinning correlator is obtained from a specific projection of the hadronic tensor, determined by the detector orientations. Due to the soft and collinear factorization, the collinear and back-to-back limits of the spinning correlators, normalized to the inclusive ones, reduce to a simple two-jet calculation, indicating that these ratios directly probe the hard structure of the underlying process.  In Section~\ref{sec:spinning:beyondenergy}, we extend the analysis to energy-charge correlators, showing that they generate additional spinning structures and that they are IR safe observables. We illustrate these results comparing our analytical predictions with simulated data at hadron level in the correlation of the energy and global QCD charges: electric charge, isospin and baryon number, finding a striking agreement.

\subsection{The spinning energy-energy correlator}
\label{sec:spinning:eec}

The tensor structure of the two-point energy correlator of a vector current is uniquely fixed to have the form in Eq.~\ref{eq:tensordec2point}. 
The dynamical part of the correlator is encoded in the dependence on the internal angle $z$ of the scalar distributions $ H^{\mathbb{1}}_{\cale\cale}(z)$, $H^c_{\cale\cale}(z)$ and $H^b_{\cale\cale}(z)$. It is possible to write these in terms of contractions of the covariant hadronic tensor $H^{\mu\nu}_{\cale\cale}$ 
with the metric $\eta_{\mu\nu}$, the sum of two detectors $\frac{1}{1-z}\frac{(n_1+n_2)_\mu (n_1+n_2)_\nu}{2}$, and the difference of them $\frac{1}{z}\frac{(n_1-n_2)_\mu (n_1-n_2)_\nu}{2}$ to form three singlets which we define as
\be \label{eq:singlets:ESD}
	E \equiv   -\eta_{\mu\nu}H^{\mu\nu}\, ,
\,
	S \equiv  \frac{1}{1-z}\frac{(n_1+n_2)_\mu}{2} \frac{(n_1+n_2)_\nu}{2} H^{\mu\nu}\, ,
\,
	D \equiv  \frac{1}{z}\frac{(n_1-n_2)_\mu}{2} \frac{(n_1-n_2)_\nu}{2} H^{\mu\nu}~.
\ee
Then, the spinning correlators are given as a combination of these three singlets,
\be
\begin{split}
 H^\mathbb{1}_{\cale\cale} =  E~,\quad\quad
 H^c_{\cale\cale} =  2S + D -  E~,\quad\quad
 H^b_{\cale\cale} =  2D + S -  E~.
\end{split}
\ee
At leading order, the contribution to these correlators stem from the overlap of the current with two free quark states, corresponding to the free $\alpha_s\to 0$ limit. In order for these states to contribute to the correlator the detectors should be on top of each other, so in the exact collinear limit $z=0$, or in the exact back-to-back limit  $z=1$. Normalizing the total energy to 1, the correlators are given by
\be
H^\mathbb{1}_{\cale\cale} =  \frac{1}{2} \delta(z)\,+\, \frac{1}{2} \delta(1-z)\,,\quad\quad
H^c_{\cale\cale} =  -\frac{1}{4} \delta(z)\,,\quad\quad
H^b_{\cale\cale} = -\frac{1}{4} \delta(1-z)\,.
\label{eq:born1cbcorrelators}
\ee
In the exact collinear limit $z=0$, the correlator becomes the one-point correlator of the energy-squared, $\ab{\cale^2_{n}}$. Therefore, we identify the coefficient multiplying the $\delta(z)$ term in $H^\mathbb{1}_{\cale\cale}$ as the total flux of the one-point correlator, and the coefficient multiplying $\delta(z)$ in the spinning $H^c_{\cale\cale}$ as the total flux times the $a_{\cale^2}$ parameter controlling the spinning one-point correlator, as in Eq.~\ref{eq:Honepointtensor}. 
The parameter $a_{\cale^2}$ is therefore $-1/2$, which indeed corresponds to the free-fermion case. Note that in this limit $H^b_{\cale\cale}$ vanishes, as expected, since the azimuthal angle $\phi$ is ill-defined. In the back-to-back limit $z=1$, the correlator instead shows a nontrivial dependence on $\phi$, captured by the $\delta(1-z)$ terms. These two special configurations thus represent the already discussed free-fermion points in the boundary of Fig.~\ref{fig:bdries2pt}. The discussion of the Born approximation is relevant. As we shall discuss, due to factorization of collinear and soft modes, the $z\to 0$ and $z\to 1$ limits of the ratio between the spinning correlator and the inclusive one at finite coupling is well approximated by the leading order result.

At order $\alpha_s$, the inclusive part of the hadronic tensor, including the endpoint contributions, can be found in \cite{Dixon:2019uzg}, and can be written as
\be
\begin{split}
 H^\mathbb{1}_{\cale\cale}
	&= \frac{\alpha_s}{\pi} C_F \left [  
	- \frac{3(3 z^4+2 z^3+14 z^2-12 z) + 4 ( z^4+z^3-3 z^2+15 z-9 )\log(1-z)}{8 z^5} \right.
	\\[5pt]
	& \hspace{1.5cm} + \frac{3}{8} \left [ \frac{1}{z} \right ]_+   + 
	\delta(z)\,\left[ \left(  -\frac{3}{4\epsilon}-\frac{43}{24} \right) + \left( \frac{3}{4\epsilon} + \frac{185}{96} \right) \right]
	\\[5pt]
	& \left . \hspace{1.5cm}
	- \frac{1}{2} \left [ \frac{\log(1-z)}{1-z} \right ]_+ 
	- \frac{3}{4} \left [ \frac{1}{1-z} \right ]_+  +  \left ( -\frac{\zeta_2}{2} - 1 \right )\delta(1-z)  \right ]~.
\end{split}
\label{eq:2ptinclusiveQCD}
\ee
We shall remark that this result is for the inclusive part of the hadronic tensor, it is not the energy correlator, and therefore there is no normalization by the total rate. The inclusion of the total rate, which at $\mathcal{O}(\alpha_s)$ it is given by $\frac34 C_F \frac{\alpha_s}{\pi}$, shifts the coefficients of the delta function by $-\frac12\times \frac34 C_F \frac{\alpha_s}{\pi}$. This shift is important as it makes the integral over $z$ of the $\mathcal{O}(\alpha_s)$ part of the inclusive correlator vanish, as will be discussed in Section~\ref{sec:sumrules}. While the inclusive correlator has been the subject of extensive precision studies \cite{DelDuca:2016ily,Moult:2018jzp,Dixon:2018qgp,Ebert:2020sfi,Duhr:2022yyp,Aglietti:2024xwv}, our focus in this work is on the structural and angular-momentum properties of spinning correlators. For establishing these features, it is sufficient to work at order $\alpha_s$, and we therefore employ the NLO expression in Eq.~\eqref{eq:2ptinclusiveQCD}. Extending the present analysis to higher perturbative orders would be interesting but lies beyond the scope of this work.

In the second line of Eq.~\ref{eq:2ptinclusiveQCD} we choose to explicitly separate the coefficient of $\delta(z)$ in two terms. While the first term comes from the plus distribution, the second one comes from the contributions where two detectors are located on top of each other and in a single particle in the final states $q\bar{q}$ and $q\bar{q}g$. Namely, the second term comes from the $\mathcal{O}(\alpha_s)$ calculation of the hadronic matrix element of the one-point correlator of $\ab{\cale^2_n}$. Of course, this is not an IR-safe observable, thus there is an explicit dependence on $1/\epsilon$. This IR divergence is canceled with the one coming from the plus distribution, as it should be due to 
Kinoshita–Lee–Nauenberg (KLN) theorem. Intuitively, at this order while the $\ab{\cale^2_n}$ term measures the squared of the parton energies, $E_q^2+E_g^2$, the plus distribution term measures the product of them in the $z\to 0$ limit, $2E_qE_g$, so the sum gives the total energy of the parent quark, which is of course a conserved quantity by the collinear splitting. It is conceptually important to keep this separation. In a would-be full nonperturbative calculation, there is no plus distribution since the $1/z$ divergence is cut-off by the mass gap, and therefore there is no respective term in the delta function as it gets resummed away. The term coming from $\ab{\cale^2_n}$, however, does survive as it measures the average energy-squared of the hadrons, which is of course IR sensitive and therefore depends on non-perturbative dynamics, but otherwise it is well defined and can certainly be measured experimentally.

The calculation of $S$ and $D$ at this order leads to the following expressions for the spinning correlators,
\be\label{eq:Hb:Hc:QCD}
\begin{split}
H^c_{\cale\cale} (z) &= \frac{\alpha_s}{\pi} C_F \left [  
\frac{1}{16} \frac{(3 z^4+4z^3+30z^2 -36z)- 4 (2 z^2- 12 z+9) \log (1-z)}{z^5}  \right .
\\[5pt]
&\left . \hspace{1.5cm} - \frac12 \cdot\frac38 \left [\frac{1}{z} \right ]_+ 
+ \left (\frac{1}{4} - \frac{1}{2}\cdot \frac{13}{96} \right ) \delta(z) \right ]~,
	\\[5pt]
		H^b_{\cale\cale} (z) &
	= \frac{\alpha_s}{\pi} C_F \left [  \frac{1}{8} \frac{ z \left(3 z^2+z+6\right)+ 2 \left(z^3 + z^2 - z+3\right) \log (1-z)}{z^4} \right .
	\\[5pt]
	&\left . \hspace{1.5cm} + \frac12 \cdot\frac{1}{2} \left [\frac{\log(1-z)}{1-z} \right ]_+ +  \frac12 \cdot\frac{3}{4} \left [\frac{1}{1-z} \right ]_+ 
	-\frac{1}{2}\left (  - \frac{\zeta_2}{2} - 1 \right ) \delta(1-z)
	\right ]~.
\end{split}
\ee
The contribution to $\delta(z)$ in $H^c_{\cale\cale}$ comes from a contribution proportional to $\frac{1}{2}\cdot \frac{13}{96}$ in $D$ (where $\frac{13}{96}$ is the contribution in the inclusive correlator of Eq.~\ref{eq:2ptinclusiveQCD}) and finite contributions of $-1/12$ and $+1/6$ in $D$ and $S$, respectively. These add up to zero in $H^b_{\cale\cale}$ which makes the $\delta(z)$ contribution term to vanish. 

The first term in the spinning correlators is finite in the collinear $z\to 0$ and back-to-back $z\to 1$ limits. The enhancement of the correlator in these limits is entirely given by the terms in the second line, corresponding to the plus distributions, which have an associated contribution to the delta functions. 
The coefficient of these terms is precisely given by the endpoint divergences of the inclusive correlator, times minus one-half. The $-1/2$ factor is in fact the ratio of the delta function coefficients of the LO result in Eq.~\ref{eq:born1cbcorrelators}. This is immediately understood by considering the soft and collinear factorization of the amplitude, which can be written as
\be
\langle p_{1,1}\dots p_{1,k_1}p_{2,1}\dots p_{n,k_n}p_{s,1}\dots p_{s,k_s}| J^\mu |0\rangle 
\,=\, 
\langle \alpha_s| \mathcal{S}|0\rangle \prod_{i=1}^n\sum_{\mathcal{O}}\langle \alpha_i |\mathcal{O}|0\rangle \langle q_1\dots q_n|J^\mu |0\rangle
\label{eq:factorizationFF}
\ee
where $\la{\alpha_i} = \la{p_{i,1}\dots p_{i,k_i}}$ and $\la{\alpha_s} = \la{p_{s,1}\dots p_{s,k_s}}$, and where $\mathcal{S}$ is the soft operator consisting in a product of Wilson lines and $\mathcal{O}$ an operator associated with the hard parton in the direction $q_i$, see e.g. \cite{Catani:1999ss,Feige:2013zla,Agarwal:2021ais}. The subset of momenta $p_{i,j}$ is considered to be collinear with $q_i$, $p_{i,j}\cdot q_i/(q_i^0)^2\sim \theta^2$ with $\theta\ll 1$, and the momenta $p_{s,j}$ are soft. 
Thus, when acted with energy detectors, at leading order in $\theta$, the energy detector operator $\cale_n$ annihilates all collinear sectors except the one where $\vec{n}$ coincides with $\vec{q}_i$. 

Given that the soft sector contains soft quanta with energy fraction $\theta^2\ll 1$, the energy operator $\cale_n$ annihilates the soft sector as well. It is important to remark that there is a hidden assumption in this statement, and that is that the number of soft quanta is less than $\sim 1/\theta^2$. In QCD, the energy of soft radiation is $\sim\Lambda_{QCD}/Q$, while the number of soft particles scales as $\sim \text{exp}\sqrt{\frac{16N_c}{b}\ln\frac{Q}{\Lambda}}$, with $b=\frac{11}{3}N_c-\frac23n_f$, see e.g.  \cite{Dokshitzer:1991wu}. Thus, at this order $\cale_n$ indeed annihilates the soft sector at high energies~\footnote{This argument depends of course on the dynamics. A relevant exception is black hole evaporation, where the entire total energy is radiated by soft quanta, and the softness of the radiation is indeed compensated by the large number of quanta emitted}.
Furthermore, $\cale_n$ acts homogeneously on the state $\la\alpha_n$, measuring the total energy of the parent parton.  This is equivalent to the argument behind the IR-safety of this observable for well separated detectors, as one can therefore just act the energy detectors on the hard matrix element $\langle q_1\dots q_n|J^\mu |0\rangle$.

The normalized spinning correlators $a_{\cale\cale}^{(2,0)}(z)$ and $a_{\cale\cale}^{(2,2)}(z)$ in Eq.~\ref{eq:en1en2spinningexplicit} were defined as the ratio of the spinning correlators with respect to the inclusive one~\footnote{Since the correlators are distributions, one should understand this definition for fixed $z$ in $0<z<1$. In the endpoints, they can be defined as the ratio of the $\delta(z)$ and $\delta(1-z)$ coefficients of the correlators. It should be noted that there is no meaning in the integral of $a_{\cale\cale}^{(2,m)}$, only the integral of $H^\mathbb{1}_{\cale\cale}\times a_{\cale\cale}^{(2,0)}$, or equivalently of $H^c_{\cale\cale}$ and $H^b_{\cale\cale}$, is meaningful.}. 
In the basis specified by the choice in Eq.~\ref{eq:twopointcanonicalbasis}, see Fig~\ref{fig:twopointcorrdiagram}, the normalized spinning correlators are given by those in Eq.~\ref{eq:definitions:c20c22}. 
Due to the factorization of the collinear sector, their collinear and back-to-back endpoint behaviour at this order is fixed by the hard matrix element $\la{q_1q_2}J^\mu\ra{0}$, which coincides with the leading order result in Eq.~\ref{eq:born1cbcorrelators}. Thus, the spinning functions $a_{\cale\cale}^{(2,0)}(z)$ and $a_{\cale\cale}^{(2,2)}(z)$ probe directly the spinning structure of the energy flux determined by the hard process, independently of infrared effects. The explicit form is given by
\be \label{eq:EEC:bc:QCD}
\begin{split}
	a_{\cale\cale}^{(2,0)}(z) &= \frac{3z(12-20z+7z^2+2z^3)+(36-78z+48z^2-4z^3) \log (1-z) }{4 (2 z-3) \left [ 3 (2-3 z) z+2 (2 z^2 - 6 z +3) \log (1-z) \right ] }~,
	\\[5pt]
	a_{\cale\cale}^{(2,2)}(z) &= \frac{z \left [ z (-2 z^2 + 5 z - 6)+(-4 z^2 + 8z  - 6) \log (1-z) \right ]}{2 (2 z-3) \left [ 3 (2-3 z) z+2 (2 z^2 - 6 z+3) \log (1-z) \right ]}~.
\end{split}
\ee
Note that since the LO contribution vanishes in the bulk $0<z<1$, the  $\mathcal{O}(\alpha_s)$ term of the spinning correlators gives the $\mathcal{O}(\alpha_s^0)$ contribution to the normalized spinning correlators of Eq.~\ref{eq:EEC:bc:QCD}. Since the $\mathcal{O}(\alpha_s^2)$ term of the inclusive energy-energy correlator has been recently computed in \cite{Dixon:2018qgp}, the calculation of the spinning part would allow to determine the bulk of $a_{\cale\cale}^{(2,0)}$ and $a_{\cale\cale}^{(2,2)}$ at order $\alpha_s$.

The collinear behaviour of the spinning correlators in Eq.~\ref{eq:Hb:Hc:QCD} is controlled by the lightray OPE in  Section~\ref{sec:collinearspinning}. The $m=0$ correlator is governed by the leading twist operators \cite{Chen:2021gdk,Chen:2022jhb}, and in the $z=0$ limit, the ratio in Eq.~\ref{eq:EEC:bc:QCD} reproduces the spinning $a_{\cale^2}=-1/2$. 
The $m=2$ correlator instead leads to a constant in the $z\to 0$ limit, and therefore the ratio in Eq.~\ref{eq:EEC:bc:QCD} vanishes. This is because, in the collinear limit and at this order in the perturbative expansion, the contribution from the leading twist-2 gluon helicity-flip operator in  Eq.~\ref{eq:twist2m2gluon} vanishes. Therefore, the collinear limit is controlled by a twist 4 operator of the schematic type $\bar\psi GG \psi$ acting on $qg$ states. The leading twist gluon helicity-flip operator does appear at the next order in the coupling. The fact that the subleading twist operator can dominate for $z\ll 1$ but $z\geq \mathcal{O}(\alpha_s)$ seems an interesting feature that merits further study once the higher order calculation becomes available. 

\begin{figure}
	\centering
	\includegraphics[width=0.6\linewidth]{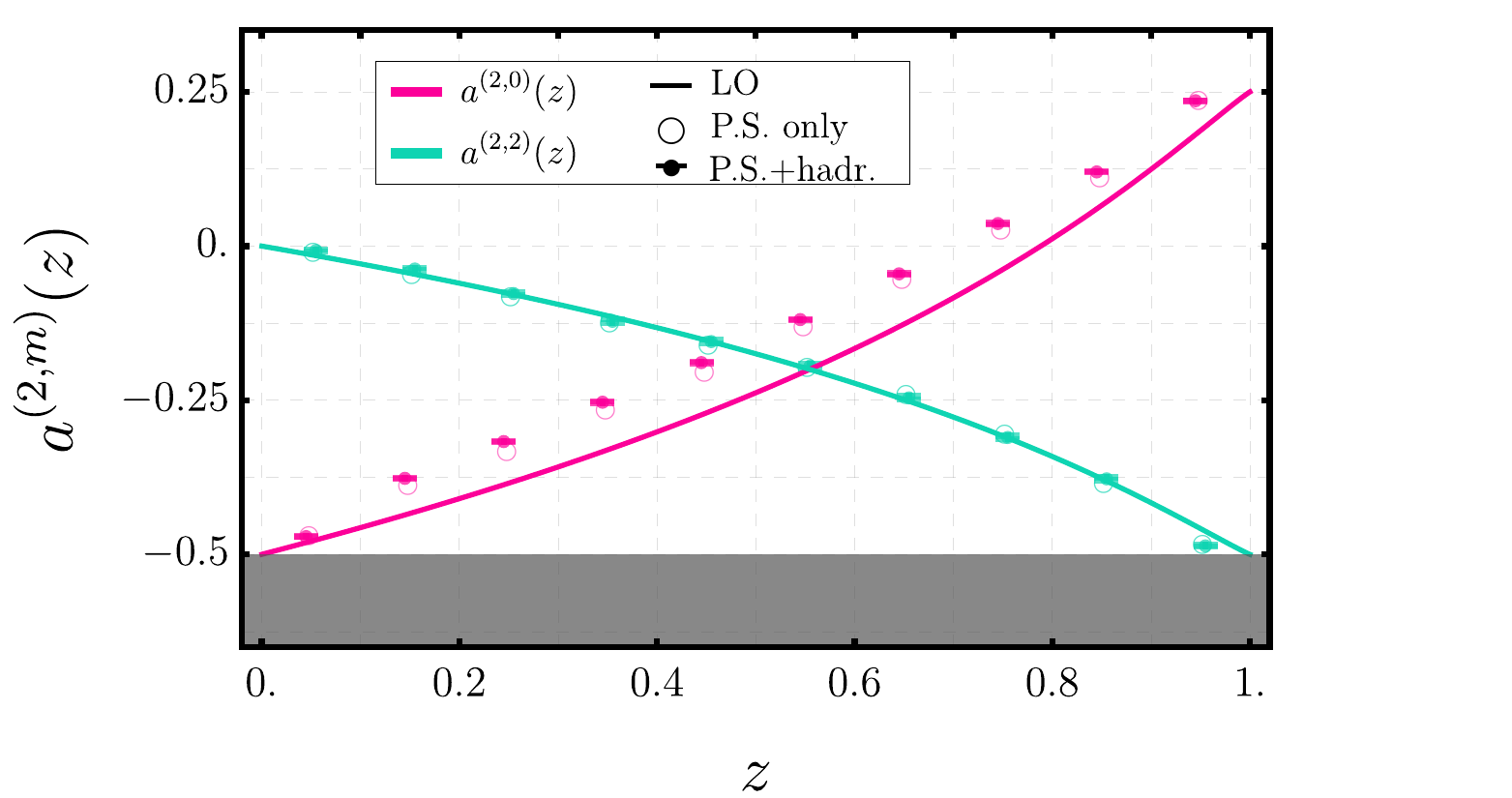}
	\caption{\small Spinning Energy-Energy correlators as a function of $z$ in QCD for an unpolarized vector current.  The solid line is the analytical result in Eq.~\ref{eq:EEC:bc:QCD}, while the data points are the result of a simulation of parton shower only (open circles) and including hadronization (solid line) via \PT. In pink and blue, the spinning correlators $a^{(2,0)}_{\cale\cale}$ and $a^{(2,2)}_{\cale\cale}$, respectively.}
	\label{fig:candbplot}
\end{figure}

The comparison of the result in Eq.~\ref{eq:EEC:bc:QCD} with the simulation of the correlator after parton shower and hadronization effects is shown in Fig.~\ref{fig:candbplot}. We simulate an $e^+e^-$ collision in \verb*|MadGraph| \cite{Alwall:2014hca} followed by the parton shower and hadronization via \PT~\cite{Bierlich:2022pfr}. For each pair of hadrons in an event, we compute $\vec{n}_\pm=\frac12(\vec{n}_1\pm\vec{n}_2)$ and get the polar and azimuthal Euler angles as $\cos\Theta = \hat{z} \cdot \hat{n}_+ $ and 
$\sin\phi = \hat{n}_-\cdot(\hat{z} \times \hat{n}_+)$ 
with $\hat{z}=(0,0,1)$ and $\hat{v}=\vec{v}/|\vec{v}|$. Note that due to the symmetry among detectors, there is a $\vec{n}_1\leftrightarrow \vec{n}_2$ symmetry in the correlator so the sign of $\vec{n}_-$ is irrelevant.  We reweight by the hadron energies accordingly to extract the inclusive and the spinning energy correlators, and sum over all pairs and over all events. Taking the appropriate ratio between the spinning and inclusive correlator leads to the result in Fig.~\ref{fig:candbplot}. 
The data points are the result at the parton shower only (open circles) and hadron level (solid) from the \PT \, simulation,
while the solid lines are the LO calculation in Eq.~\ref{eq:EEC:bc:QCD}. We see that even at this order the agreement with the simulation is quite remarkable, due to the cancellation of infrared effects. This is to be compared with the order $\alpha$ accuracy of the inclusive energy-energy correlator, whose comparison with data shows order $\sim 30\%$ deviations, and both higher order and non-perturbative effects should be taken into account to get an accurate prediction \cite{Schindler:2023cww}. 

Since, as already mentioned, the LO contribution to the spinning correlators  vanishes, the normalized ratios in Eq.~\ref{eq:EEC:bc:QCD} are effectively $\mathcal{O}(\alpha_s)$ predictions. The leading missing correction is parametrically of order $\alpha_s/\pi$, in accordance to the difference between the LO and the parton shower only simulation. We checked numerically that the difference between $e^+e^-\to jjj$ and $e^+e^-\to jjjj$ samples matched to the parton shower is small, which indicates that the bulk of higher order corrections is captured by the next order. 
Extending the inclusive calculation in~\cite{Dixon:2018qgp} to the spinning case is necessary in order to provide an accurate prediction for the spinning correlators.

In order to estimate the size of non-perturbative corrections, Fig.~\ref{fig:candbplot} shows the comparison between parton shower only and the full simulation including hadronization. The hadronization effect is at most a few percent in the whole range. The robustness has the structural origin already discussed: the ratios compare the two-point correlators at fixed angular separation $z$, weighted by different spherical harmonics that depend on the Euler angles, so the numerator and denominator probe the same angular scales of the QCD dynamics. 
Moreover, due to factorization, the collinear region weights the different spinning correlators by the global orientation of the jet, and the leading non-perturbative correction is common in both numerator and denominator.
It is instructive to contrast this with the ratios involving the projected $N$-point correlators,  for which non-perturbative corrections have been shown to be sizable and possibly underestimated by Monte Carlo simulations~\cite{Lee:2024esz}. In the projected correlators, the additional detectors are integrated over all smaller angles than the largest one of the configuration, denoted by $x_L$. Therefore, additional detectors probe parametrically distinct angular scale, in contrast to the discussed spinning correlators. 

Finally, statistical uncertainties of the spinning correlators are larger than the statistical uncertainties of the inclusive correlator. As is transparent from Fig.~\ref{fig:candbplot}, the spinning correlators are not positive definite and their variance is larger than the inclusive ones. So reaching a given precision on spinning correlators requires a correspondingly larger sample. 

It should be emphasized that the spinning correlators shown in Fig.~\ref{fig:candbplot} are accessible in current $e^+e^-$ data.

\subsection{The flow of the spinning correlators}
\label{sec:flow:spinningcorrelator}

The spinning correlators in QCD in Eq.~\ref{eq:EEC:bc:QCD} and Fig.~\ref{fig:candbplot} are supposed to be valid only at \textit{at high energies} where $q^2$ injected in the hadronic current in Eq.~\ref{eq:Npointhadronictensor} is much larger than the hadronization scale $\Lambda_{QCD}^2$. 
It is only in this regime that the vector current $J_\mu$ interpolates with free fermion states, and therefore the result is Eq.~\ref{eq:born1cbcorrelators} correct.

In the opposite limit, that due to the mass gap in QCD is $q^2\to 4m_\pi^2$, the vector current $J^\mu$ does not interpolate with free fermion states but rather with free scalars, the pions. 
As emphasized in Section~\ref{sec:twopointspinning}, fermions and scalars imply a very different angular momentum configuration for energy fluxes, see Fig.~\ref{fig:bdries2pt}. 
This implies that the spinning part of the correlator flows from the one corresponding to free fermions to the one corresponding to free scalars as the theory flows from the UV to the IR. This effect was observed and the flow explicitly computed recently for the case of the spinning one-point energy correlator in \cite{Riembau:2025wjc}. Here we show that this flow between boundaries is a generic feature of $N$-point spinning energy correlators of theories with different matter content in the UV and IR.

In the Born approximation, the correlators for the free two pion state are
\be
H^\mathbb{1}_{\cale\cale} =  \frac12 \delta(z)\,+\, \frac12 \delta(1-z)\,,\quad\quad
H^c_{\cale\cale} =  \frac12 \delta(z)\,,\quad\quad
H^b_{\cale\cale} = \frac12 \delta(1-z)\,,
\ee
independent of the pion mass and to be compared with the case of high energy states in Eq.~\ref{eq:born1cbcorrelators}. 
Since QCD is a gapped theory, this is the full correlator at low energies with a single contribution at the vertices of parameter space for $z=0$ and $z=1$. 
Above the three pion threshold, the three pion state contributes to the correlator. Since it goes though an interaction of the type $\epsilon^{\mu\nu\rho\sigma}p_1^\nu p_2^\rho p_3^\sigma$, it gives a contribution located at the WZW vertex in Fig.~\ref{fig:bdries2pt} for $0<z<1$.

\begin{figure}
	\centering
	\includegraphics[width=0.45\linewidth]{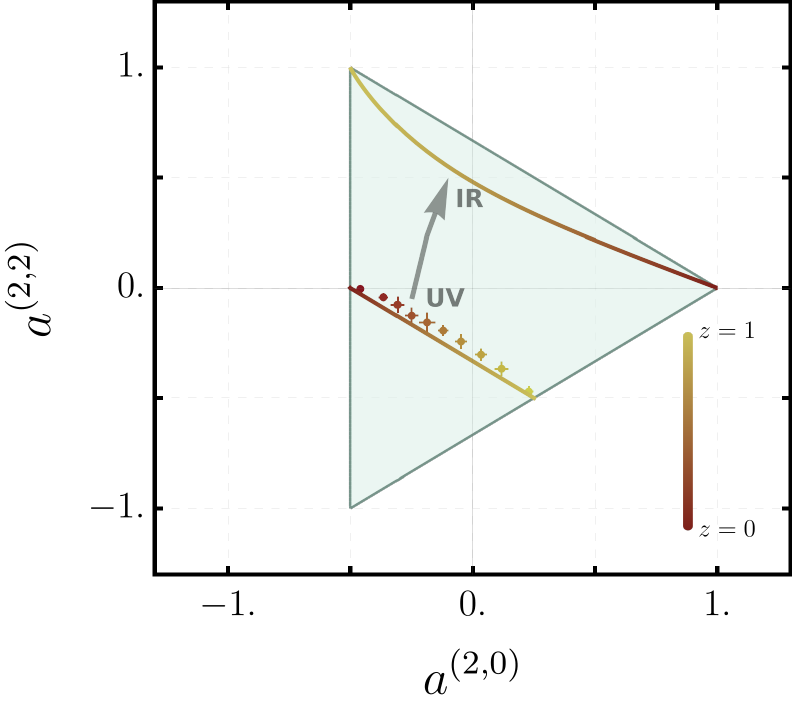}
	\caption{\small Allowed space of spinning correlators, as in  Fig.~\ref{fig:bdries2pt}. We plot the correlators in QCD. At large energies (UV), the line is equivalent to the plot in Fig.~\ref{fig:candbplot}. At low energies (IR), the current overlaps instead with scalars, the pions, which live in the vertices of the upper region. Photon radiation links the exact collinear and back-to-back Born limits. Thus, going from the UV to the IR induces a flow on the functional form of the spinning correlator, interpolating between the two.}
	\label{fig:candbplot_triang}
\end{figure}

This is shown in Fig.~\ref{fig:candbplot_triang}, where we plot the correlators in Eq.~\ref{eq:EEC:bc:QCD} in the allowed space by the positivity, see Fig.~\ref{fig:bdries2pt}. Again, this is done in the basis specified by the choice in Eq.~\ref{eq:twopointcanonicalbasis}, see Fig~\ref{fig:twopointcorrdiagram}. The line and the points near the ``UV" label in Fig.~\ref{fig:candbplot_triang} correspond to the LO calculation and the simulation in Fig.~\ref{fig:candbplot}. 
In particular, the line near UV is exactly straight specified by the relation $a^{(2,0)} + \frac{3}{2} a^{(2,2)} = -\frac{1}{2}$ at the LO.
The line near the ``IR" label denotes, instead, the contribution from the $\pi^+\pi^-\gamma$ states, neglecting the effect of the pion mass. For finite $\alpha_{em}$, the correlators are continuous from $z=0$ to $z=1$, even if possibly very small, for instance at $z\sim 1/2$ and near threshold, or at $z\ll m_\pi^2/Q^2$.

In summary, in QCD, when flowing from the UV to the IR, the spinning correlators, keeping track of the angular momentum of the energy fluxes, flow from the ``UV" line to the ``IR" line in  Fig.~\ref{fig:candbplot_triang}. Perturbations around the ``UV" line are controlled by the strong coupling $\alpha_s$, and are therefore amenable for higher order calculation. Perturbations around the ``IR" line come from the three pion and four pion states. The three pion state only dominates on the $\omega$ and $\phi$ resonances, so at these energies the line receives a large contribution from the WZW vertex in the other side of the allowed space in the triangle. Four pion states are expected to give large deviations with respect to the two pion state due to intermediate resonances. Similar to the results in \cite{Riembau:2025wjc} for the one-point energy correlator, we expect the resulting trajectory as a function of $q^2$ to be a non-monotonic flow and sensitive to resonances and thresholds. 

For completeness, we report the $\mathcal{O}(\alpha_{em})$ contributions to the inclusive and the spinning correlators. While the inclusive one is given by
\be \label{eq:sqed:eec:trace}
\begin{split}
H^\mathbb{1}_{\mathcal{E}\mathcal{E}}
	&=\frac{\alpha_{em}}{\pi} \left [
	- \frac{ 2z (4 z^3+3 z^2-15 z+54) + 3(z^4+z^3+4 z^2-28 z+36)\log(1-z)}{24z^5}
	\right .
	\\[5pt]
	& \hspace{1.5cm}
	+ \frac{1}{12} \left [ \frac{1}{z} \right ]_+  + \frac{85}{288} \delta (z)
	\\[5pt]
	&\left .  \hspace{1.5cm}  -\frac{1}{4} \left [ \frac{1}{1-z} \right ]_+ 
	-\frac{1}{8} \left [ \frac{\log(1-z)}{1-z} \right ]_+ + \left ( -\frac{\zeta_2}{8} - \frac{1}{4} \right ) \delta(1-z)
	\right ]~,
\end{split}
\ee
the spinning ones are given by
\be
\begin{split} \label{eq:sqed:eec:HcHb}
	H^c_{\cale\cale} (z)  &= \frac{\alpha_{em}}{\pi} \left [
	\frac{- z (z (z (2 z+3)+33)-54) + \left(9 z^2-60 z+54\right) \log (1-z)}{24 z^5} \right .
	\\[5pt]
	&\left .  \hspace{1.5cm} + \frac{1}{12} \left [ \frac{1}{z} \right ]_+ + \frac{53}{576} \delta (z) \right ]~,
	\\[5pt]
	H^b_{\cale\cale} (z) &=  \frac{\alpha_{em}}{\pi}\left [
	- \frac{1}{8} \frac{2 z^3+z^2+6z +\left ( z^3+z^2 -2z+6\right) \log (1-z) }{z^4} \right .
	\\[5pt]
	&\left . \hspace{1.5cm}  -\frac{1}{4} \left [ \frac{1}{1-z} \right ]_+ - \frac{1}{8} \left [ \frac{\log(1-z)}{1-z} \right ]_+ 
	+ \left ( -\frac{\zeta_2}{8} - \frac{1}{4} \right ) \delta(1-z) \right ]~.
\end{split}
\ee
The appropriate ratios lead to the spinning functions for the free two pion state at low energies,
\be
\begin{split}
	a^{(2,0)}_{\cale\cale}(z) &= \frac{z(36-52z+15z^2+3z^3)+(36-70z+38z^2-3z^3)\log(1-z)}{-4 z (6 z^2 - 23 z+18) +2 \left [ z(z-8) (3 z-8)-36 \right ] \log (1-z)}~,
	\\[5pt]
	a^{(2,2)}_{\cale\cale}(z) &= \frac{z \left [ - z(z-2) (z-3)  + (-3 z^2 + 8z - 6) \log (1-z) \right ]}{-2 z (6 z^2 - 23 z+18) + \left [ z(z-8) (3 z-8)-36 \right ] \log (1-z)}~.
\end{split}
\ee

\subsection{Spinning beyond energy}
\label{sec:spinning:beyondenergy}

While we centered the discussion around energy correlators, due to their theoretical and experimental simplicity and cleanness, one can of course consider the correlator of any other type of detector. A particularly relevant set is given by detectors measuring the charges of global currents, defined to act on multiparticle states as
\be
\mathcal{Q}_n |\alpha \rangle \,=\, \sum_{i\in \alpha} q_i \delta^{(2)}(\Omega_i-\Omega_n)\,|\alpha\rangle~,
\label{eq:Qnonalpha}
\ee
where $q_i$ is given by the charge of the measured particle. Since the charge is associated to a conserved current, its associated operator does not develop anomalous dimensions and we can express and measure it in terms of hadrons while express it and compute it in terms of partons. Moreover, a large set of correlators of global charges can be argued to be perturbatively safe under soft and collinear radiation, as argued in \cite{Riembau:2024tom}. These involve correlators of the global symmetries of QCD, namely electric charge $\mathcal{Q}$, but also isospin $\mathcal{I}$ and baryon number $\mathcal{B}$. 
In particular, a class of perturbatively IR safe correlators is the energy-charge correlator $\ab{\cale_{n_1} \mathcal{Q}_{n_2}}$. 
The reason for this can be readily seen taking into account the factorization in Eq.~\ref{eq:factorizationFF}. Both $\cale_n$ and $\mathcal{Q}_n$ annihilate the soft sector, the latter because soft radiation carries zero charge on average, and the contribution from the soft sector averages to zero after being inclusive over all states in the theory. 
Moreover, they both act homogeneously on the collinear sectors, measuring the energy and charge of the parent parton, similarly when acting on the same collinear sector, see \cite{Riembau:2024tom} for more details \footnote{As shown recently in \cite{Monni:2025zyv}, a similar argument holds for the charge-charge correlator in the back-to-back limit. A way to see that it is IR safe is to realize that the correlator factorizes as the convolution of two one-point charge correlators, which as argued are IR safe}. 
In the following we consider the calculation of the spinning correlator between the energy and one of the global charges, like $\ab{\cale_{n_1} \mathcal{Q}_{n_2}}$. The reason is three-fold. First, to explicitly show that it is IR-safe. Second, it will unveil a different spinning structure with respect to the energy-energy correlator. Third, it will be useful for the discussion of sum rules in the next Section \ref{sec:sumrules}.

Consider the chiral current sourced by an electroweak gauge boson. The structure of the hadronic tensor associated with the two-point energy-charge correlator has more structure with respect to the one of energy-energy correlator. This is due to the fact that the two detectors are now distinct. The hadronic tensor is given by 
\be
\label{eq:tensordec2pointEQ}
\begin{split}
	H^{ab}_{\cale\mathcal{Q}}\,=\, &H^{\mathbb{1}}_{\cale\mathcal{Q}}(z)\,\frac{\delta^{ab}}{3} 
	\,\,+H^+_{\cale\mathcal{Q}}(z) \frac{i}{\sqrt{1-z}}\epsilon^{abc}\frac{n_1^c+n_2^c}{2}
	\,+H^-_{\cale\mathcal{Q}}(z) \frac{i}{\sqrt{z}}\epsilon^{abc}\frac{n_1^c-n_2^c}{2}\\[5pt]
	&+ H^c_{\cale\mathcal{Q}}(z)\left( \frac{1}{1-z}\frac{n_1^a+n_2^a}{2}\frac{n_1^b+n_2^b}{2}-\frac{\delta^{ab}}{3}\right)
+ H^b_{\cale\mathcal{Q}}(z)\left( \frac{1}{z}\frac{n_1^a-n_2^a}{2}\frac{n_1^b-n_2^b}{2}-\frac{\delta^{ab}}{3}\right)
	\\[5pt]
	&+ H^x_{\cale\mathcal{Q}}(z) \frac{1}{\sqrt{z(1-z)}}\left[ \frac{n_1^a+n_2^a}{2}\frac{n_1^b-n_2^b}{2}+\frac{n_1^a-n_2^a}{2}\frac{n_1^b+n_2^b}{2} \right]~.
\end{split}
\ee
The $H^{\mathbb{1}}_{\cale\mathcal{Q}}$ controls the inclusive part of the correlator, while $H^c_{\cale\mathcal{Q}}(z)$ and $H^b_{\cale\mathcal{Q}}(z)$ are the spinning parts associated with $J=2$, $|m|=0,2$, fully equivalent to the structures already existing in the energy-energy correlator in Eq.~\ref{eq:tensordec2point}, as they are symmetric under $\vec{n}_1\leftrightarrow \vec{n}_2$. 
The term in the last line of Eq.~\ref{eq:tensordec2pointEQ} controls instead the $J=2$, $|m|=1$ part of the correlator. It is antisymmetric under the exchange $\vec{n}_1\leftrightarrow \vec{n}_2$, equivalent to the $\phi\to \phi+\pi$ in the coordinates of Fig.\ref{fig:twopointcorrdiagram}. It vanishes for the energy-energy correlator while it is non-zero whenever the detectors are distinct. We can consider therefore the spinning correlator $H^x_{\cale\mathcal{Q}}$ to be odd under the exchange of detectors, i.e. $H^x_{\cale\mathcal{Q}}=-H^x_{\mathcal{Q}\cale}$, while the other two $J=2$ structures are symmetric.

Moreover, in the first line of Eq.~\ref{eq:tensordec2pointEQ} we show how in this case there are also $J=1$ tensors\footnote{As will be discussed in Section~\ref{sec:sumrules}, when considering finite masses for the hadrons there is also a $J=1$ component in the energy-energy correlator, which is however $\Lambda_{QCD}^2/Q^2$ suppressed.}. 
The term even under $\vec{n}_1\leftrightarrow \vec{n}_2$ is associated with the $m=0$ projection, while the odd term is associated with the $|m|=1$ component. The normalizations are such that, under the coordinate choice of Eq.~\ref{eq:twopointcanonicalbasis} or Fig.\ref{fig:twopointcorrdiagram}, the tensor does not depend on the internal angle $z$ and all the dependence on it is in the spinning correlators $H^\pm_{\cale\mathcal{Q}}(z)$.
The additional $J=2$ spinning correlator $H^x_{\cale\mathcal{Q}}(z)$ can be written in terms of a singlet combination of the hadronic tensor as
\be\label{eq:Hx:singlet}
H^x_{\cale\mathcal{Q}}(z) = \frac{1}{2\,\sqrt{z(1-z)}}
\left[ \frac{(n_1+n_2)_\mu}{2}\frac{(n_1-n_2)_\nu}{2}+\frac{(n_1-n_2)_\mu}{2}\frac{(n_1+n_2)_\nu}{2} \right ]H^{\mu\nu}~\,,
\ee
while the other structures are projected with Eq.~\ref{eq:singlets:ESD}. The $J=1$ spinning correlators $H^\pm_{\cale\mathcal{Q}}(z)$ can instead be written as
\be\label{eq:Hpm:singlets}
H^+_{\cale\mathcal{Q}}(z) =  \frac{-i}{2\sqrt{1-z}}\epsilon_{\rho\mu\nu\sigma}q^\rho \frac{(n_1+n_2)^\sigma}{2} H^{\mu\nu}~,\quad
H^-_{\cale\mathcal{Q}}(z) =  \frac{-i}{2\sqrt{z}}\epsilon_{\rho\mu\nu\sigma}q^\rho \frac{(n_1-n_2)^\sigma}{2} H^{\mu\nu}~.
\ee
If we were to measure such correlator in an unpolarized state, then the $J=1$ part of the correlator would vanish due to the antisymmetry of the tensor. However, the chiral nature of the electroweak interactions implies that the source for the chiral currents $J^\mu$ is in general polarized and therefore accessible experimentally. For instance, an example of a system where this could be measured, together with the one-point charge correlators as pointed out in \cite{Riembau:2024tom}, is in $W$ boson decays. An environment where this measurement could be performed at the LHC is in semileptonic $t\bar{t}$ decays, where the sample of relatively clean hadronically decaying $W$ bosons at LHC is reaching the level of hadronic $Z$ decays at LEP, and allows for precision studies, see e.g. \cite{Kats:2015cna,Kats:2023gul}. In the future, the $WW$ threshold run of a potential lepton collider like the FCC-ee would be a cleaner environment. 

At Born level, normalizing the total energy to 1, the correlators are given by
\be
H^\mathbb{1}_{\cale\mathcal{Q}} =  \frac{1}{4}(q_q+q_{\bar{q}^\prime})\left ( \delta(z)\,+\, \delta(1-z) \right )~,\quad
H^c_{\cale\mathcal{Q}} =  -\frac{1}{8}(q_q+q_{\bar{q}^\prime}) \delta(z)~,\quad
H^b_{\cale\mathcal{Q}} = -\frac{1}{8}(q_q+q_{\bar{q}^\prime}) \delta(1-z)~,
\label{eq:born1cbcorrelatorsEQ}
\ee
with $q_q$ and $q_{\bar{q}^\prime}$ being the charges of the quark and antiquark  of the chiral current. This result is equivalent to the one for the energy-energy correlator up to the replacement of $\frac12\to q_q,q_{\bar{q}^\prime}$ where $1/2$ represents the quark energy\footnote{Note the extra factor of $1/2$ due to considering now a chiral current}.  Instead, the $J=2$, $|m|=1$ correlator vanishes, $H^x_{\cale\mathcal{Q}}=0$. It should be noted that the correlators are proportional to both the total energy and the total charge $q_q+q_{\bar{q}^\prime}$ of the state. 
The $J=1$ components are computed via Eq.~\ref{eq:Hpm:singlets} and the Born level contributions are given by
\be
H^+_{\cale \mathcal{Q}}\,=\, -(q_q-q_{\bar{q}^\prime})\frac14 \delta(z)\,\,,\,\,\,\,\,
H^-_{\cale \mathcal{Q}}\,=\, -(q_q-q_{\bar{q}^\prime})\frac14 \delta(1-z)\,,
\ee
that is, they are only nonvanishing in the collinear and back-to-back limit as can be expected from the form of the tensor structure. Importantly, they are instead proportional to the difference of quark charges. This implies that even neutral chiral currents, like the ones the $Z$ boson couples to, lead to nonvanishing $J=1$ correlators if the source is polarized. These should be accessible in $e^+e^-$ collisions at the $Z$ pole.

At NLO, the calculation of the spinning energy-charge correlators is, as was mentioned, IR safe and the soft and collinear $1/\epsilon$ divergences cancel. For the chiral current, the inclusive correlator is given by
\begin{equation}\label{eq:eqc:Htr}
	\begin{split}
		H^{\mathbb{1}}_{\cale\mathcal{Q}}(z)
		&= (q_q+q_{\bar{q}'})\frac{\alpha_s}{\pi} C_F \left [
\frac{z(6-z(13z+15))-6\left(z(z^2+z+3)-1\right)\log(1-z)}{24 z^4} \right . \\[5pt]
		&  \hspace{3.cm} +  \frac{1}{6}\left[\frac{1}{z}\right]_+
		+  \delta(z)\frac{13}{144}
		\\[5pt]
		&\left . 
		\hspace{3.cm} 
		-\frac38\left[\frac{1}{1-z}\right]_+ 
		-\frac14\left[\frac{\log(1-z)}{1-z}\right]_+
		+\delta(1-z)\left(-\frac14 \zeta_2-\frac38\right)
		\right ]~,
	\end{split}
\end{equation}
where we have normalized the total energy to 1. 
The spinning $J=2$ structures in Eq.~\ref{eq:tensordec2pointEQ}  are given by
\begin{equation}\label{eq:eqc:HcHbHx}
	\begin{split}
		H^{c}_{\cale\mathcal{Q}}(z)
		&= (q_q+q_{\bar{q}'})\frac{\alpha_s}{\pi} C_F \left [
		\frac{z(z(4z+9)-6)+6(2z-1)\log(1-z)}{48 z^4} -  \frac{1}{12}\left[\frac{1}{z}\right]_+ + \delta(z)\frac{7}{144}
		\right ]~,\\[5pt]
	H^{b}_{\cale\mathcal{Q}}(z)
	&= (q_q+q_{\bar{q}'})\frac{\alpha_s}{\pi} C_F \left [ 
	\frac{3z(3z+2)+6(z^2+z+1)\log(1-z)}{48 z^3} \right . \\[5pt]
	&\left . \hspace{3cm}+\frac{3}{16}\left[\frac{1}{1-z}\right]_+ + \frac18\left[\frac{\log(1-z)}{1-z}\right]_+-\delta(1-z)\frac12 \left(-\frac14 \zeta_2-\frac38 \right)
	\right ]~,\\[5pt]
	H^{x}_{\cale\mathcal{Q}}(z)
	&= (q_q+q_{\bar{q}'})\frac{\alpha_s}{\pi} C_F \left [
	\frac{z(2-z)+2(1-z)\log(1-z)}{16 z^3\sqrt{z(1-z)}} - \delta(z)\frac{7\pi}{120}
	\right ]~.
	\end{split}
\end{equation}
The $J=2$ correlators are proportional to both the total energy and total charge. We also observe the collinear factorization as in the case of the energy-energy correlators: the coefficients of the plus distribution of the spinning $H^{c}$ and $H^{b}$ are $-1/2$ times the distributions of the inclusive correlator, as is dictated by the ratio between the correlators at Born level in Eq.~\ref{eq:born1cbcorrelatorsEQ}. While we are considering a chiral current, it should be noted that the inclusive and spinning $J =2$ correlators in Eqs.~\ref{eq:eqc:Htr} and~\ref{eq:eqc:HcHbHx} also exist for a vector current and the overall size of them is twice the case for the chiral current. Nonetheless, being proportional to $q_q+q_{\bar{q}'}$, they would vanish for the electromagnetic current. 
Finally, the spinning $J=1$ structures in Eq.~\ref{eq:tensordec2pointEQ} for the chiral current are given by
\bea\label{eq:eqc:Hpm}
\nonumber
 H^+_{\mathcal{E}\mathcal{Q}} (z) \!\!\!\! &=&\!\!\!\!  ( q_q - q_{\bar{q}'} ) \frac{\alpha_s}{\pi} C_F
  \bigg [
  - \frac{1}{48} \frac{z(12-12z+(4\sqrt{1-z}-3)z^2)+6(2-3z+z^2)\log(1-z)}{z^4 \sqrt{1-z} }
 \\[5pt]
  &&\hspace{1.5cm} 
    + \frac{1}{12} \left [ \frac{1}{z} \right ]_+ + \delta(z)  \frac{1}{288} \bigg ]~,
\\[10pt]
\nonumber
 H^-_{\mathcal{E}\mathcal{Q}} (z) \!\!\!\!&=&\!\!\!\!  ( q_q - q_{\bar{q}'}) \frac{\alpha_s}{\pi} C_F
 \bigg [ 
\frac{1}{16} \frac{z \left(3 z^2+2 \left(\sqrt{z}+1\right) (z+2)\right)+2 \left(z^3+\left(\sqrt{z}+1\right) \left(z^2+2\right)\right) \log (1-z)}{\left(\sqrt{z}+1\right) z^{7/2}}
  \\[5pt]
  \nonumber
  &&\hspace{1.5cm} 
  + \frac{3}{16} \left [ \frac{1}{1-z} \right ]_+ + \frac{1}{8} \left [ \frac{\log(1-z)}{1-z} \right ]_+
  - \delta(1-z) \frac12 \left ( - \frac{1}{4}\zeta_2 - \frac{3}{8} \right )  + \delta(z) \frac{\pi}{15} \bigg ]~.
\eea

\begin{figure}
	\centering
	\includegraphics[width=0.6\linewidth]{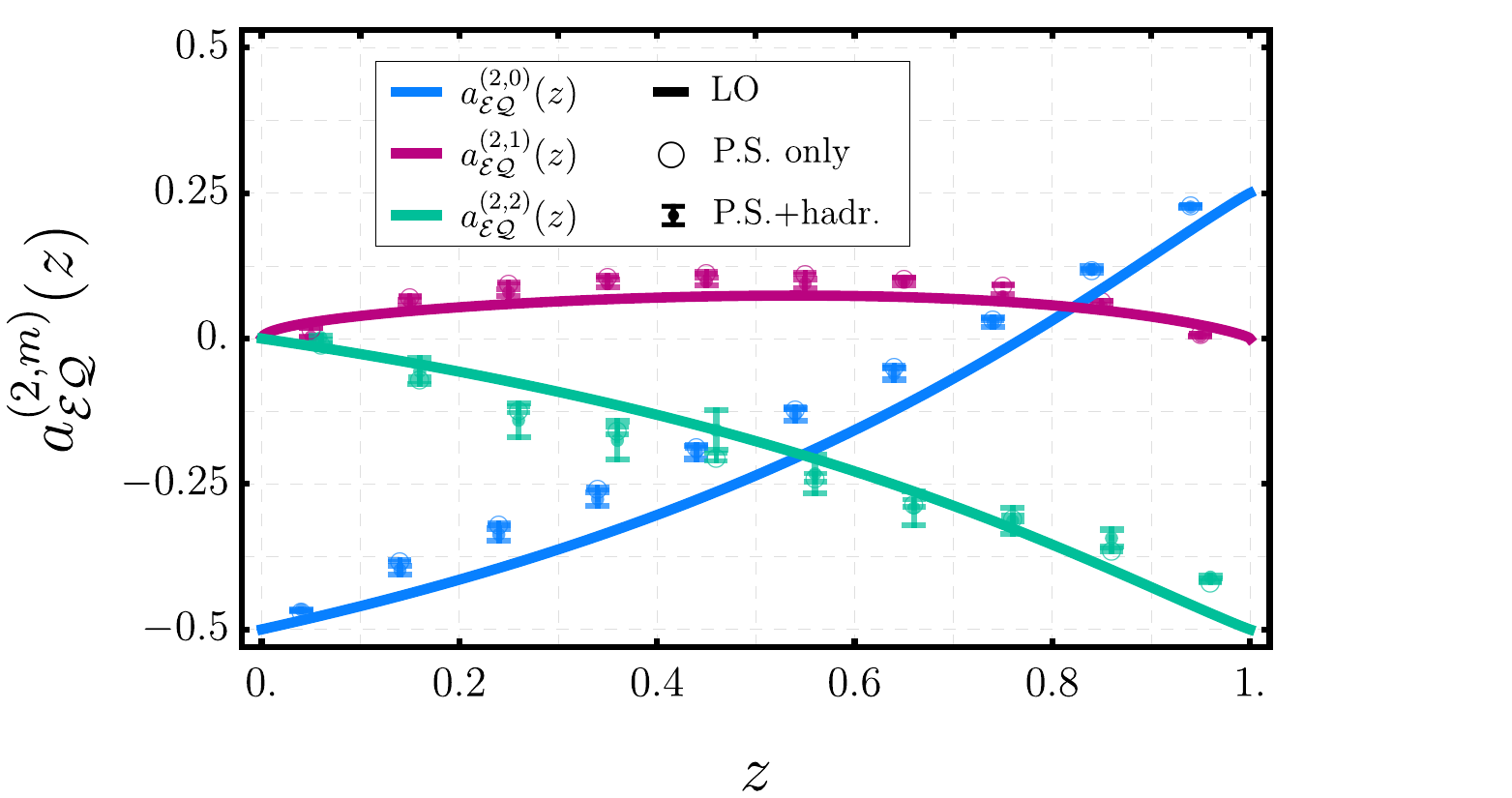}
	\caption{\small $J=2$ components of the Spinning Energy-Charge correlator as a function of the internal angle $z$, normalized to the inclusive correlators. The solid line is the analytical result in Eq.~\ref{eq:a2mEQ}, while the data points are the result of a simulation of parton shower and hadronization via \PT. In blue, purple and green, the spinning correlators $a^{(2,0)}_{\cale\mathcal Q}$, $a^{(2,1)}_{\cale\mathcal Q}$ and $a^{(2,2)}_{\cale\mathcal Q}$, respectively.}
	\label{fig:EQplotJ2}
\end{figure}

Using the convention in Eq.~\ref{eq:twopointcanonicalbasis} for the detector directions, corresponding to the angles in Fig.~\ref{fig:twopointcorrdiagram}, we obtain the spinning coefficients $a^{(2,0)}_{\mathcal{E}\mathcal Q}$ and $a^{(2,2)}_{\mathcal{E}\mathcal Q}$ being defined as in Eq.~\ref{eq:definitions:c20c22}, while $a^{(2,1)}_{\mathcal{E}\mathcal Q}$ is given by $a^{(2,1)}_{\mathcal{E}\mathcal Q}=H^x_{\cale \mathcal Q}/H^{\mathbb{1}}_{\cale \mathcal Q}$. Since all correlators are proportional to the total energy and total charge, they drop from the ratios and they only depend on the internal angle $z$. These ratios are given by

\be
\begin{split}
	a^{(2,0)}_{\mathcal{E}\mathcal{Q}} (z) &=\frac{z(z(8-7z)-4)-2(2+z(-5+4z))\log(1-z)}{4 z (2 - 7 z + 2 z^2) + 8(1 - 4 z +2 z^2) \log(1-z)}~,
	\\[5pt]
	a^{(2,1)}_{\mathcal{E}\mathcal{Q}} (z) &=-\frac{3\sqrt{z(1-z)}( z(z-2)-2(1-z)\log(1-z) )}{2 z (2 - 7 z + 2 z^2) + 4(1 - 4 z +2 z^2) \log(1-z)}~,
	\\[5pt]
	a^{(2,2)}_{\mathcal{E}\mathcal{Q}} (z) &=\frac{z( z(2+z)+2\log(1-z)) }{2 z (2 - 7 z + 2 z^2) + 4(1 - 4 z +2 z^2) \log(1-z)}~.
	\label{eq:a2mEQ}
\end{split}
\ee
We compare these analytic results with a numerical simulation in Fig.~\ref{fig:EQplotJ2}, where the blue, purple and green solid lines correspond to the expressions in Eq.~\ref{eq:a2mEQ} for $a^{(2,0)}_{\cale\mathcal Q}$, $a^{(2,1)}_{\cale\mathcal Q}$ and $a^{(2,2)}_{\cale\mathcal Q}$, respectively. The dots are the result of a simulation of an $e^+\nu_e$ collision in \verb*|MadGraph| \cite{Alwall:2014hca} followed by the parton shower and hadronization via \PT~\cite{Bierlich:2022pfr}, with the error bars being purely a statistical error of the simulation. 
We find that the analytical predictions for the spinning energy–charge correlators are in relatively good agreement with hadron-level simulations, confirming that these observables are infrared safe. The residual discrepancy between the analytical result and the simulation of $a^{(2,0)}_{\cale\mathcal Q}$ is compatible to the one for the energy-energy case $a^{(2,0)}_{\cale\cale}$ in Fig.~\ref{fig:candbplot}. 
We thus expect that the difference is controlled by the same higher order calculations.

A comment on statistical uncertainties is in order. For a dataset of a fixed size, the statistical uncertainty associated to the spinning energy-charge correlators is larger than the statistical uncertainty associated with the spinning energy-charge correlator. This is because convergence not only requires the angular averages to stabilize, but also the charge-weighted measurements of the correlator at each bin. For this reason the simulation used for Fig.~\ref{fig:EQplotJ2} contains now 20 million events. Still, this could be improved: given that the soft sector is annihilated by the correlator, imposing a lower energy cut on the hadrons does not affect the mean value provided the cut is sufficiently mild. In practice, this allows for a substantial improvement in the statistical precision.

It is clear from Fig.~\ref{fig:EQplotJ2} that non-perturbative corrections are small across all range of $z$. This is due to the fact that we are projecting on correlators with even $J$, which 
are sensitive to the parity-even part of the correlator. 
As a result, these observables are insensitive to charge-asymmetric contributions and, in particular, to differences in the quark electric charges.
The same spinning components of the energy-isospin correlator $\ab{\cale_{n_1} \mathcal{I}_{n_2}}$ lead to very similar results, as it can be checked in the simulation. 
This implies that the correlator is insensitive to the baryon number of both quarks and final-state hadrons, as baryon charge is given by the difference between electric charge and isospin~\footnote{Corrections to this picture are induced by kaons and weak interactions, which break isospin symmetry, and induce a nonvanishing stangeness density in the final state, see \cite{Riembau:2024tom} for a discussion on this point in the context of the one-point correlators.}.

The situation for the spinning $J=1$ correlators is quite different. 
The spinning correlators $H^+_{\mathcal{E}\mathcal{Q}}$, $H^-_{\mathcal{E}\mathcal{Q}}$ in Eq.~\ref{eq:tensordec2pointEQ} control the parity-odd structures and are proportional to the difference of quark charges, as manifest in Eq.~\ref{eq:eqc:Hpm}. 
As a result, the $J=1$ components of the spinning energy-isospin correlator vanish. This prediction can be explicitly verified in the numerical simulation, in which it holds true at the per-cent level. This is the same order of magnitude and compatible with the size of isospin breaking effects in the one-point isospin density observed in \cite{Riembau:2024tom}. 
For the nonvanishing parts of the correlator, which are sensitive to the baryon number of the final-state hadrons, large nonperturbative corrections are observed. 
Nevertheless, the functional form of the correlators $a^{(1,0)}_{\mathcal{E}\mathcal{Q}}$ and $a^{(1,1)}_{\mathcal{E}\mathcal{Q}}$ shows qualitative agreement with the simulated data, and similar behavior is found  for the spinning energy-baryon correlator. A detailed investigation of the origin and structure of these nonperturbative effects in energy–charge and energy–baryon correlations is left for future work.

\section{Generic correlators and their network of sum rules}
\label{sec:sumrules}

Correlators obey a network of sum rules that relate integrals of higher point correlators to lower point ones~\cite{Korchemsky:2019nzm,Dixon:2019uzg,Kologlu:2019mfz,Chang:2022ryc}.
These stem from the definition of a correlator and hold nonperturbatively for generic detector insertions. 
In the following, we generalize the results to $N$-point correlators while clarifying the assumptions under which they apply. We show that sum rules do more than relating inclusive observables to total fluxes: when expressed in terms of spinning correlators, they act independently in each angular-momentum channel, providing a direct relation between the spinning components of $N$- and $(N-1)$-point correlators.

In gapped theories like QCD, one may consider a generic detector $\mathcal{D}_{n}$, which acts on multiparticle asymptotic states (i.e. hadrons) as 
\be
\mathcal{D}_{n} |\alpha \rangle \,=\, \sum_{j\in \alpha} d_j \delta^{(2)}(\Omega_j-\Omega_n)\,|\alpha\rangle~,
\label{eq:Dnonalpha}
\ee
where $d_j$ denotes an appropriate weight made out of a product of quantum numbers and energy of the particle $j$. 
Consider the $N$-point correlator of generic detectors, given by $\langle \mathcal{D}^1_{n_1}\mathcal{D}^2_{n_2}\cdots \mathcal{D}^N_{n_N} \rangle$. 
Integrating the detector $\mathcal{D}^1_{n_1}$ over  $\vec{n}_1$ leads to the correlator  $\langle \left(\int d\Omega_{n_1}\mathcal{D}^1_{n_1}\right)\mathcal{D}^2_{n_2}\cdots \mathcal{D}^N_{n_N} \rangle$. The definition in Eq.~\ref{eq:Dnonalpha} implies that the integral over the detector gives an operator that, when acted on multiparticle states, simply measures the total charge of the state,
\be
\left( \int d\Omega_{n}\mathcal{D}^1_{n}\right) |\alpha \rangle \,=\, \sum_{j\in \alpha} d_j  \,|\alpha\rangle~.
\label{eq:Dnonalphaintegrated}
\ee
As a result, the integral of the $N$-point correlator is related to the $(N-1)$-point correlator through the sum rule
\be
\int d\Omega_{n_1} \langle \mathcal{D}^1_{n_1}\mathcal{D}^2_{n_2}\cdots \mathcal{D}^N_{n_N} \rangle \,=\, \ab{\mathcal{D}^1} \langle \mathcal{D}^2_{n_2}\cdots \mathcal{D}^N_{n_N} \rangle~,
\label{eq:nptton-1}
\ee
where $\ab{\mathcal{D}^1}$ is the expected total flux of $\mathcal{D}^1$ integrated over the full sphere, $\ab{\mathcal{D}^1} = \int d\Omega_n \ab{\mathcal{D}^1_n} = \int d\Omega_n \frac{\la{\Psi}\mathcal{D}^1_n\ra{\Psi}}{\ab{\Psi|\Psi}}$. 
Crucially, this identity includes the points of coincident detector directions  $\vec{n}_1=\vec{n}_j$. 
When $\mathcal{D}^1$ corresponds to a conserved charge, the total flux is fixed by the quantum numbers of the source, making the sum rule particularly useful.

A key implication of the sum rules is that reconstructing a fully differential $(N-1)$ correlators requires knowledge of the fully differential $N$-point one. 
In particular, each spin-$J$ part of the $(N-1)$-point correlator is connected to the spin-$J$ part of the $N$-point one. 
In consequence, obtaining the one-point correlator requires information from the spinning two-point correlator. 
Measuring the two-point correlator inclusively over the Euler angles only determines the total flux, and the angular distribution of the flux density is lost. This is general and trivially true for any higher points.

\smallskip

When the integrated detector measures the energy, an additional sum rule can be derived under a certain assumption. Consider a detector that measures the three momentum $\vec{P}_n$. In this case, the sum rule in Eq.~\ref{eq:nptton-1} is given by
\be
\int d\Omega_{n_1} \langle \vec{P}_{n_1}\mathcal{D}^2_{n_2}\cdots \mathcal{D}^N_{n_N} \rangle \,=\, \ab{\vec{P}} \langle \mathcal{D}^2_{n_2}\cdots \mathcal{D}^N_{n_N} \rangle\,.
\label{eq:nptton-1_3momentum}
\ee
In the rest frame, where the injected momentum in the operator is $q^\mu=(Q,\vec{0})$, one has that $\ab{\vec{P}}=0$. 
Assuming massless particles in the final state, which in QCD corresponds to massless hadrons satisfying $E_i\simeq |\vec{p}_i|$, 
the three-momentum operator $\vec{P}_n$ can be interpreted as $\vec{n}\mathcal{E}_{n}$, namely the energy flux in the direction of the calorimeter. Therefore, the sum rule in Eq.~\ref{eq:nptton-1_3momentum} can be written as
\be
\int d\Omega_{n_1} \vec{n}_1 \langle \mathcal{E}_{n_1}\mathcal{D}^2_{n_2}\cdots \mathcal{D}^N_{n_N} \rangle \,=\, 0~,
\label{eq:energyextrasumrule}
\ee
which provides an additional sum rule for correlators involving the energy detector. 
Finite hadron masses may induce violation of the sum rule of order $\Lambda_{QCD}^2/Q^2$. 
In fact, in many cases the sum rule holds much better than what this naive estimate suggests. Since the violation of the sum rule is proportional to $\int d\Omega_{n}\vec{n}\ab{\cale_n}$, it is proportional to the $J=1$ component of the one-point correlator. This implies that, while the contribution may be non-zero, for an unpolarized source it does vanish. Observable deviations may nevertheless arise, for instance in polarized $W$ boson decays, in particular in the $c\bar{s}$ channel.

In the following we neglect the finite mass corrections and focus instead the consequences of Eq.~\ref{eq:energyextrasumrule}. 
By dotting this relation with $\vec{n}_i$ for $i\neq 1$ it yields $N-1$ independent sum rules, given by
\be
\int d\Omega_{n_1} (1-2z_{1i}) \langle \mathcal{E}_{n_1}\mathcal{D}^2_{n_2}\cdots \mathcal{D}^i_{n_i}\cdots  \mathcal{D}^N_{n_N} \rangle \,=\, 0\,.
\label{eq:energyextraSRdotted}
\ee
Using energy conservation, $\ab{\cale}=E$, it is convenient to combine these $N-1$ sum rules and the $N-1$ sum rules of the type in Eq.~\ref{eq:nptton-1} into the set of $2(N-1)$ relations
\be
\begin{split}
\int d\Omega_{n_1}\,z_{1i} \langle \mathcal{E}_{n_1}\mathcal{D}^2_{n_2}\cdots \mathcal{D}^N_{n_N} \rangle &= \frac{E}{2} \langle \mathcal{D}^2_{n_2}\cdots \mathcal{D}^N_{n_N} \rangle~,
\\[5pt]
\int d\Omega_{n_1}\,(1-z_{1i}) \langle \mathcal{E}_{n_1}\mathcal{D}^2_{n_2}\cdots \mathcal{D}^N_{n_N} \rangle &= \frac{E}{2} \langle \mathcal{D}^2_{n_2}\cdots \mathcal{D}^N_{n_N} \rangle~.
\end{split}
\ee
These relations are particularly useful as they explicitly eliminate contact interactions at $z_{1i}=0$ and $z_{1i}=1$. As a result, if we know the bulk contribution to the correlator $\langle \mathcal{E}_{n_1}\mathcal{D}^2_{n_2}\cdots \mathcal{D}^N_{n_N} \rangle$, determined by configurations with separated detectors $n_1\neq n_2\neq\dots\neq n_N$, the $2(N-1)$ contact terms at $\delta(z_{1i})$ and $\delta(1-z_{1i})$ are fixed by the sum rules. This is not a trivial information, as infrared singularities are localized precisely in the collinear and back-to-back limits, and for infrared-safe observables their cancellation in perturbation theory requires higher-loop contributions to these endpoint terms. 
A similar set of sum rules was discussed in \cite{Korchemsky:2019nzm,Dixon:2019uzg,Kologlu:2019mfz}, but connecting the two-point correlator with the total energy. 

The fundamental sum rule underlying these results is Eq.~\ref{eq:energyextrasumrule}, and that the form in Eq.~\ref{eq:energyextraSRdotted} is a particularly convenient one to isolate the endpoint singularities. Other projections of Eq.\ref{eq:energyextrasumrule} are possible and may be useful to isolate certain kinematic configurations. For instance, by dotting Eq.~\ref{eq:energyextrasumrule} with $\eps^{abc}n^b_{i}n^c_j$, one obtains
\be
\int d\Omega_{n_1} \vec{n}_1\cdot(\vec{n}_j\times \vec{n}_k) \langle \mathcal{E}_{n_1}\mathcal{D}^2_{n_2}\cdots \mathcal{D}^j_{n_j}\cdots \mathcal{D}^k_{n_k}\cdots \mathcal{D}^N_{n_N} \rangle \,=\, 0\,.
\ee
This implies a sum rule for planar configurations of detectors. The planar configuration is also enhanced due to three-jet events, see \cite{Gao:2024wcg}. It would be interesting to explore whether it may be possible to determine or constrain divergences of planar configurations in a similar way that the form in Eq.~\ref{eq:energyextraSRdotted} determines higher-loop contributions at endpoints of the two-point correlator.

\subsection{Relations among two and one point correlators}

The sum rule of Eq.~\ref{eq:nptton-1} allows to connect, in particular, the two-point correlator with the one-point correlator. Given the explicit decomposition discussed earlier, this relation can be made explicit at the level of individual spinning structures. In general, the hadronic tensor of the one-point correlator of a vector current is given by
\be
H^{ab}_{\mathcal{D}}\,=\, H^\mathbb{1}_{\mathcal{D}}\frac{\delta^{ab}}{3}\,+\, H^{(c)}_{\mathcal D} \left( n^an^b-\frac13\delta^{ab} \right)
\,- i \,
H^{(b)}_{\mathcal D} \epsilon^{abc}n^c\,,
\ee
where the $J=1$ term is retained as it is nonvanishing for electroweak currents. 
The decomposition for the two-point hadronic tensor $H^{ab}_{\mathcal D^i\mathcal D^j}$ admits an analogous form as in Eq.~\ref{eq:tensordec2pointEQ}. We denote by $\mathcal{N}$ the normalization by the total rate, which is common for the one- and two-point correlator. 
Doing so leads to~\footnote{
	The hadronic tensor $H^{ab}_{\cale\cale}$ contains terms at most quadratic in $\vec{n}_2$ (similarly in $\vec{n}_1$). It is straightforward to show that an arbitrary function $f(z)$ (with $z=\frac{1}{2}(1-\vec{n}_1 \cdot \vec{n}_2)$) satisfies $\int d\Omega_{n_2}\, n_2^a f(z) = n_1^a \int d\Omega_{n_2}\, (1-2z) f(z)$ and $\int d\Omega_{n_2}\, n_2^a n_2^b f(z) =
			\delta^{ab}  \int d\Omega_{n_2}\, 2z (1-z) f(z) 
			+ n_1^a n_1^b  \int d\Omega_{n_2}\, [1- 6z (1-z)] f(z)$. 
Similar integral relations apply to the integration over $\vec{n}_1$. For an odd function under the exchange of two detectors, the integration over $\vec{n}_1$ leads to an overall opposite sign to the case with integrating over $\vec{n}_2$.}
\begin{equation}\label{eq:sumrulesgeneric}
\begin{split}
\ab{\mathcal{D}^i} \ab{\mathcal{D}^j}
&=
\frac{1}{\mathcal N}\int dz\, \frac13 H^\mathbb{1}_{\mathcal{D}^i\mathcal{D}^j}~,
\\[5pt]
\ab{\mathcal{D}^i} H^{(c)}_{\mathcal{D}^j}
&=
\int dz \left[\left(1-\frac{3z}{2}\right)H^{(c)}_{\mathcal{D}^i\mathcal{D}^j}+\left(1-\frac{3(1-z)}{2}\right)H^{(b)}_{\mathcal{D}^i\mathcal{D}^j}
+ 3\sqrt{z(1-z)}H^{(x)}_{\mathcal{D}^i\mathcal{D}^j}\right]~,
\\[5pt]
\ab{\mathcal{D}^i} H^{(b)}_{\mathcal{D}^j}
&=
\int dz \left[ \sqrt{1-z}H^{+}_{\mathcal{D}^i\mathcal{D}^j}
+
\sqrt{z}H^{-}_{\mathcal{D}^i\mathcal{D}^j}\right]~,
\end{split}
\end{equation}
where the different $H_{\mathcal{D}^i\mathcal{D}^j}$ are distributions that depend on the internal angle $z$. 
The correlators $H^{(x)}_{\mathcal{D}^i\mathcal{D}^j}$ and $H^{-}_{\mathcal{D}^i\mathcal{D}^j}$ are odd under the exchange of the detectors,  $H^{(x)}_{\mathcal{D}^i\mathcal{D}^j}=-H^{(x)}_{\mathcal{D}^j\mathcal{D}^i}$ and $H^{-}_{\mathcal{D}^i\mathcal{D}^j}=-H^{-}_{\mathcal{D}^j\mathcal{D}^i}$, while the rest are even. Then, \ref{eq:sumrulesgeneric} includes all sum rules connecting the two-point correlator with the one-point.

It is worth emphasizing that the sum rules preserve total angular momentum. The trace part of the one-point correlator is fixed by the integral of the trace part of the two-point correlator, while the $J=1$ and $J=2$ components of the one-point map independently with the $J=1$ and $J=2$ components of the two-point correlator, respectively. As a consequence, the singlet $H^\mathbb{1}_{\mathcal{D}^i\mathcal{D}^j}$ maps to the total flux. 
The angular structure of the one-point correlator, determined by the parameter $a_{\mathcal{D}} = H^{(c)}_{\mathcal{D}}/\mathcal N$, necessarily requires information from the spinning two-point correlator.

The last two lines of Eq.~\ref{eq:sumrulesgeneric} encode two distinct sum rules. 
Since the left hand side is not symmetric under the exchange of detectors, any difference between, for instance, $\ab{\mathcal{D}^i} H^{(c)}_{\mathcal{D}^j}$ and $\ab{\mathcal{D}^j} H^{(c)}_{\mathcal{D}^i}$, must originate from the antisymmetric terms $H^{(x)}_{\mathcal{D}^i\mathcal{D}^j}$, $H^{-}_{\mathcal{D}^i\mathcal{D}^j}$:
\begin{equation}\label{eq:sumrulesdifference}
\begin{split}
\ab{\mathcal{D}^i} H^{(c)}_{\mathcal{D}^j}-\ab{\mathcal{D}^j} H^{(c)}_{\mathcal{D}^i}
&=
6 \int dz \sqrt{z(1-z)}H^{(x)}_{\mathcal{D}^i\mathcal{D}^j}~, 
\\[5pt]
\ab{\mathcal{D}^i} H^{(b)}_{\mathcal{D}^j}-\ab{\mathcal{D}^j} H^{(b)}_{\mathcal{D}^i}
&= 2
\int dz \sqrt{z} H^{-}_{\mathcal{D}^i\mathcal{D}^j}~.
\end{split}
\end{equation}
That is, differences in the flux densities measured by two detectors are controlled entirely by the odd terms of their two-point correlator.
Consistently, these structures vanish for the same type of detectors $H_{\mathcal{D}^i\mathcal{D}^i}^{x}=H_{\mathcal{D}^i\mathcal{D}^i}^{-}=0$. 

An important consequence arises when these relations are applied to charge–charge correlators. While the one-point charge correlator is infrared safe, the corresponding two-point correlator is not. The sum rule in Eq.~\ref{eq:sumrulesgeneric} connects the IR safe one-point correlator to the IR-sensitive two-point correlator, meaning that the sum rule must annihilate the IR-unsafe part of the two-point correlator. 
This can be understood from factorization, as in Eq.~\ref{eq:factorizationFF}. Whenever the detectors act on different collinear sectors, or in the collinear and the soft sector, the correlator is IR-safe for the same reasoning that the one-point charge correlator is IR-safe, see \cite{Riembau:2024tom}. Sensitivity to IR come from the cases where both detectors act either on the soft sector or in the same collinear sector. 
When integrating over $z$, which crucially includes the coincident limit $z=0$ where the detectors are on top of each other, sums over all pairings eliminate the soft contribution while correctly reproducing the charge squared of the parent parton in the collinear case.

\begin{figure}
	\centering
	\includegraphics[width=0.4\linewidth]{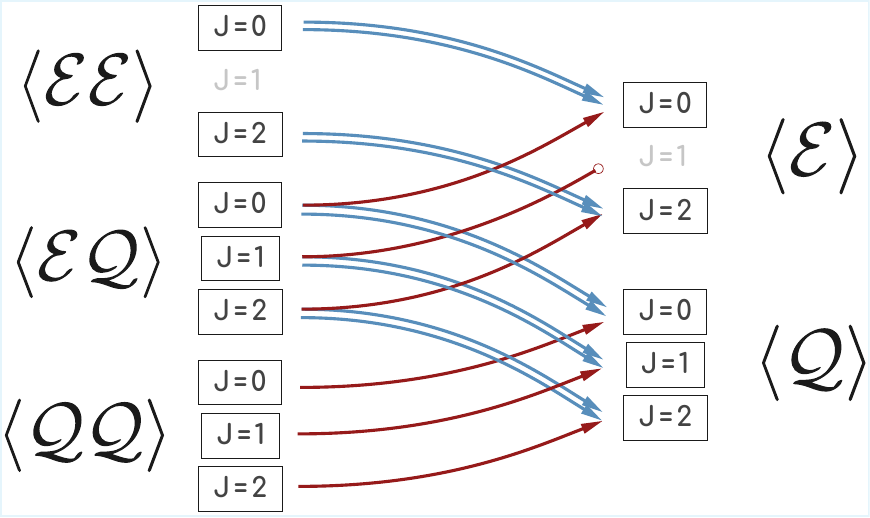}
	\caption{\small Representative of the network of sum rules between the different two-point $\langle \cale \cale \rangle$, $\langle \cale \mathcal Q\rangle $ and $\langle \mathcal Q\mathcal Q\rangle$ spinning correlators and the one-point $\langle \cale \rangle$ and $\langle \mathcal Q \rangle$ spinning correlators. Blue arrows indicate the integration over an energy detector, red ones over a charge detector. The doubling of blue arrows are associated to the bonus sum rule due to momentum conservation.}
	\label{fig:diagSR}
\end{figure}

When one of the detectors is the energy flow operator, $\mathcal{D}^i=\cale$, additional sum rules follow from the momentum conservation in Eq.~\ref{eq:energyextrasumrule}. They can be written as
\be
\int d\Omega_{n_1} z \ab{\cale_{n_1}\mathcal{D}_{n_2}} = \frac{\ab{\cale}}{2}\ab{\mathcal{D}_{n_2}}~,
\hspace{0.6cm}\text{and}\hspace{0.6cm}
\int d\Omega_{n_1} (1-z) \ab{\cale_{n_1}\mathcal{D}_{n_2}} = \frac{\ab{\cale}}{2}\ab{\mathcal{D}_{n_2}}~,
\ee
where $\ab{\cale}$ is the total energy flux. Together with the sum rules obtained by integrating the energy detector, these relations can be expressed in terms of the scalars controlling the one-point correlator,
\begin{equation}\label{eq:sumrules:integrateE}
\begin{split}
\ab{\mathcal{D}}
&=
\frac{1}{\ab{\cale}}\frac{1}{\mathcal{N}}\int dz\, \mu (z) \frac13 H^\mathbb{1}_{\cale \mathcal{D}}~,
\\[5pt]
H^{(c)}_{\mathcal{D}}
&=
\frac{1}{\ab{\cale}}\int dz\, \mu(z)\, \left[\left(1-\frac{3z}{2}\right)H^{(c)}_{\cale \mathcal{D}}+\left(1-\frac{3(1-z)}{2}\right)H^{(b)}_{\cale \mathcal{D}}
 - 3\sqrt{z(1-z)}H^{(x)}_{\cale \mathcal{D}} \right]~,
\\[5pt]
H^{(b)}_{\mathcal{D}}
&=
\frac{1}{\ab{\cale}}\int dz\, \mu(z)\, \left[ \sqrt{1-z}H^{+}_{\cale\mathcal{D}} - \sqrt{z} H^{-}_{\cale\mathcal{D}}\right]~,
\end{split}
\end{equation}
where $\mu(z)$ is equal to either $2 z$ or $2(1-z)$ for the momentum conservation sum rules, and $\mu(z)=1$ for the sum rules of Eq.~\ref{eq:sumrulesgeneric}. 
The relative minus signs in front of odd terms in Eq.~\ref{eq:sumrules:integrateE} are consistent with the explicit decomposition in Eqs.~\ref{eq:eqc:HcHbHx} and~\ref{eq:eqc:Hpm} for the energy-charge correlator. 
One may alternatively consider to integrate the detector $\mathcal{D}$. While the inclusive part of the sum rule does not provide an extra information, the spinning part, however, reads
\begin{equation}\label{eq:sumrules:integrateD}
\begin{split}
a_{\cale}
&=\frac{1}{\ab{\mathcal D} \ab{\cale }}
\frac{1}{\mathcal{N}}\int dz\, \left[\left(1-\frac{3z}{2}\right)H^{(c)}_{\cale \mathcal{D}}+\left(1-\frac{3(1-z)}{2}\right)H^{(b)}_{\cale \mathcal{D}}
 + 3\sqrt{z(1-z)}H^{(x)}_{\cale \mathcal{D}}\right]~,
\\[5pt]
0
&=
\int dz\, \left[
\sqrt{1-z}H^{+}_{\cale\mathcal{D}} + \sqrt{z}H^{-}_{\cale\mathcal{D}}\right]~,
\end{split}
\end{equation}
where $a_\cale$ is the parameter controlling the one-point energy correlator, defined as $H^{c}_\cale/H^\mathbb{1}_\cale$, and $\ab{\mathcal D}$ is assumed to be nonzero. As indicated in Eq.~\ref{eq:sumrulesdifference}, any nonzero difference between $a_\cale$ and the corresponding parameter $a_{\mathcal D}$ for $\mathcal D$ should be given by the integral of $H^{(x)}_{\cale \mathcal{D}}$. Finally, the second sum rule in Eq.~\ref{eq:sumrules:integrateD} follows from assuming the $J=1$ component of the one-point energy correlator to be zero, as previously discussed. Thus, it provides a bonus relation between the integrals of the two components of the $J=1$ part of the two-point $\ab{\cale \mathcal D}$ correlator. 
As a summary, Fig.~\ref{fig:diagSR} illustrates the network of sum rules relating the various structures of the two-point correlators $\langle \cale \cale \rangle$, $\langle \cale \mathcal Q\rangle $ and $\langle \mathcal Q\mathcal Q\rangle$ to the coefficients that control the one-point correlators $\langle \cale \rangle$ and $\langle \mathcal Q \rangle$.

That endpoints can be obtained from the bulk contributions is well known for the case of the inclusive part of the energy-energy correlator \cite{Dixon:2019uzg,Korchemsky:2019nzm,Kologlu:2019mfz,Chang:2022ryc}. The same is true for the inclusive part of the energy-charge correlator due to momentum conservation, and one has
\be
H^{\mathbb{1}}_{\cale\mathcal{D},0}= \frac{E\ab{\mathcal{D}}}{2} - \int dz\,(1-z)\,  H^{\mathbb{1}}_{\cale\mathcal{D},\text{reg}}(z)~,
\hspace{0.4cm}\text{and}\hspace{0.4cm}
H^{\mathbb{1}}_{\cale\mathcal{D},1}= \frac{E\ab{\mathcal{D}}}{2} - \int dz\,z\,  H^{\mathbb{1}}_{\cale\mathcal{D},\text{reg}}(z)~.
\ee
where $H^{\mathbb{1}}_{\cale\mathcal{D},\text{reg}}$ denotes the regular part of the correlator, while $H^{\mathbb{1}}_{\cale\mathcal{D},0}$ and $H^{\mathbb{1}}_{\cale\mathcal{D},1}$ the coefficients of the $\delta(z)$ and $\delta(1-z)$, respectively. 
By contrast, correlators of two conserved charges that do not include the energy, the only sum rule is given by
\begin{equation}
	\begin{split}
		\ab{\mathcal{D}}\ab{\mathcal{D}^\prime}\,=\, \int dz\, H^{\mathbb{1}}_{\mathcal{D}\mathcal{D}^\prime,\text{reg}}(z)\,+\, H^{\mathbb{1}}_{\mathcal{D}\mathcal{D}^\prime,0}+H^{\mathbb{1}}_{\mathcal{D}\mathcal{D}^\prime,1}~,
	\end{split}
\end{equation}
so the only quantity fixed by the bulk contribution is the sum of the endpoint contributions. 

At the nonperturbative level, in QCD, collinear divergences are not regularized by a plus distribution but instead by the generation of a mass gap. Thus, there is no $\delta(1-z)$ contribution and the coefficient of $\delta(z)$ is the one-point correlator $\ab{\cale^2}$. For the energy-energy correlator, this leads to the sum rule
\be
E^2 \,=\, \ab{\cale^2} \,+\, \int_\epsilon^1 dz\, H_\text{bulk}(z)\,
\ee
which expresses the fact that the total energy squared is either carried by the individual hadrons or spread in the correlations. This relation differs qualitatively from the other sum rules connecting the two-point correlators to the one-point ones, and it can be equally generalized to other types of detectors. 

The consistency conditions are expanded to the spinning correlators. 
For the energy-energy correlator, the spinning structure $H^c_{\cale\cale}$ has no back-to-back boundary term, $H^c_{\cale\cale,1} =0$, and the structure $H^b_{\cale\cale}$ has no collinear boundary term, $H^b_{\cale\cale,0} =0$. Under these conditions,  
the sum rules in Eq.~\ref{eq:sumrules:integrateE} uniquely determine the remaining boundary terms. This can be explicitly verified in perturbative QCD using the result presented in Sections~\ref{sec:spinning:eec},~\ref{sec:flow:spinningcorrelator}, and~\ref{sec:spinning:beyondenergy}. 

These results highlight that sum rules constrain spinning correlators in close analogy with the inclusive case. Crucially, retaining the full angular dependence is necessary to determine lower-point correlators in each angular momentum channel.


\section{Conclusions}
\label{sec:conclusions}

Most existing studies on energy and charge correlators focus on their inclusive structure. The full angular dependence of the correlator encodes additional, previously unexplored physical information of both the source and of the angular structure of the final state, information which is otherwise integrated out at the inclusive level. 
In this work, we have presented a systematic framework that retains this information, organizing it into \textit{spinning correlators}, classified by their angular momentum quantum numbers and directly accessible both theoretically and experimentally. 

A central result of this work is that spinning correlators are highly constrained by unitarity and energy positivity. 
We demonstrated that the space of spinning correlators forms a bounded, multidimensional region, whose boundary is saturated by states with definite spin projections along the detector axis, while generic theories populating the interior correspond to convex combinations of such extremal configurations. The bounds on spinning correlators are the extension of the classic Hofman–Maldacena bounds for one-point energy correlators to sources with arbitrary spin and to general $N$-point energy correlators. Our presented framework provides a unified and physically transparent interpretation of positivity constraints in terms of angular momentum structure. 
As an explicit example, within the allowed two-dimensional triangular bounded space of spinning two-point correlators, we showed that weakly coupled quarks at fixed energy trace a straight-line trajectory, while free pions interpolate along a curved path between extremal vertices, with the remaining vertex saturated by WZW interactions. 
This picture suggests that, as the theory flows from the perturbative to the hadronic regime, spinning correlators evolve along nontrivial trajectories, making them sensitive probes of confinement and hadronization dynamics.

We illustrated these ideas with two-point correlators in perturbative QCD, where we computed all independent spinning correlators for a spin-one current, including their endpoint contributions. 
We presented new set of observables, given by the ratios of spinning correlators to the inclusive one, that they can be accurately computed in perturbation theory and precisely measured. These ratios are direct probes of the hard structure of the underlying process. 
We also presented the spinning energy-charge correlator and showed that the different $J=0$ and $J=2$ components of the spinning correlator have small non-perturbative corrections. Therefore, the perturbation theory result gives a good description of the simulated data at hadron level. 
In addition, we derived general sum rules relating $N$-point and $(N-1)$-point correlators, extending well-known inclusive relations to the fully differential ones. These sum rules hold independently for each spin component. 

The presented spinning correlators can be experimentally measured for any spinning source, notably in hadronically decaying electroweak vectors and the top quark. Existing $e^+e^-$ data can be used in order to measure the spinning energy correlators of Fig.~\ref{fig:candbplot} and the $J=1$ energy-charge correlators. 

The spinning energy correlators proposed here form a natural and systematic set of observables that classify the full kinematical dependence of energy and charge correlations. They open several promising directions for future work, ranging from the exploration of positivity and unitarity constraints in quantum field theory, to higher-order and nonperturbative calculations, and to precision experimental studies at current and future colliders.

\section*{Acknowledgments}
MR and MS thank Gauthier Durieux, Chul Kim, Taehyun Kwon, Francesco Riva for useful discussions. 
MS was supported by National Research Foundation of Korea (NRF) under Grant Number RS-2024-00450835.

\appendix

\section{Spinning sum rules in QCD}
\label{app:sec:spinning:sumrules:QCD}
In this Appendix, we explicitly work out the spinning sum rules by reproducing the one-point correlator from the two-point one in QCD.
\subsection{Energy-energy and energy-charge correlators}

For the the energy-energy correlator for the vector current, using Eqs.~\ref{eq:2ptinclusiveQCD} and~\ref{eq:Hb:Hc:QCD}, the one-point energy correlator at the order of $\alpha_s$ is obtained through the sum rules,
\begin{equation}\label{eec:sumrules:QCD:checked}
	\begin{split}
		H^\mathbb{1}_\mathcal{E} &=\int dz\, \mu(z)H^\mathbb{1}_{\cale\cale} = 1+\frac{\alpha_s}{\pi}~,
		\\[5pt]
		a_{\mathcal{E}} H^\mathbb{1}_{\mathcal{E}}&= \int dz\, \mu(z) \left [ \left ( 1- \frac{3}{2}z \right ) H^c_{\cale\cale} (z) +\left ( 1- \frac{3}{2}(1-z) \right ) H^b_{\cale\cale} (z)  \right ] 
		= -\frac12 + \frac{\alpha_s}{\pi} ~,
	\end{split}
\end{equation}
where $\mu(z) = \{ 1,\, 2z,\, 2(1-z) \}$. Thus, the sum rules determine the $a_\cale$ coefficient of the spinning one-point energy correlator,
\begin{equation}
	a_{\mathcal{E}} 
	= \frac{- \ds\frac{1}{2} + \ds\frac{\alpha_s}{\pi}}{1 + \ds\frac{\alpha_s}{\pi}}
	\approx - \frac{1}{2} + \frac{3\alpha_s}{2\pi}~.
\end{equation}

Consider now the energy-charge correlator for the chiral current sourced by an electroweak gauge boson. This correlator will connect the one-point correlators of both energy and charge via the sum rules. The integration of the inclusive part in Eq.~\ref{eq:eqc:Htr} over $z$ gives rise to
\begin{equation}\label{eq:Heqintegral}
	\begin{split}
		H^\mathbb{1}_{\mathcal{E}} \langle \mathcal{Q} \rangle  =
		H^\mathbb{1}_{\mathcal{Q}} \langle \mathcal{E} \rangle  =
		\int_0^1 dz\, \mu(z) H^\mathbb{1}_{\mathcal{E}\mathcal{Q}}
		=  \frac{q_q + q_{\bar{q}'}}{2}\left( 1 +  \frac{\alpha_s}{\pi}\right)~,
	\end{split}
\end{equation}
where the unit of $\langle \cale \rangle = E = 1$ is assumed and the total charge $\langle \mathcal{Q} \rangle = q_q + q_{\bar{q}'}$. The factor 1/2 with respect to Eq.~\ref{eq:2ptinclusiveQCD} is due to considering a chiral current. 
The final $H^\mathbb{1}_{\mathcal{D}} $ agrees with the total contribution to the one-point energy/charge correlator for the chiral current. For the energy, it is consistent with Eq.~\ref{eec:sumrules:QCD:checked} as expected. 
Even though the bonus sum rule arises when integrating over the energy, due to the integrand being symmetric under $\mathcal{E} \leftrightarrow \mathcal{Q}$, Eq.~\ref{eq:Heqintegral} holds for $\mu(z) = \{ 1,\, 2z,\, 2(1-z) \}$. 
Using the result in Eq.~\ref{eq:eqc:HcHbHx}, the sum rules in Eqs.~\ref{eq:sumrules:integrateD} and~\ref{eq:sumrules:integrateE}  for the spinning correlators are evaluated to be
\begin{equation}
	\begin{split}
		&
		\int_0^1 dz\,  \left [ \left ( 1- \frac{3}{2}z \right )H^c_{\mathcal{E}\mathcal{Q}}(z)
		+ \left ( 1- \frac{3}{2}(1-z) \right )H^b_{\mathcal{E}\mathcal{Q}} (z) \pm 3 \sqrt{z(1-z)} H^x_{\mathcal{E}\mathcal{Q}}(z) \right ]
		\\[10pt]
		& \hspace{0.5cm}= \begin{cases} 
			(a_{\mathcal{E}}H^\mathbb{1}_{\mathcal{E}}) \langle \mathcal{Q} \rangle =
			\ds\frac{q_q + q_{\bar{q}'}}{2}  \left( \displaystyle-\frac12+\frac{\alpha_s}{\pi} \right)\\[10pt]
			(a_{\mathcal{Q}}H^\mathbb{1}_{\mathcal{Q}}) \langle \mathcal{E} \rangle =
			\ds\frac{q_q + q_{\bar{q}'}}{2}  \left( \displaystyle-\frac12+\frac{\alpha_s}{2\pi} \right)
		\end{cases}~,
	\end{split}
\end{equation}
where $+$ ($-$) corresponds to the integration over the charge (energy) detector. 
We see that the spinning one-point energy and charge correlators $a_\cale$ and $a_{\mathcal Q}$ at order $\alpha_s$ are correctly reproduced~\cite{Riembau:2024tom}. 
Finally, we consider the second moment for the energy-charge correlator coming from the momentum conservation. 
The sum rule in Eq.~\ref{eq:sumrules:integrateE} implies that the integration over energy satisfies
\begin{equation}
	\begin{split}
		& \int_0^1 dz\, \mu(z) \left [ \left ( 1- \frac{3}{2}z \right )H^c_{\mathcal{E}\mathcal{Q}}(z)
		+ \left ( 1- \frac{3}{2}(1-z) \right )H^b_{\mathcal{E}\mathcal{Q}} (z) - 3 \sqrt{z(1-z)} H^x_{\mathcal{E}\mathcal{Q}}(z) \right ]
		\\[7pt]
		&\hspace{1cm}= \left ( q_q + q_{\bar{q}'}\right ) \ds\frac12 \displaystyle\frac{\alpha_s}{2\pi} 
		=  (a_{\mathcal{Q}}H^\mathbb{1}_{\mathcal{Q}}) \langle \mathcal{E}\rangle~,
	\end{split}
\end{equation}
while the similar relation when integrating the charge, i.e. with the plus sign, does not hold. 
The $J=1$ component of the one-point charge correlator, using the result in Eq.~\ref{eq:eqc:Hpm}, its integral can evaluated to be
\begin{equation}\label{eq:hphmEQcheck}
	\begin{split}
		 \int dz \left [ \sqrt{1-z} H^+_{\mathcal{E}\mathcal{Q}} (z)  \pm \sqrt{z} H^-_{\mathcal{E}\mathcal{Q}} (z) \right ]~
		= 0~,
	\end{split}
\end{equation}
the contribution with a negative sign is the one associated to the integration over the charge detector, leading to $\langle \mathcal{Q} \rangle H^{(b)}_{\mathcal{E}}$, and it is zero since the one-point energy correlator has no $J=1$ component. That also the positive sign, associated with $\langle \mathcal{E} \rangle H^{(b)}_{\mathcal{Q}}$, vanishes is, as far as we can tell, accidental. It is of course compatible with the fact that $ H^{(b)}_{\mathcal{Q}}=0$ and the contribution to the parity-odd coefficient $b_\mathcal{Q}$ of the one-point correlator at one loop comes from the overall normalization. In fact, the integrands $\sqrt{1-z} H^+_{\mathcal{E}\mathcal{Q}}$ and $\sqrt{z} H^-_{\mathcal{E}\mathcal{Q}}$ in Eq.~\ref{eq:hphmEQcheck} vanish separately. 
In the case of integrating the energy detector, associated with the plus sign in Eq.~\ref{eq:hphmEQcheck}, the bonus relation holds. 

For the completeness, we also derive the one-point energy correlator through the sum rules from the $\mathcal{O}(\alpha_{em})$ contributions to the energy-energy correlator for $\pi^+\pi^-\gamma$ states, given in Eqs.~\ref{eq:sqed:eec:trace} and~\ref{eq:sqed:eec:HcHb}.
\begin{equation}
	\begin{split}
		H^{\mathbb{1}}_\mathcal{E} &=\int dz H^{\mathbb{1}}_{\cale\cale} = \frac{3\alpha}{4\pi}~,
		\\[5pt]
		a_\cale H^{\mathbb{1}}_\cale&= \int dz \left [ \left ( 1- \frac{3}{2}z \right ) H_{\cale\cale} ^c (z) +\left ( 1- \frac{3}{2}(1-z) \right ) H_{\cale\cale} ^b (z)  \right ] 
		= 0~.
	\end{split}
\end{equation}
We check that this vanishing one-loop correction to the spinning correlator is consistent with the direct computation of the one-point energy correlator acting on $\pi^+\pi^-\gamma$ states.

\subsection{Charge-charge correlator}
We can close the loop by showing that the perturbative calculation of the charge-charge correlator satisfies a similar set of sum rules relating it to the same one-point observable. While the charge-charge correlator receives large non-perturbative corrections, the fact that the sum rules relate it with one-point correlators, which are IR-safe, implies that the non-perturbative corrections should be annihilated by the sum rules, which means that their decomposition in an orthogonal basis in $z$ should have no component in the directions given by the sum rules. 

We continue working out the charge-charge correlator for the chiral current as in the decay of a polarized electroweak boson and check explicitly its sum rules.
Only parity even parts for the chiral current are presented. 
The inclusive charge-charge correlator is given by 
\begin{equation}\label{eq:qcd:qqc:trace}
	\begin{split}
		H^\mathbb{1}_{\mathcal{Q}\mathcal{Q}} &= 
		q_q q_{\bar{q}'} \frac{4\alpha_s}{\pi}C_F \left [
		- \frac{1}{8} \frac{z (3 z+2) + 2 \left(z^2+z+1\right) \log (1-z)}{z^3} 
		\right .
		\\[5pt]
		&\left . \hspace{3.0cm}
		- \frac{1}{4} \left [ \frac{\log(1-z)}{1-z} \right ]_+ 
		- \frac{3}{8} \left [ \frac{1}{1-z} \right ]_+  +  \left ( - \frac{\zeta_2}{4} - \frac{1}{4} \right ) \delta(1-z)
		\right ]
		\\[5pt]
		&\hspace{3.0cm} +\delta(z) (q_q^2 + q_{\bar{q}'}^2)  \frac{4\alpha_s}{\pi}C_F \frac{3}{32}~,
	\end{split}
\end{equation}
where $q_q$ and $q_{\bar{q}'}$ are the conserved global charges of the quark $q$ and anti-quark $q'$, respectively.
The coefficient functions for spinning tensors in charge-charge correlator are given by
\begin{equation}\label{eq:qcd:qqc:HcHb}
	\begin{split}
		H^c_{\mathcal{Q}\mathcal{Q}} (z) &= q_q q_{\bar{q}'} \frac{4\alpha_s}{\pi} C_F 
		\left [ 
		\frac{1}{16}  \frac{ z (z+2)+2 \log (1-z)}{ z^3}  \right ]
		+  (q_q^2 + q_{\bar{q}'}^2) \frac{4\alpha_s}{\pi} C_F \frac{3}{64}  \delta(z)~,
		\\[5pt]
		H^b_{\mathcal{Q}\mathcal{Q}} (z) &= q_q q_{\bar{q}'} \frac{4\alpha_s}{\pi} C_F
		\left [
		\frac{1}{8} \frac{z+(z+1) \log (1-z)}{ z^2} \right .
		\\[5pt]
		&\hspace{3cm}  \left . 
		+ \frac{3}{16} \left [ \frac{1}{1-z} \right ]_+ +  \frac{1}{8} \left [ \frac{\log(1-z)}{1-z} \right ]_+
		-  \frac{1}{2} \left ( - \frac{\zeta_2}{4} - \frac{1}{4} \right ) \delta(1-z)  \right ]~.
	\end{split}
\end{equation}
The total flux of the one-point charge correlator can be derived from the sum rule for the inclusive charge-charge correlator in Eq.~\ref{eq:qcd:qqc:trace}. 
The integration of the two-point charge correlator over $z$, including the Born contribution 
$H^{\mathbb{1}}_{\mathcal{Q}\mathcal{Q}} = \frac{1}{2} (q_q^2 + q_{\bar{q}'}^2) \delta(z) + \frac{1}{2}2 q_q q_{\bar{q}'} \delta(1-z)$, gives rise to, up to the order of $\alpha_s$,
\begin{equation}
	\begin{split}
		H^{\mathbb{1}}_{\mathcal{Q}} (q_q + q_{\bar{q}'})  &=
		\int_0^1 dz\, H^{\mathbb{1}}_{\mathcal{Q}\mathcal{Q}}
		= \frac{1}{2}  \left ( q_q + q_{\bar{q}'}\right )^2\left( 1 +  \frac{\alpha_s}{\pi} \right)~. 
	\end{split}
\end{equation}
Consistently, even though the two-point correlator only has terms proportional to either $q_q q_{\bar{q}'}$ or $q_q^2 + q_{\bar{q}'}^2$, the sum rule is proportional to the charge squared $\left ( q_q + q_{\bar{q}'}\right )^2$. 
Similarly, the coefficient function for the spinning tensor in the one-point charge correlator can be obtained using the sum rule for those in the charge-charge correlator in Eq.~\ref{eq:qcd:qqc:HcHb}.  
\begin{equation}
	\begin{split}
		a_{\mathcal{Q}} H^{\mathbb{1}}_{\mathcal{Q}} \sum_{i=q,\bar{q}'} q_i  &=
		\int_0^1 dz\,  \left [ \left ( 1- \frac{3}{2}z \right )H_{\mathcal{Q}\mathcal{Q}}^c(z)
		+ \left ( 1- \frac{3}{2}(1-z) \right )H_{\mathcal{Q}\mathcal{Q}}^b (z) \right ]
		\\[5pt]
		&= \frac{1}{2}  \left ( q_q + q_{\bar{q}'}\right )^2\left( -\frac12 +  \frac{\alpha_s}{2\pi}\right)~.
	\end{split}
\end{equation}
Taking ratio of two sum rules recovers the one-point charge correlator parameter for the chiral current in~\cite{Riembau:2024tom}
\be
a_{\mathcal{Q}} 
= \frac{ -\ds\frac{1}{2}+ \ds\frac{\alpha_s}{2\pi}}{1 + \ds\frac{\alpha_s}{\pi}}
\approx - \frac{1}{2} + \frac{\alpha_s}{\pi}~.
\ee

{\small
\bibliography{lit}{}
\bibliographystyle{JHEP}}

\end{document}